\documentclass[a4paper,11pt]{article}
\pdfoutput=1

\usepackage[margin=0.8in]{geometry}

\usepackage[utf8]{inputenc}
\usepackage[english]{babel}
\usepackage{csquotes} 
\usepackage[T1]{fontenc}

\usepackage{pifont}

\usepackage{amsmath,amssymb,amsthm}
\usepackage{mathtools}

\usepackage[
    labelfont=bf %
]{caption}
\usepackage{float}
\usepackage{subcaption}
\usepackage{tikz}
\usetikzlibrary{decorations.markings,arrows,patterns,patterns.meta,shapes}

\usepackage{pgfplots}
\pgfplotsset{compat=1.18}
\pgfdeclareplotmark{donut}{
    \pgfseteorule
    \pgfpathellipse{\pgfpoint{0cm}{0cm}}
             {\pgfpoint{3pt}{0cm}}
             {\pgfpoint{0cm}{1pt}}
     \pgfpathellipse{\pgfpoint{0cm}{0cm}}
             {\pgfpoint{4pt}{0cm}}
             {\pgfpoint{0cm}{2pt}}
    \pgfusepath{fill}
}
\pgfdeclareplotmark{donut*}{
     \pgfpathellipse{\pgfpoint{0cm}{0cm}}
             {\pgfpoint{4pt}{0cm}}
             {\pgfpoint{0cm}{2pt}}
    \pgfusepath{fill}
}

\usepackage[outline]{contour}
\contourlength{0.2em}

\usepackage[compat=0.6]{yquant}
\useyquantlanguage{groups}

\usepackage{url}

\usepackage{mathrsfs}
\usepackage{dsfont}
\usepackage{setspace}
\usepackage[export]{adjustbox}
\usepackage{makecell}
\usepackage{tqft}
\usepackage[inline]{enumitem}

\usepackage{graphicx}

\usepackage{xcolor}
\definecolor{darkblue}{RGB}{0,0,128}
\definecolor{darkgreen}{RGB}{0,150,0}

\definecolor{dark-red}{rgb}{0.4,0.15,0.15}
\definecolor{dark-blue}{rgb}{0.15,0.15,0.4}
\definecolor{dark-green}{rgb}{0.15,0.4,0.15}
\definecolor{medium-blue}{rgb}{0,0,0.5}

\definecolor{drawing-blue}{HTML}{c1ecff}
\definecolor{drawing-yellow}{HTML}{f6dab5}
\definecolor{drawing-red}{HTML}{e5bdd2}

\usepackage[%
	backend=biber,
	style=alphabetic,
	url=false,
	isbn=true,
	maxbibnames=10,
	backref, 
]{biblatex}
\usepackage[pdfusetitle]{hyperref}
\hypersetup{
    breaklinks,
    colorlinks,
	linkcolor={dark-blue},
	citecolor={dark-green},
	urlcolor={medium-blue},
    filecolor=red,
}

\usepackage[nameinlink,capitalise]{cleveref}
\crefname{figure}{Figure}{Figures}

\crefname{conjecture}{Conjecture}{Conjectures}
\Crefname{conjecture}{Conjecture}{Conjectures}

\newtheorem*{theorem*}{Theorem}

\usepackage{thmtools}
\usepackage{thm-restate}

\usepackage{braket}
\usepackage{dsfont}
\usepackage[normalem]{ulem}
\usepackage{algorithm}
\usepackage{algpseudocode}

\newcommand{\idty}{\mathds{1}}

\let\emptyset\varnothing

\usepackage{stmaryrd}

\DeclarePairedDelimiter\floor{\lfloor}{\rfloor}

\newcommand{\sansserif}[1]{%
  \ifmmode
  \mathsf{#1}%
  \else
   \textsf{#1}%
  \fi
}

\usepackage[mathscr]{euscript}

\usepackage{mathtools}

\DeclarePairedDelimiterXPP\bigo[1]{O}{(}{)}{}{#1}
\DeclarePairedDelimiterXPP\littleo[1]{o}{(}{)}{}{#1}
\DeclarePairedDelimiterXPP\bigomega[1]{$\Omega$}{(}{)}{}{#1}
\DeclarePairedDelimiterXPP\bigtheta[1]{$\Theta$}{(}{)}{}{#1}

\DeclarePairedDelimiter\paren{(}{)}
\DeclarePairedDelimiter\genSet{\langle}{\rangle}

\newcommand{\smdash}{\hspace{0pt}-\hspace{0pt}}

\graphicspath{{../}{./}{./tikz}{figs/}}

\renewcommand{\sec}[1]{\hyperref[sec:#1]{Section~\ref*{sec:#1}}}
\newcommand{\app}[1]{\hyperref[app:#1]{Appendix~\ref*{app:#1}}}
\newcommand{\ssec}[1]{\hyperref[ssec:#1]{Subsection~\ref*{ssec:#1}}}
\newcommand{\fig}[1]{\hyperref[fig:#1]{Figure~\ref*{fig:#1}}}
\newcommand{\tab}[1]{\hyperref[tab:#1]{Table~\ref*{tab:#1}}}
\newcommand{\lemm}[1]{\hyperref[lemm:#1]{Lemma~\ref*{lemm:#1}}}
\newcommand{\propos}[1]{\hyperref[propos:#1]{Proposition~\ref*{propos:#1}}}
\newcommand{\thm}[1]{\hyperref[thm:#1]{Theorem~\ref*{thm:#1}}}
\newcommand{\alg}[1]{\hyperref[alg:#1]{Algorithm~\ref*{alg:#1}}}
\newcommand{\defn}[1]{\hyperref[defn:#1]{Definition~\ref*{defn:#1}}}

\newcommand{\be}{\begin{equation}}
\newcommand{\ee}{\end{equation}}
\newcommand{\bea}{\begin{eqnarray}}
\newcommand{\eea}{\end{eqnarray}}

\newcommand{\bem}{\begin{multline}}
\newcommand{\eem}{\end{multline}}

\usepackage{array}
\usepackage{booktabs}
\usepackage{makecell}
\usepackage{diagbox}
\usepackage{multicol}
\usepackage{multirow}

\usepackage{siunitx}

\newcolumntype{?}{!{\vrule width 1pt}}

\usepackage{authblk}

\title{Tour de gross: A modular quantum computer based on bivariate bicycle codes}
\author[1]{Theodore J. Yoder}
\author[1]{Eddie Schoute}
\author[1]{Patrick Rall}
\author[1]{Emily Pritchett}
\author[1]{Jay M. Gambetta}
\author[1]{Andrew W. Cross}
\author[1]{Malcolm Carroll}
\author[1]{Michael E. Beverland\thanks{%
    Authors in reverse alphabetical order.
}}

\affil[1]{IBM Quantum}

\date{}

\begin{document}
\maketitle

\begin{abstract}
We present the \emph{bicycle architecture}, a modular quantum computing framework based on high-rate, low-overhead quantum LDPC codes identified in prior work.
For two specific bivariate bicycle codes with distances 12 and 18, we construct explicit fault-tolerant logical instruction sets and estimate the logical error rate of each instruction under circuit noise.
We develop a compilation strategy adapted to the constraints of the bicycle architecture, enabling large-scale universal quantum circuit execution. 
Integrating these components, we perform end-to-end resource estimates demonstrating that an order of magnitude larger logical circuits can be implemented with a given number of physical qubits on a bicycle architecture than on surface code architectures.  
We anticipate further improvements through advances in code constructions, circuit designs, and compilation techniques.

\end{abstract}

\clearpage
\tableofcontents

\section{Introduction}
\label{sec:intro}

Quantum computing offers the possibility to solve certain problems of interest, such as period-finding \cite{shorfactoring}, quantum chemistry \cite{BauerChemistry,MottaEStructure}, and dynamic simulation \cite{utility,ShinjoDynamics}, more efficiently than the best known algorithms on classical computers.
The discovery that quantum computers can operate reliably despite noise~\cite{shor1996fault} sparked the field of fault-tolerant quantum computing~\cite{roffe2019quantum,gottesman2022opportunities,campbell2024series}.
Translating fault-tolerant theory into practice requires a \emph{fault-tolerant architecture} specifying qubit connectivity, allowed operations with their error rates and durations, and consistent, explicit procedures for fault-tolerant error correction and logical operations.
In this work, we propose a novel fault-tolerant architecture --  {\it the bicycle architecture}  --  that builds on recent advancements in quantum Low-Density Parity-Check (LDPC) codes \cite{bravyi2024high}.

\begin{figure}[h!]
    \begin{subcaptionblock}{0.4\linewidth}
        \includegraphics[width=\linewidth]{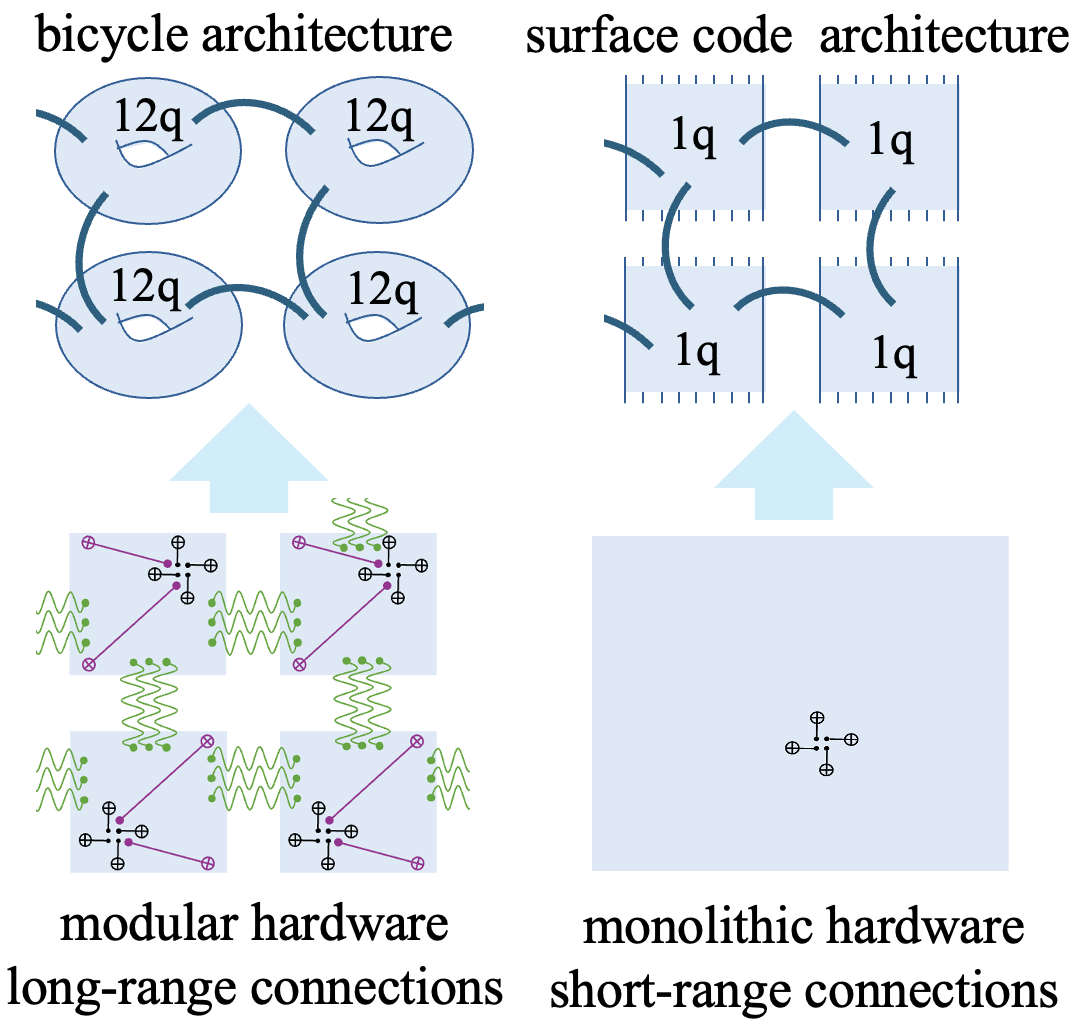}%
        \caption{Quantum architectures}%
        \label{fig:quantumArchitectures}
    \end{subcaptionblock}%
    \hfill%
    \begin{subcaptionblock}{0.58\linewidth}
        \begin{tikzpicture}
            \begin{loglogaxis}[
                xlabel={logical qubits},
                ylabel={logical T count},
                grid=major,
                scatter/classes={
                    gross={mark=donut*,red},
                    twogross={mark=donut,blue},
                    surface={mark=square*,black}
                }
                 , legend style={
                    cells={anchor=east},
                    legend pos=outer north east,
                },
                legend image code/.code={
                    \draw[only marks] plot coordinates {(0cm,0cm)};
                },
                xmin=10,
                xmax=3e4,
                ymin=1,
                ymax=1e17,
                width=\linewidth,
                height=2.6in,
            ]

            \addplot[no marks] coordinates {(3e4, 10)} node[anchor=east] {\contour{white}{classically simulable}};
            
            \newcommand\patternopacity{0.25}
            \addplot[fill=gray, fill opacity=\patternopacity, draw=none, pattern=dots] coordinates {
                (1e20,1e20)
                (75,1e20)
                (75,75)
                (1e20,75)
            };
            \addplot[no marks] coordinates {(3e4, 1e4)} node[anchor=east] {\contour{white}{unexplored classically}};
    
            \addplot[fill=blue, fill opacity=\patternopacity, pattern={Dots[xshift=1.5pt, yshift=1.5pt]}, draw=none] coordinates {
                (100, 1e20)
                (100, 1e12)
                (101, 7.2e6)
                (230, 2.4e6)
                (1e20, 2.4e6)
                (1e20, 1e20)
            };
            \addplot[no marks] coordinates {(3e4, 3e7)} node[anchor=east] {\contour{white}{scientific toy models}};
    
            \addplot[pattern=north west lines, fill opacity=\patternopacity, opacity=\patternopacity] coordinates {
                (1e20, 1e20)
                (1014, 1e20)
                (1014, 7.5e10)
                (3078, 4e8)
                (50000, 3.48e8)
                (1e20, 3.48e8)
            };
            \addplot[no marks] coordinates {(3e4, 7e9)} node[anchor=east] {\contour{white}{cryptography}};
        
            \addplot[fill=red, fill opacity=\patternopacity, pattern=north east lines, opacity=\patternopacity] coordinates {
                (1e20,1e20)
                (135, 1e20)
                (135, 3.3e14)
                (1366, 6.4e10)
                (2142, 4.24e10)
                (1e20, 4.24e10)
            };
            \addplot[no marks] coordinates {(3e4, 1e12)} node[anchor=east] {\contour{white}{quantum chemistry}};
        
        
            \addplot[
                ->,
                thick,
                shorten >=3pt,
                scatter,
                scatter src=explicit symbolic,
            ] coordinates {
                (45, 7) [surface]
                (121, 13.595802668072071)  [gross]
            } node [pos=0.5,label={5k, $10^{-3}$}] {};
        
            \addplot[
                ->,
                thick,
                shorten >=4pt,
                scatter,
                scatter src=explicit symbolic,
            ] coordinates {
                (42, 2.5e6) [surface]
                (704, 3471247.948411036) [twogross]
            } node [pos=0.75, label={\contour{white}{50k, $10^{-3}$}}] {};

            \addplot[
                ->,
                thick,
                shorten >=4pt,
                scatter,
                scatter src=explicit symbolic,
            ] coordinates {
                (21, 3.8e5) [surface]
                (110, 222089.62968351436) [gross]
            } node [pos=0.5, label={-90:5k,  $10^{-4}$}] {};

            \addplot[
                ->,
                thick,
                shorten >=4pt,
                scatter,
                scatter src=explicit symbolic,
            ] coordinates {
                (27, 1.2e15) [surface]
                (440, 7015434087082015.0) [twogross]
            } node [pos=0.25, label={-90:\contour{white}{$q=50$k, $p=10^{-4}$}}] {};
        
            \addplot[
                ->,
                thick,
                shorten >=4pt,
                scatter,
                scatter src=explicit symbolic,
            ] coordinates {
                (416, 7.9e13) [surface]
                (6886, 448297625597020.8) [twogross]
            } node [pos=0.5, label={\contour{white}{500k,  $10^{-4}$}}] {};
    
            \addplot[
                very thick,
                mark=asterisk,
            ] coordinates {
                (100, 19036800) 
            }%
            node [pos=0.0, label={\contour{white}{TFIM}~~~~~}] {};

            \end{loglogaxis}
        \end{tikzpicture}
        \caption{Logical capability estimates}%
        \label{fig:logicalCapabilities}
    \end{subcaptionblock}
\caption{%
    \subref{fig:quantumArchitectures} The bicycle architecture uses a long-range-connected modular hardware with quantum LDPC codes, in contrast to the conventional surface code architecture based on short-range, monolithic hardware.
    \subref{fig:logicalCapabilities} Comparison of logical capabilities at physical qubit counts ($q$: 5k, 50k, 500k) and error rates ($p$: $10^{-3}$ and $10^{-4}$). 
    The bicycle architecture (gross: filled red ellipse; two-gross: hollow blue ellipse) can reliably execute circuits with an order of magnitude more logical qubits than the conventional surface code architecture using the same physical resources ($q$, $p$)
    and with a similar number of logical T gates (black square, connected with arrow) .
}
\label{fig:logical-circuit-capabilities}
\end{figure}

At the hardware level, the bicycle architecture differs from conventional surface code architectures by integrating long-range qubit connectivity,
see \cref{fig:quantumArchitectures}.
The advantages of this capability are two-fold.
Firstly, long-range qubit connectivity facilitates modularity~\cite{monroe2014large,brecht2016multilayer,bombin2021interleaving}, allowing for individual modules to be scalably optimized.
In contrast to monolithic architectures, these 
modules can be interconnected or swapped out as needed without compromising overall performance.
Secondly, long-range connections capitalize on recent progress in low-overhead quantum LDPC codes which appear impractical with short-range connections alone \cite{bravyi2010tradeoffs,delfosse2021bounds} despite recent theoretical progress \cite{pattison2023hierarchical,gidney2025constant}.

By employing bivariate bicycle codes \cite{kovalev2013quantum,panteleev2021degenerate,bravyi2024high}, the bicycle architecture significantly reduces estimated qubit requirements for quantum computers executing specified algorithms, as shown in \cref{fig:logicalCapabilities}.
Relative to conventional surface code architectures, it provides approximately an order of magnitude more logical qubits for the same physical resources, thereby enabling access to a broader range of applications.
(The application regions in \cref{fig:logicalCapabilities} reflect coarse resource estimates and are expected to improve with continued algorithmic advances.)
In \sec{results}, we show that scientifically interesting problems, such as transverse-field 
Ising model (TFIM) simulation, can be accessed with as few as $8138 \approx \text{8k}$ physical qubits if sufficiently low physical error rates $p \approx 7 \times 10^{-4}$ can be achieved.

Two types of long-range couplers are necessary to realize the bicycle architecture. 
Within a module, each qubit is coupled to a small number of other qubits anywhere in the module. The length of these connections is a function of the module size~\cite{bravyi_future_2022}. In the bicycle architecture presented here, modules contain less than 1000 physical qubits with the longest intramodule couplers tens of lattice sites long using appropriate qubit placement on a 2-dimensional grid \cite{poole2024architecture}. To create inter-module entanglement, long-range couplers between modules are needed, the number of which scales with the code distance.  Implementing long-range couplers in real physical systems presents substantial challenges; however, there is a substantial history of research in this area for superconducting qubits~\cite{campagne2018deterministic,chou2018deterministic,kurpiers2018deterministic,leung2019deterministic,magnard2020microwave,zhong2021deterministic,storz2023loophole,niu2023low} and, moreover, state transfer with fidelity $98.8\%$ between transmon qubits over 0.6 meters was recently demonstrated~\cite{lcoupler2025}.

We establish criteria we believe desirable of any fault-tolerant architecture proposal in \sec{criteria} before defining the bicycle architecture in \sec{bicycle-arch}.
In \sec{ft-ops}, the fault-tolerant implementation of each \emph{bicycle instruction} in the bicycle architecture is described along with a performance analysis. 
In \sec{compiling}, we explore techniques for compiling algorithms to this universal set of bicycle instructions. 
\sec{results} compares resource estimates needed to implement example logical circuits with our architecture with the conventional surface code architecture.
These estimates are informative for evaluating our approach's long-term feasibility and for identifying aspects that can be improved.
We conclude in \sec{conclusion} by highlighting major opportunities that could have significant impact if solved.

It is an exciting time for quantum error correction, with emerging hardware capabilities aligning with theoretical breakthroughs leveraging long-range connectivity.  
While already promising, we expect the bicycle architecture---like surface code architectures over the past decades---to continue advancing, with this work serving as a foundational baseline.

\subsection{Fault-tolerant architecture criteria}
\label{sec:criteria}

For a fault-tolerant architecture to be considered a reliable platform for quantum computing, we propose the following sufficient criteria for scalable fault-tolerance:
\begin{enumerate}[noitemsep,label={(\roman*)}]
 
\item {\it Fault-tolerant} -- logical errors are suppressed enough for meaningful algorithms to succeed.

\item {\it Addressable} -- individual logical qubits can be prepared or measured throughout the computation.   

\item {\it Universal} -- a universal set of quantum instructions can be applied to the logical qubits.

\item {\it Adaptive}  -- measurements are real-time decoded and can alter subsequent quantum instructions.

\item {\it Modular} -- the hardware is distributed across a set of replaceable modules connected quantumly.

\item {\it Efficient} --  meaningful algorithms can be executed with reasonable physical resources. 

\end{enumerate}

The last criterion is crucial to use a quantum computer to solve large scale problems.
How efficiently a fault-tolerant architecture scales is affected by how practically all the other criteria are satisfied, including choice of hardware platform (solid state, atomic, photonic, etc), the choice of error-correcting code family (surface code, quantum LDPC, etc.), the specific methods of performing logical gates and measurements, the efficiency of decoding and speed of the control stack, and the probability a module is fabricated successfully. 

The surface code emerged early as a promising approach to fault-tolerance due to its high error threshold, straightforward decoding, and crucially, its implementation using simple 2D connectivity.
This makes it amenable to a wide range of hardware platforms, and many small surface codes have been demonstrated \cite{weight_four_IBM,Marques_2021,Krinner_2022,Zhao_2022,Google2022suppressingquantumerrorsscaling,chen2022calibrated,sundaresan2023demonstrating,Google2024quantumerrorcorrectionsurface,Bluvstein_2023,Berthusen_2024,ELQ2025}.
Surface code architectures fulfill the first four criteria effectively, and if some long-range connections are allowed, they can be made to satisfy the modularity condition too~\cite{nickerson2013topological,li2016hierarchical,ramette2024fault,jacinto2025network}.
However, they falter with the efficiency criterion due to an unfavorable physical-to-logical qubit ratio.

The bicycle architecture proposed here meets all the proposed criteria.
(i) We consider two bivariate bicycle codes, \emph{gross} and \emph{two-gross}, each encoding 12 logical qubits. 
The two-gross code offers better error suppression but needs double the number of physical qubits. 
These codes, feasible on moderately connected qubit lattices with some long-range connections \cite{bravyi2024high}, allow parallel syndrome measurement circuits which helps ensure low logical error rates. 
(ii) While addressing individual logical qubits was an early issue for quantum LDPC codes, it is now known how to generalize lattice surgery to measure arbitrary logical Pauli operators~\cite{CKBB22,bravyi2024high,cowtan2024ssip,zhang2025time,CHRY24,WY24,swaroop2024universal,ide2024fault,cowtan2025,he2025extractors} by introducing for each code a set of auxiliary qubits that we refer to as a logical processing unit (LPU). 
(iii) Universal operations are applied to logical qubits by teleporting T states synthesized in T factory modules, similar to surface code architecture proposals~\cite{GSJ2024cultivation,bravyi2005universal,knill2005quantum}.
(iv) Fast classical processing, including decoding algorithms~\cite{RelayPaper} implemented in classical hardware such as ASICs or FPGAs, can modify the execution of logical operations in real time, making T state cultivation and injection possible~\cite{terhal2015quantum}. 
(v) Generalized lattice surgery~\cite{CKBB22} approaches can be extended~\cite{CHRY24,swaroop2024universal} to enable logical entangling operations between quantum LDPC codes in separate modules. 
(vi) The bicycle architecture has approximately an order of magnitude lower qubit overhead than surface codes with comparable error suppression \cite{bravyi2024high}. 
Additional qubit overhead for logical operations within and between codes remains considerably smaller than that of the codes themselves \cite{CHRY24,WY24}.

\subsection{The bicycle architecture}
\label{sec:bicycle-arch}

\begin{figure}[h]
\centering
    \begin{subcaptiongroup}
	\includegraphics{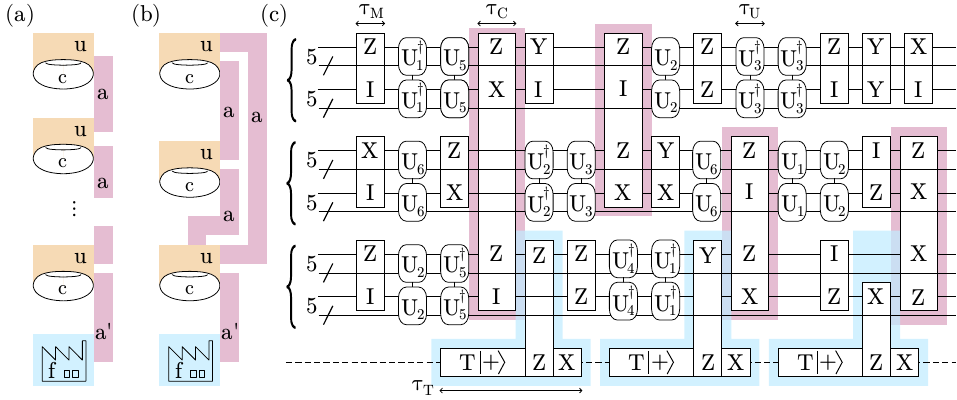}
    \phantomcaption\label{fig:linearConnectivity}
    \phantomcaption\label{fig:triangleConnectivity}
    \phantomcaption\label{fig:bicycleInstructions}
    \end{subcaptiongroup}
    \captionsetup{subrefformat=parens}
\caption{
\textbf{Module components.}
Different module connectivity can be considered, such as \subref{fig:linearConnectivity} a linear arrangement and \subref{fig:triangleConnectivity} a triangular arrangement.
Each physical qubit is part of exactly one of the pictured components: a code $c$, an LPU $u$, a factory $f$, a code-code adapter $a$, or a code-factory adapter $a'$.
\textbf{Bicycle Circuit.} \subref{fig:bicycleInstructions}
An example circuit expressed in the universal set of bicycle instructions.
Each code module hosts $k=12$ logical qubits (black wires).
Multi-qubit Pauli measurements (square boxes) are allowed:
(i) between the first and seventh qubits within a module (duration $\tau_{\mathrm M}$), or
(ii) between the first and seventh qubits of two connected modules (duration $\tau_{\mathrm C}$; red highlight).
A T factory can inject a T state on the first or seventh qubit of any adjacent code module over time $\tau_{\mathrm T} \geq \tau_{\mathrm C}$ (blue highlight), during which the factory is occupied.
\emph{Shift automorphisms} can be applied, each acting as $U_i \otimes U_i$ on the 12 logical qubits in a code where $U_i$ is a six-qubit unitary (duration $\tau_U$).
}\label{fig:gross-architecture}
\end{figure}

We propose the \emph{bicycle architecture}, built from a set of connected code modules and T factory modules, each of which requires physical qubits and connectivity as illustrated in \cref{fig:linearConnectivity,fig:triangleConnectivity}.
Classical communication is assumed to be free and noiseless.
Between modules, entangling components can be long-range, supporting various types of module connectivity.

We consider two choices of code module, the gross code and the two-gross code,
which each store 12 logical qubits,
and a set of compatible T factory modules.
\Cref{tab:qubit-counts} lists the physical qubit count for each component for each module type.
We estimate these qubit counts in \sec{ft-ops}.

\begin{table}[htbp]
    \centering
    \begin{tabular}{lrrrrrrr}\toprule
                    &       &       &       &   \multicolumn{2}{c}{$f$} &  \multicolumn{2}{c}{$a'$}   \\ \cmidrule(r){5-6}  \cmidrule(l){7-8}
        code type   & $c$   & $u$   & $a$   &    $p=10^{-3}$ & $p=10^{-4}$  &  $p=10^{-3}$ & $p=10^{-4}$ \\  \midrule
        gross       & 288   &  90   &  22   &     454  &  810               &   29 & 13     \\
        two-gross   &  576  &  158  & 34    &     463  &  18,600            &   29 & 49    \\
        \bottomrule
    \end{tabular}
    \caption{\textbf{Module component physical qubit counts.} 
    For the code module, $c$,
    the LPU, $u$,
    the factory, $f$,
    the code-code adapter, $a$,
    and the code-factory adapter, $a'$.
    Large T factory qubit counts $f$ for $p = 10^{-4}$ result from our choice to use distillation there, as cultivation protocols are not yet well benchmarked in this regime~\cite{GSJ2024cultivation} (see \sec{T-factories}).
    }
    \label{tab:qubit-counts}
\end{table}

The bicycle architecture supports a set of logical operations
called \emph{bicycle instructions}, $\mathcal I$.
We fix a basis for the logical operators of each code module, and label the logical qubits in each module 1 through 12.
We call qubit 1 the \emph{pivot qubit},
which has an important role as an ancilla for the synthesis of unitary operations on each module.
Now, the bicycle instructions are:
\begin{description}[noitemsep,font=\normalfont\itshape]
    \item[Shift automorphisms:]
    Twelve instructions, consisting of six code automorphisms and their inverses
    $\{\bar U_1,$ $\ldots,$ $\bar U_6,$ $\bar U^\dagger_1,$ $\ldots,$ $\bar U^\dagger_6\}$,
    which act as $\bar U \otimes \bar U$, with each $\bar U$ acting independently on qubits 1--6 and qubits 7--12 respectively.
    The action of $\bar U_i$ is some CNOT circuit,
    explicitly specified in \app{automorphisms}.
    Any element of the group of 36 automorphism unitaries,
    $\mathcal A \coloneqq \genSet*{\bar U_1, \bar U_2, \ldots, \bar U_6}$,
    is the product of at most two generators for both gross and two-gross codes.

    \item[In-module measurements:]
    Fifteen instructions, consisting of measurements of the non-trivial Paulis in the group
    $\mathcal M \coloneqq \genSet*{\bar X_1, \bar Z_1,\bar X_7, \bar Z_7}$.\footnote{This is a minor abuse of notation. Despite being in the group $\genSet*{\bar X_1,\bar X_7,\bar Z_1,\bar Z_7}$, non-Hermitian operators like $\bar X_1\bar Z_1=-i\bar Y_1$ cannot be measured. What we mean is that we can measure the related Hermitian operators, such as $\bar Y_1$.} 
    
    \item[Inter-module measurements:]
    Thirty-six instructions, consisting of measurements of 36 Pauli products from the set $(\genSet*{\bar X_1,\bar Z_7}\cup \genSet*{\bar X_7,\bar Z_1})^{\otimes2}$ where the two Paulis in the tensor product operate non-trivially on logical qubits in different, but connected, code modules.

    \item[T injection:]
    Six instructions, consisting of preparing a T state, $T\ket{+} \coloneqq e^{i \frac{\pi}{8}Z} \ket{+}$,
    in a factory module, then an inter-module $\bar{Z}_\text{T}\otimes \bar{P}$ measurement across the factory and a connected code module, then a destructive $\bar{X}_\text{T}$ factory measurement.
    There are six cases: $\bar{P} \in \set{\bar X_1, \bar Y_1, \bar Z_1, \bar X_7, \bar Y_7, \bar Z_7}$.
\end{description}

We provide the time requirements and logical error rates of the bicycle instructions in \tab{logical_ops} and illustrate them with an example circuit in \cref{fig:bicycleInstructions}. 
As Calderbank-Shor-Steane (CSS) codes, the bivariate bicycle codes in this work enable fault-tolerant preparation of all logical qubits in $\ket{0}^{\otimes 12}$ or all in $\ket{+}^{\otimes 12}$, and destructive readout of all logical qubits in the $X$ basis or all in the $Z$ basis. 
We assume no additional time or noise is incurred by these CSS preparation and measurement operations, similar to the surface code~\cite{litinski2019game}, and omit them from \cref{tab:logical_ops}.

\begin{table}[htb]
    \centering
    \begin{tabular}{lclrrr}
        \toprule
           & &  &  & \multicolumn{2}{c}{$P_i$ (logical error rate)} \\ \cmidrule{5-6}
        instruction & $i$  &  module &     $\tau_i$ (timesteps)      & $p=10^{-3}$ & $p=10^{-4}$ \\  \midrule
        idle & $I$   & gross code      & $8$ & $10^{-8.8 \pm 0.2}$ & $10^{-14.8 \pm 0.4 }$ \\
             && two-gross code  & $8$ & $10^{ -20.1 \pm 0.5 }$ & $10^{ -38.3 \pm 0.9}$ \\
        shift automorphism & $U$ &   gross code      &  $14$ & $10^{ -6.4 \pm 0.2 }$ & $10^{ -12.2 \pm 0.5 }$ \\
           && two-gross code  & $14$ & $10^{ -14.5 \pm 0.4 }$ & $10^{ -37 \pm 1 }$ \\
        in-module meas. & $M$ & gross code         & $120$ & $10^{ -5.0 \pm 0.1 }$ & $10^{ -9.0 \pm 0.2 }$ \\
        && two-gross code  & $216$ & [$10^{-11}$] & [$10^{-20}$] \\
        inter-module meas. & $C$ & gross code      & $120$ &  $10^{-2.7 \pm 0.1 }$ & $10^{ -7.3 \pm 0.3  }$ \\  
          && two-gross code &   $216$ & [$10^{-9}$] & [$10^{-18}$] \\
        T injection & $T$    
                        
                        &  $10^{-3}$ gross factory & $351 + 120$ &  $10^{-5.5} + P_C$ & --  \\
                        && $10^{-3}$ two-gross factory &  $2167 + 216$ &  $10^{-7.7} + P_C$ & --  \\
                        && $10^{-4}$ gross factory &  $73+ 120$ &  -- & $10^{-7.4} + P_C$  \\
                        && $10^{-4}$ two-gross factory & $407 + 216$ &  -- & $10^{-24.4} + P_C$  \\
        \bottomrule
    \end{tabular}
    \caption{\textbf{Bicycle instruction properties.} 
        Each idle or bicycle instruction $i \in \{I, U, M, C, T\}$ is characterized by a duration $\tau_i$ where a timestep is a physical gate and a logical error rate $P_i$, defined as the probability that the instruction fails under physical circuit noise with error rate $p$ (using preliminary data from \cref{sec:ft-ops}).
        We assume each physical operation takes the same time, which we refer to as a timestep.
        We assume independent logical errors and say that an execution fails if any logical error occurs.
        Error bars are obtained from a bootstrapping approach based on Monte Carlo uncertainty. 
        Values in square braces are assumed without data, leaving the justification that these accuracies can be achieved to future work. 
        See \cref{app:logical-op-simulations}.
    }
    \label{tab:logical_ops}
\end{table}

\section{Fault-tolerant implementation of the bicycle architecture}
\label{sec:ft-ops}

In this section we provide explicit implementations of the bicycle architecture in \fig{gross-architecture} using two specific bivariate bicycle codes, the gross code and the two-gross code.
The approaches we present here represent our current proposals, but they will inevitably evolve over time.

The bicycle architecture is based heavily around prior work on quantum LDPC surgery \cite{CHRY24,WY24,ide2024fault,swaroop2024universal} and can be considered an instantiation of the more general extractor architecture proposed by He et al.~\cite{he2025extractors}. However, the present work and Ref.~\cite{he2025extractors} have complementary objectives. Whereas Ref.~\cite{he2025extractors} is focused on designing at a high-level an architecture applicable to any quantum LDPC code, here we focus on benchmarking an explicit and practical scheme for a specific set of example codes.

To accomplish our present objective, we incorporate quantum LDPC surgery ideas into an architecture based on bivariate bicycle codes with concrete realizations of all the necessary components. This features a smaller and more capable (compared with Ref.~\cite{CHRY24}) surgery construction for one of our codes---the $[[144,12,12]]$ gross code---and an entirely new construction for a larger code---the $[[288,12,18]]$ two-gross code. This is complemented by explicit gadgets for connecting bivariate bicycle code modules and magic state factories based on the surface code using the concepts of bridges and adapters from Refs.~\cite{CHRY24,swaroop2024universal}. 
These connections are implemented at the physical level by Bell pair generation, as in the model considered by Ref.~\cite{jacinto2025network}. 

In what follows, we first review the bivariate bicycle codes, specifying the two codes of interest: the gross and two-gross codes in \sec{gross-review}.
Then, in \sec{shift_automorphisms}, \sec{in-module}, \sec{inter-module} and \sec{T-factories}, we specify our proposed implementation of each of the bicycle instructions in these codes.
The culmination of this section is the validation of the fault-tolerance and performance of each bicycle instruction in \sec{instruction-validation} under circuit noise with error rate $p$, before briefly highlighting opportunities for improvement to our approaches in \sec{future-directions-ft}.

\subsection{Review of bivariate bicycle codes}\label{sec:gross-review}

\begin{figure}[t]
\centering
\begin{subfigure}{\textwidth}
\centering
\usetikzlibrary{positioning,automata}
\usetikzlibrary{arrows,calc}
\usetikzlibrary{fit}


\definecolor{zcheck}{RGB}{0,203,147}
\definecolor{xcheck}{RGB}{255,111,147}

\definecolor{zlabel}{RGB}{170,255,220}
\definecolor{xlabel}{RGB}{255,220,200}

\tikzstyle{none}=[fill=none,draw=none,shape=circle]
\tikzstyle{Xcheck}=[fill={rgb,255: red,255; green,111; blue,147}, draw=none, shape=rectangle, inner sep=0pt, minimum width=3.5mm, minimum height=3.5mm]
\tikzstyle{Zcheck}=[fill={rgb,255: red,0; green,203; blue,147}, draw=none, shape=rectangle, inner sep=0pt, minimum width=3.5mm, minimum height=3.5mm]
\tikzstyle{Lqubit}=[fill={rgb,255: red,0; green,102; blue,160}, draw=none, shape=circle, xshift = 0cm, inner sep=0pt, minimum width=4.3mm]
\tikzstyle{Rqubit}=[fill={rgb,255: red,230; green,159; blue,0}, draw=none, shape=circle, xshift = 0cm, inner sep=0pt, minimum width=4.3mm]

\tikzstyle{thick}=[-, draw=black, fill=none, line width=1pt]
\tikzstyle{thick arrow}=[->, line width=2pt]
\tikzstyle{ultra thick}=[-, line width=2pt]
\tikzstyle{thin}=[-, line width=0.3pt]
\tikzstyle{bracket}=[-, line width=0.3pt, draw={rgb,255: red,191; green,191; blue,191}]

\pgfdeclarelayer{nodelayer}
\pgfdeclarelayer{edgelayer}
\pgfdeclarelayer{toplayer}
\pgfsetlayers{edgelayer,nodelayer,toplayer}


\begin{tikzpicture}
	\begin{pgfonlayer}{nodelayer}
		\node [style=Zcheck] (0) at (-1, -1) {};
		\node [style=Lqubit] (1) at (0, -1) {};
		\node [style=Rqubit] (2) at (-1, 0) {};
		\node [style=Xcheck] (3) at (0, 0) {};
		\node [style=none] (4) at (0.5, 0) {};
		\node [style=none] (5) at (0, 0.5) {};
		\node [style=none] (6) at (-1, 0.5) {};
		\node [style=none] (7) at (-1.5, 0) {};
		\node [style=none] (8) at (-1.5, -1) {};
		\node [style=none] (9) at (-1, -1.5) {};
		\node [style=none] (10) at (0, -1.5) {};
		\node [style=none] (11) at (0.5, -1) {};
		\node [style=Xcheck] (12) at (1.5, 0) {};
		\node [style=none] (13) at (2.75, 0) {$= X\text{\space check}$};
		\node [style=Lqubit] (14) at (1.5, -1) {};
		\node [style=none] (15) at (2.75, -1) {$= L\text{\space qubit}$};
		\node [style=Rqubit] (16) at (-4.5, 0) {};
		\node [style=none] (17) at (-3.25, 0) {$= R\text{\space qubit}$};
		\node [style=Zcheck] (18) at (-4.5, -1) {};
		\node [style=none] (19) at (-3.25, -1) {$= Z\text{\space check}$};
	\end{pgfonlayer}
	\begin{pgfonlayer}{edgelayer}
		\draw [style=thin] (7.center) to (4.center);
		\draw [style=thin] (11.center) to (8.center);
		\draw [style=thin] (9.center) to (6.center);
		\draw [style=thin] (5.center) to (10.center);
	\end{pgfonlayer}
\end{tikzpicture}
\vspace{-15pt}
\caption{A unit cell}\label{fig:unit_cell}
\end{subfigure}

\begin{subfigure}{\textwidth}
    \centering
    \vspace{15pt}
    \input{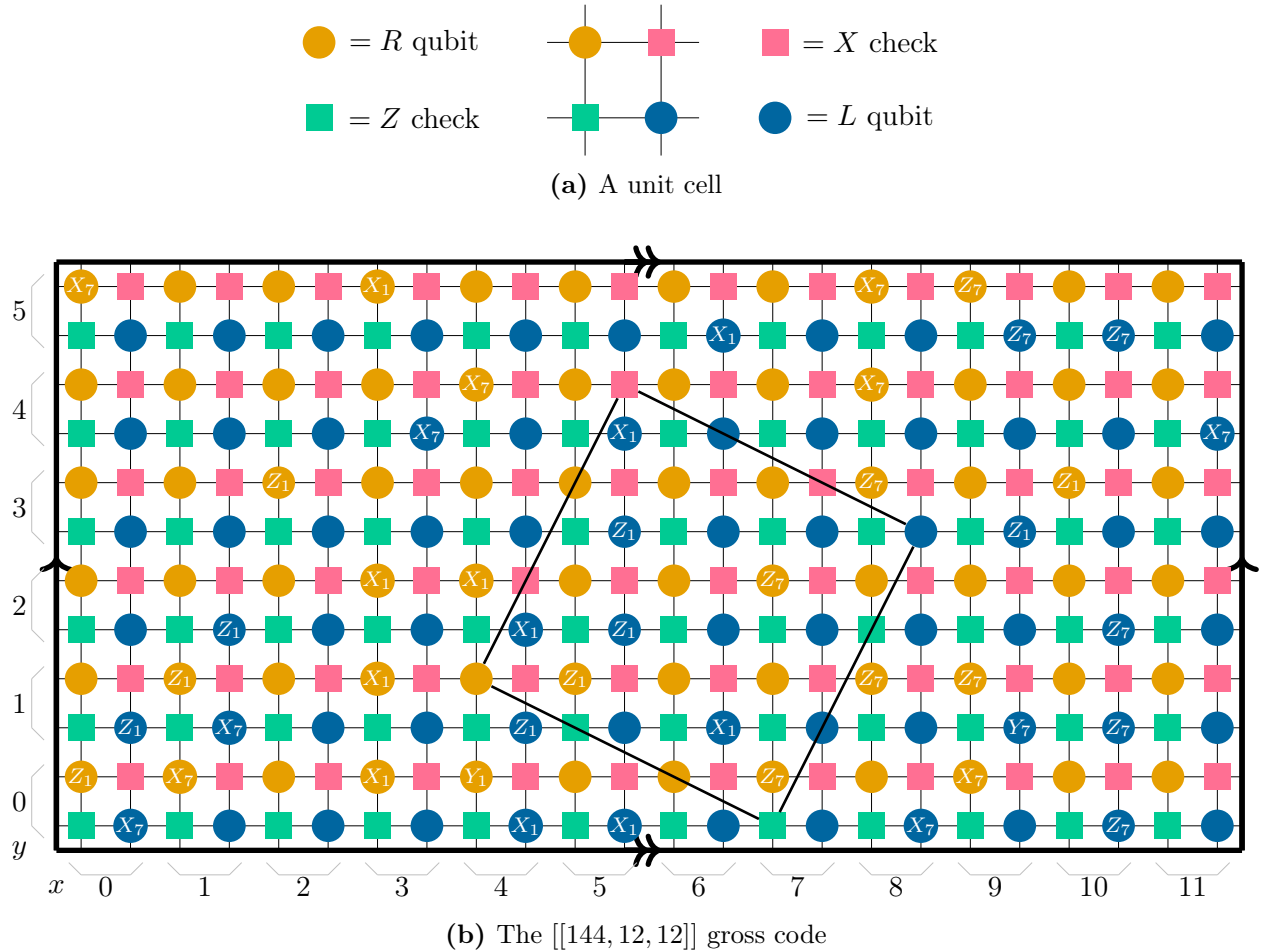}
    \vspace{-5pt}
    \caption{The $[[144,12,12]]$ gross code}\label{fig:gross_code}
\end{subfigure}
\caption{\subref{fig:unit_cell} We can build our BB codes from unit cells, which each contain two checks and two qubits. A line drawn between a qubit and an $X$($Z$) check indicates that the check acts non-trivially on that qubit with Pauli $X$($Z$).
\subref{fig:gross_code} The gross code is built by assembling unit cells in a $12\times 6$ grid on the surface of the torus. Only four example long-range connections are drawn, but these can be extrapolated to all unit cells by translational symmetry so that each qubit and check participates in two long-range connections. Qubits in logical operators $\bar X_1$, $\bar X_7$, $\bar Z_1$, and $\bar Z_7$ are labeled $X_1$, $X_7$, $Z_1$, and $Z_7$, respectively. Where $\bar X_1$ and $\bar Z_1$ (resp.~$\bar X_7$ and $\bar Z_7$) overlap, the qubit is labeled $Y_1$ (resp.~$Y_7$). These logical operators and their shifts generate the full 12-qubit logical Pauli group.
}
\label{fig:BB_codes}
\end{figure}

Bivariate bicycle (BB) codes \cite{kovalev2013quantum,panteleev2021degenerate,bravyi2024high} are a generalization of the toric code \cite{kitaev2003fault,dennis2002topological} that make use of longer range connectivity unconfined to two dimensions to achieve better code parameters such as rate and code distance $d$. 
Here, we take a more geometrically-minded approach to describing this family of codes,\footnote{Note, however, that our description covers only those BB codes with a toric layout \cite{bravyi2024high}. This is a special case but sufficient for our present work.} while \app{basis} contains a description using polynomials similar to Ref.~\cite{bravyi2024high}.

In analogy to the horizontal and vertical edge qubits in the toric code, we define two types of qubits, which we call left ($L$) and right ($R$). As CSS codes, BB codes also have two types of checks, $X$ and $Z$. To build the code, first arrange one $L$ qubit, one $R$ qubit, one $Z$ check, and one $X$ check in a unit cell as shown in \fig{unit_cell}. Both checks in the cell act on both qubits of the cell.

Place an $\ell\times m$ grid of unit cells on the surface of a torus. The example in \fig{gross_code} has $\ell=12$ and $m=6$. This makes a total of $2\ell m$ qubits and $2\ell m$ checks. The $X$ check in each cell is connected additionally to the $L$ qubit of the cell to the north and the $R$ qubit of the cell to the east. The $Z$ check in each cell is connected additionally to the $L$ qubit of the cell to the west and the $R$ qubit of the cell to the south. So far, all connections can be drawn on the surface of the torus without intersecting. Indeed, we have just described a standard two-dimensional toric code in which each check acts on just four nearby qubits.

\paragraph{Gross and two-gross codes.}
To obtain more general BB codes, we add additional long-range connections to every check. The codes we use here add two long-range connections per check. 
A connection is specified by the translation in the two directions on the torus that map the unit cells containing its endpoints. 
We connect each $X$ check to the $L$ qubit in the cell $a_{\rightarrow}$ steps east and $a_{\uparrow}$ steps north and to the $R$ qubit in the cell $b_{\rightarrow}$ steps east and $b_{\uparrow}$ steps north.\footnote{Steps here count the integer number of unit cells traversed. For instance, unit cells 1 step away in any compass direction are the neighboring unit cells.} 
To ensure commutation of the checks, we must then connect each $Z$ check to the $L$ qubit in the cell $-b_{\rightarrow}$ steps east and $-b_{\uparrow}$ steps north (equivalently, $b_{\rightarrow}$ steps west and $b_{\uparrow}$ steps south) and to the $R$ qubit in the cell $-a_{\rightarrow}$ steps east and $-a_{\uparrow}$ steps north. 
In this work, we focus on two codes identified in Ref.~\cite{bravyi2024high}. 
The $[[144,12,12]]$ \emph{gross code} in \fig{gross_code} and the $[[288,12,18]]$ \emph{two-gross code} in \app{basis}, \fig{two_gross_code} both are defined with $a_{\rightarrow}=-b_{\uparrow}=3$ and $a_{\uparrow}=b_{\rightarrow}=-1$, but differ in size parameters with $\ell=12,m=6$ for the gross code and $\ell=12,m=12$ for the two-gross code.

\begin{figure}[t]
\centering%
\begin{subcaptiongroup}%
    \includegraphics[width=\textwidth]{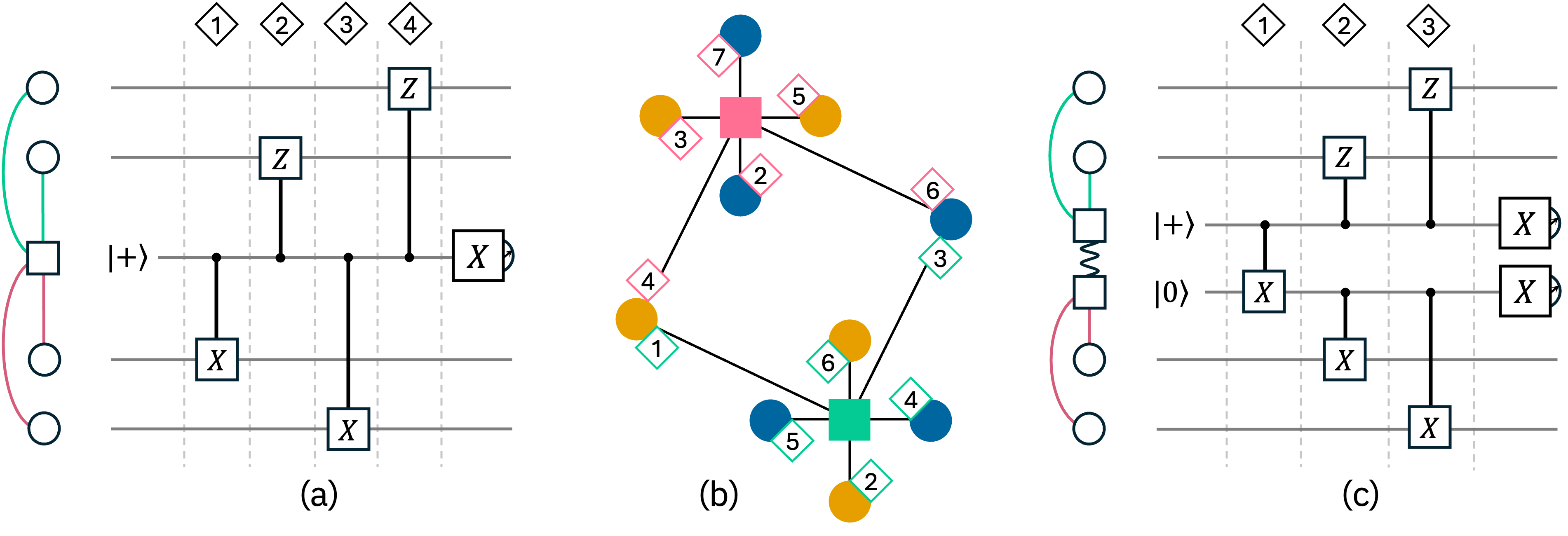}
    \phantomcaption\label{fig:syndromeCircuit}
    \phantomcaption\label{fig:tannerGraph}
    \phantomcaption\label{fig:bellCheck}
\end{subcaptiongroup}%
\caption{\subref{fig:syndromeCircuit} An example of the standard circuit for measuring a Pauli check using just one check qubit prepared in $\ket{+}$ and measured out in the $X$ basis along with controlled-$X$ and controlled-$Z$ gates. We depict a Tanner (sub)graph on the left in which data qubits are drawn as circles, check qubits drawn as squares, and edges are color coded red and green to indicate Pauli $X$ and $Z$ support of the check, respectively. The connectivity of the Tanner graph matches the connectivity required for gates in the circuit. 
\subref{fig:tannerGraph} A representative Tanner subgraph of a BB code as in \fig{gross_code} with edges labeled with the timestep in which gates are scheduled. It should be understood that the $Z$ check qubit is initialized in timestep 0 and measured in timestep 7. Similarly, the $X$ check qubit is initialized in timestep 1 and measured in timestep 8. While the $X$ check is measured, the $Z$ check is reinitialized for the next syndrome cycle. From just this subgraph, the schedule for the whole BB code can be extrapolated by translational symmetry.
\subref{fig:bellCheck} The same check measurement from (a) can be performed in fewer timesteps (indexed across the top) and reduced check node degree by preparing a Bell state instead, indicated by a wavy line between check nodes. The check measurement is the parity of the two $X$ measurements. We sometimes refer to a check measured this way as a Bell check. 
}\label{fig:check_circuits}
\end{figure}

\subsection{Idles and shift automorphisms}\label{sec:shift_automorphisms}

Here we briefly describe the idle and shift-automorphism instructions which each involve only the qubits and connectivity of a single BB code module, as drawn for instance in \fig{gross_code}.

When operating a BB code as a fault-tolerant memory, we must repeatedly measure the check operators.
We refer to one cycle that extracts all the code's check operators as the idle instruction.
We do this by associating one qubit, a check qubit, to each check node. 
Two-qubit gates are used to write the eigenvalue of a check onto a check qubit, which is subsequently measured out. 
See \cref{fig:syndromeCircuit} for an example of this canonical circuit. 

To measure many checks at once, it is ideal if the circuits for measuring individual checks are parallelized as much as possible. 
This parallelization was performed for BB code memories in Ref.~\cite{bravyi2024high}, and we convey the schedule from that work in \cref{fig:tannerGraph}. 
Suitably applied across the whole code to measure all the checks once, this circuit defines a syndrome measurement cycle. 
If we assume gates, qubit initializations, and measurements all take the same unit of time, one timestep, it takes $8C_\text{mem}+1$ timesteps to perform $C_\text{mem}$ syndrome cycles using this schedule.  

\paragraph{Idle instruction parameters} 
The idle instruction specifies the underlying overhead of implementing the BB code itself.
It uses a qubit for each of the $2\ell m$ checks and, adding in the $2\ell m$ data qubits, acts as a circuit on a total of $c=4\ell m$ qubits (288 and 576 qubits for the gross and two-gross codes, respectively). 
Since for large $C_\text{mem}$, the average time of a syndrome cycle approaches $8$ timesteps, this is the time reported in \tab{logical_ops} for a fault-tolerant idle operation of a single code module. 

Some logical operations within a single BB code module can be performed by simply permuting qubits according to code automorphisms \cite{grassl2013leveraging}. 
In BB codes, the group of automorphisms \cite{eberhardt2024logical} includes the translationally uniform shifts of all qubits across the surface of the tori visualized in \fig{gross_code} and \fig{two_gross_code}. 
For instance, shifting all qubits $\delta x\in[\ell] \coloneqq \set{1, \ldots, \ell}$ unit cells east and $\delta y\in[m]$ cells north is a valid shift automorphism. 
A shift automorphism is implemented at the physical level by first swapping all $L$ and $R$ data qubits with their adjacent check qubits and then swapping them again into another adjacent data qubit position of the same $L$ or $R$ type. 
The first swap can be done in the north, south, east, west directions as well as along the two non-local edges in the toric layouts and the second swap can be done always in two ways, giving us a total of $6\times2=12$ different shift automorphism instructions. 

\paragraph{Shift automorphism instruction parameters} 
To implement a shift automorphism, we have a particular circuit making use of the same qubits and connectivity used for idle instructions. 
Swaps are only ever performed between data qubits and ancilla qubits initialized in $\ket{0}$, so that each swap takes two CNOT gates. 
To offer some error detection, the ancilla qubits are measured out after each swap, which also serves to reinitialize them, up to an unapplied but tracked Pauli correction, for the next round of swaps. 
The total time for the two rounds of swaps is 6 timesteps, counting the 4 CNOTs and 2 measurements used. Afterward, we perform a syndrome cycle, adding another 8 timesteps. A shift automorphism therefore uses a total of 14 timesteps as reported in \tab{logical_ops}. 
In both gross and two-gross codes, any logically non-trivial product of shift automorphism instructions can be expressed as a product of at most two shift automorphisms. 
See \app{automorphisms} for more details.

\subsection{In-module measurements}\label{sec:in-module}

In-module logical measurements are done via quantum LDPC surgery \cite{CKBB22,CHRY24,WY24,ide2024fault,swaroop2024universal}.
Similar to lattice surgery for the surface code \cite{horsman2012surface}, quantum LDPC surgery works by temporarily deforming the original quantum LDPC code into another quantum LDPC code in which the logical Pauli operator being measured is a product of stabilizer generators. 
This deformation lasts for $C$ syndrome cycles to suppress time-like logical errors that may corrupt the logical measurement result. For the gross code we take $C=10$ while for the two-gross code we take $C=18$ for reasons explained in \sec{instruction-validation}. We also recommend Ref.~\cite{he2025extractors} for a more thorough review of surgery and recent developments.

The in-module measurements use a set of ancillary qubits and connectivity, the logical processing unit (LPU).\footnote{Our LPU constructions are also examples of partial extractors \cite{he2025extractors}.} 
The particular construction we use here assumes the qubit connectivity illustrated in \fig{LPU_summary}. 
The LPU and BB code together form a fixed-connectivity system constructed so that different components can be switched on to measure different logical operators.
Using the LPU, the entire set of possible logical measurements corresponds to elements of $\langle\bar X_1,\bar X_7,\bar Z_1,\bar Z_7\rangle$, or in other words, any logical Pauli on two specific logical qubits, 1 and 7. 
Interleaving these logical measurements with the shift automorphisms from \sec{shift_automorphisms} enables a much larger set of logical measurements.\footnote{For example, as in Ref.~\cite{CHRY24} we can use the logical measurements enabled by the LPU and shift automorphisms together to implement the full Clifford group on 11 out of the 12 logical qubits in a module.} In \cref{fig:gross_code,fig:two_gross_code}, we label the supports of logical operators $\bar X_1,\bar X_7,\bar Z_1$, and $\bar Z_7$ in the bases we have chosen for the gross and two-gross codes, respectively.

{%
\setlength{\tabcolsep}{3pt}
\begin{figure}[t]
\begin{subfigure}{0.5\textwidth}
        \adjustbox{Clip=115pt 50pt 15pt 0}{
            \newcommand{\mysize}{9mm}
\newcommand{\sizediff}{1.5mm}
\newcommand{\edgethickness}{1.2pt}
\newcommand{\nodethickness}{1pt}

\definecolor{BBcolorL}{RGB}{170,170,170}
\definecolor{LPUcolorL}{RGB}{250,224,140}

\definecolor{BBcolor}{RGB}{0, 0, 0}
\definecolor{LPUcolor}{RGB}{243, 182, 0}

\tikzstyle{none}=[fill=none,draw=none,shape=circle, minimum size=\mysize, xshift = 0cm, inner sep=0pt]

\tikzstyle{qubit}=[fill=white, draw=BBcolor, shape=circle, line width=\nodethickness, minimum size=\mysize, xshift = 0cm, inner sep=0pt]
\tikzstyle{LPU qubit}=[fill=white, draw=LPUcolor, shape=circle, line width=\nodethickness, minimum size=\mysize, xshift = 0cm, inner sep=0pt]
\tikzstyle{check}=[fill=white, draw=BBcolor, shape=rectangle, line width=\nodethickness, minimum size=\mysize, xshift = 0cm, inner sep=0pt]
\tikzstyle{LPU check}=[fill=white, draw=LPUcolor, shape=rectangle, line width=\nodethickness, minimum size=\mysize, xshift = 0cm, inner sep=0pt]
\tikzstyle{background qubit}=[fill=white, draw=BBcolorL, shape=circle, line width=\nodethickness, minimum size=\mysize, xshift = 0cm, inner sep=0pt]
\tikzstyle{background check}=[fill=white, draw=BBcolorL, shape=rectangle,  line width=\nodethickness, minimum size=\mysize, xshift = 0cm, inner sep=0pt]
\tikzstyle{small qubit}=[fill=white, draw=BBcolor, shape=circle, line width=\nodethickness, minimum size=\mysize-\sizediff, xshift = 0cm, inner sep=0pt]
\tikzstyle{small LPU check}=[fill=white, draw=LPUcolor, shape=rectangle, line width=\nodethickness, minimum size=\mysize-\sizediff, xshift = 0cm, inner sep=0pt]
\tikzstyle{small background qubit}=[fill=white, draw=BBcolorL, shape=circle, line width=\nodethickness, minimum size=\mysize-\sizediff, xshift = 0cm, inner sep=0pt]
\tikzstyle{bell LPU check}=[fill=white, draw=LPUcolor, shape=rectangle, line width=\nodethickness, minimum size=\mysize-\sizediff-\sizediff-\sizediff, xshift = 0cm, inner sep=0pt]

\tikzstyle{LPU edge}=[-, line width=\edgethickness]
\tikzstyle{LPU background edge}=[-, draw={rgb,255: red,191; green,191; blue,191}, line width=\edgethickness]
\tikzstyle{Z LPU edge}=[-, line width=\edgethickness, draw={rgb,255: red,0; green,203; blue,147}]
\tikzstyle{X LPU edge}=[-, line width=\edgethickness, draw={rgb,255: red,255; green,111; blue,147}]
\tikzstyle{X LPU background edge}=[-, line width=\edgethickness, draw={rgb,255: red,255; green,215; blue,225}]
\tikzstyle{Z LPU background edge}=[-, line width=\edgethickness, draw={rgb,255: red,171; green,225; blue,194}]
\tikzstyle{LPU arrow}=[->, densely dotted, line width=0.5pt, draw={rgb,255: red,128; green,128; blue,128}]
\tikzstyle{mydotted}=[-, densely dotted, draw={rgb,255: red,114; green,114; blue,114}]

\pgfdeclarelayer{nodelayer}
\pgfdeclarelayer{edgelayer}
\pgfdeclarelayer{bgnodelayer}
\pgfdeclarelayer{bgedgelayer}
\pgfsetlayers{bgedgelayer,bgnodelayer,edgelayer,nodelayer}

\begin{tikzpicture}[scale=1.75]
	\begin{pgfonlayer}{nodelayer}
		\node [style=LPU qubit] (0) at (1, 0.5) {};
		\node [style=LPU qubit] (1) at (1.075, 0.425) {$\mathcal{E}_r$};
		\node [style=LPU check] (6) at (2, 0.5) {};
		\node [style=LPU check] (7) at (2.075, 0.425) {$\mathcal{V}_r$};
		\node [style=bell LPU check] (8) at (0.25, 0.5) {};
            \node [style=bell LPU check] (92) at (-0.25, 0.5) {};
		\node [style=LPU qubit] (9) at (-1, 0.5) {};
		\node [style=LPU qubit] (10) at (-0.925, 0.425) {$\mathcal{E}_l$};
		\node [style=LPU check] (11) at (-2, 0.5) {};
		\node [style=LPU check] (12) at (-1.925, 0.425) {$\mathcal{V}_l$};
		\node [style=LPU check] (13) at (1, 1.5) {};
		\node [style=LPU check] (14) at (1.075, 1.425) {$\mathcal{U}_r$};
		\node [style=LPU qubit] (15) at (0, 2.5) {};
		\node [style=LPU qubit] (16) at (0.075, 2.425) {$B$};
		\node [style=LPU check] (17) at (0, 1.5) {};
		\node [style=LPU check] (18) at (0.075, 1.425) {$\mathcal{U}_B$};
		\node [style=LPU check] (19) at (-1, 1.5) {};
		\node [style=LPU check] (20) at (-0.925, 1.425) {$\mathcal{U}_l$};
		\node [style=qubit] (38) at (2, -1) {};
		\node [style=qubit] (39) at (2.075, -1.075) {$Z_1$};
		\node [style=check] (42) at (1, -1) {};
		\node [style=check] (43) at (1.075, -1.075) {}; 
		\node [style=check] (51) at (-1, -1) {};
		\node [style=check] (52) at (-0.925, -1.075) {}; 
		\node [style=qubit] (55) at (-2, -1) {};
		\node [style=qubit] (56) at (-1.925, -1.075) {$X_1$};
		\node [style=small qubit] (61) at (0, -1) {$Y_1$};
		\node [style=none] (66) at (-2, -1.5) {};
		\node [style=none] (67) at (-2.25, -1.5) {};
		\node [style=none] (71) at (1.75, -1.5) {};
		\node [style=none] (72) at (2, -1.5) {};
		\node [style=none] (73) at (-1.25, -1.5) {};
		\node [style=none] (74) at (-1, -1.5) {};
		\node [style=none] (75) at (1, -1.5) {};
		\node [style=none] (76) at (0.75, -1.5) {};
		\node [style=none] (77) at (-0.25, -1.5) {};
		\node [style=none] (78) at (0, -1.5) {};
		\node [style=none] (83) at (0, -1.85) {to the rest of the BB code};
		\node [style=none] (84) at (-1.5, -1.5) {};
		\node [style=none] (85) at (-1.5, -2) {};
		\node [style=none] (86) at (1.5, -1.5) {};
		\node [style=none] (87) at (1.5, -2) {};
		\node [style=none] (88) at (2.4, -0.125) {};
		\node [style=none] (89) at (-2.4, -0.125) {};
		\node [style=none] (90) at (-0.55, -0.03) {\tiny \textit{LPU}};
		\node [style=none] (91) at (-0.55, -0.22) {\tiny \textit{BB code}};
            \node [style=none] (93) at (-0.15, 0.5) {};
            \node [style=none] (94) at (-0.075, 0.5) {};
            \node [style=none] (95) at (0, 0.5) {};
            \node [style=none] (96) at (0.075, 0.5) {};
            \node [style=none] (97) at (0.15, 0.5) {};
            \node [style=none] (98) at (0,0.75) {$v_{\text{Bell}}$};
	\end{pgfonlayer}
	\begin{pgfonlayer}{edgelayer}
		\draw [style=X LPU edge, bend right=45, looseness=1.25] (15) to (11);
		\draw [style=X LPU edge, bend left=45, looseness=1.25] (15) to (6);
		\draw [style=X LPU edge] (9) to (92);
		\draw [style=X LPU edge] (0) to (8);
		\draw [style=X LPU edge] (6) to (0);
		\draw [style=X LPU edge] (9) to (11);
		\draw [style=Z LPU edge] (19) to (9);
		\draw [style=Z LPU edge] (13) to (0);
		\draw [style=Z LPU edge] (0) to (42);
		\draw [style=Z LPU edge] (9) to (51);
		\draw [style=Z LPU edge] (17) to (9);
		\draw [style=Z LPU edge] (17) to (0);
		\draw [style=Z LPU edge] (17) to (15);
		\draw [style=X LPU edge] (42) to (61);
		\draw [style=X LPU edge] (42) to (38);
		\draw [style=Z LPU edge] (55) to (51);
		\draw [style=LPU edge] (92) to (61);
		\draw [style=X LPU edge] (11) to (55);
		\draw [style=Z LPU edge] (6) to (38);
		\draw [style=LPU edge] (55) to (66.center);
		\draw [style=LPU edge] (38) to (72.center);
		\draw [style=Z LPU edge] (51) to (74.center);
		\draw [style=X LPU edge] (42) to (75.center);
		\draw [style=LPU edge] (61) to (78.center);
		\draw [style=Z LPU edge] (51) to (61);
		\draw [style=LPU arrow] (84.center) to (85.center);
		\draw [style=LPU arrow] (86.center) to (87.center);
		\draw [style=mydotted] (89.center) to (88.center);
            \draw [style=LPU edge,bend left=75,looseness=1.50] (93.center) to (94.center);
            \draw [style=LPU edge,bend right=75,looseness=1.50] (94.center) to (95.center);
            \draw [style=LPU edge,bend left=75,looseness=1.50] (95.center) to (96.center);
            \draw [style=LPU edge,bend right=75,looseness=1.50] (96.center) to (97.center);
	\end{pgfonlayer}
        \begin{pgfonlayer}{bgnodelayer}
		\node [style=background qubit] (36) at (1.75, -0.75) {};
		\node [style=background qubit] (37) at (1.825, -0.825) {};
		\node [style=background check] (40) at (0.75, -0.75) {};
		\node [style=background check] (41) at (0.825, -0.825) {};
		\node [style=background check] (49) at (-1.25, -0.75) {};
		\node [style=background check] (50) at (-1.175, -0.825) {};
		\node [style=background qubit] (53) at (-2.25, -0.75) {};
		\node [style=background qubit] (54) at (-2.175, -0.825) {};
		\node [style=small background qubit] (60) at (-0.25, -0.75) {\scriptsize\color{gray}$Y_7$};
		\node [style=none] (79) at (1.73, -0.75) {\scriptsize\color{gray}$X_7$};
		\node [style=none] (80) at (0.75, -0.75) {}; 
		\node [style=none] (81) at (-1.25, -0.75) {}; 
		\node [style=none] (82) at (-2.26, -0.75) {\scriptsize\color{gray}$Z_7$};
        \end{pgfonlayer}
        \begin{pgfonlayer}{bgedgelayer}
		\draw [style=Z LPU background edge] (11) to (53);
		\draw [style=Z LPU background edge] (9) to (49);
		\draw [style=Z LPU background edge] (0) to (40);
		\draw [style=X LPU background edge] (6) to (36);
		\draw [style=LPU background edge] (8) to (60);
		\draw [style=X LPU background edge] (49) to (60);
		\draw [style=X LPU background edge] (49) to (53);
		\draw [style=Z LPU background edge] (40) to (60);
		\draw [style=Z LPU background edge] (40) to (36);
		\draw [style=LPU background edge] (53) to (67.center);
		\draw [style=LPU background edge] (36) to (71.center);
		\draw [style=Z LPU background edge] (40) to (76.center);
		\draw [style=X LPU background edge] (49) to (73.center);
		\draw [style=LPU background edge] (60) to (77.center);
        \end{pgfonlayer}
\end{tikzpicture}%
        }%
    \caption{Logic processing unit (LPU)}%
    \label{fig:lpu}
\end{subfigure}%
\hfill%
\begin{subcaptionblock}{0.48\textwidth}
\begin{subfigure}{\linewidth}
    \centering
    \begin{tabular}{lrrrrrrrr}\toprule
        degree &$\mathcal{V}_l$&$\mathcal{E}_l$&$\mathcal{U}_l$&$B$&$\mathcal{U}_B$&$\mathcal{U}_r$&$\mathcal{E}_r$&$\mathcal{V}_r$\\ \midrule
        3&--&--&--&1 &1 &--&--&--\\
        4&--&1 &2 &10&10&2 &2 &--\\
        5&--&2 &3 &--&--&1 &8 &--\\
        6&11&13&--&--&--&--&8 &11\\
        7&--&2 &--&--&--&--&--&-- \\ \addlinespace
        LPU size & \multicolumn{8}{c}{$u = 90$ qubits}\\ \bottomrule
    \end{tabular}
    \caption{Gross LPU census}%
    \label{tab:grossCensus}
    \vspace*{0.5em}
\end{subfigure}
\begin{subfigure}{\linewidth}
    \centering
    \begin{tabular}{lrrrrrrrr}\toprule
    degree & $\mathcal{V}_l$&$\mathcal{E}_l$&$\mathcal{U}_l$&$B$&$\mathcal{U}_B$&$\mathcal{U}_r$&$\mathcal{E}_r$&$\mathcal{V}_r$\\ \midrule
    3&--&1 &4 &1 &1 &5 &--&--\\
    4&--&1 &--&16&16&1 &4 &--\\
    5&2 &8 &6 &--&--&1 &11&2 \\
    6&13&12&1 &--&--&2 &13&13\\
    7&4 &10&--&--&--&--&4 &4 \\ \addlinespace
    LPU size & \multicolumn{8}{c}{$u = 158$ qubits}\\ \bottomrule
\end{tabular}
\caption{Two-gross LPU census}%
\label{tab:twoGrossCensus}
\end{subfigure}
\end{subcaptionblock}
\caption{
\textbf{In-module qubit connectivity.} 
\app{lpu_BB} describes components of this figure in more detail.
\subref{fig:lpu} Data qubits (circles) and check qubits (squares) are gathered into sets (doubled nodes), and drawn in yellow or black depending on whether they are counted as part of the LPU or BB code, matching the colors in \fig{gross-architecture}. 
Edges represent sparse connectivity between adjacent sets with colors red or green indicating that checks act on data with Pauli $X$ or $Z$, respectively. 
Black edges indicate any Pauli type, possibly depending on what logical operator is being measured. Qubits in the BB code that are connected to the LPU support the logical operators $\bar X_1,\bar X_7,\bar Z_1,\bar Z_7$ and are labeled as in \cref{fig:gross_code}.
\subref{tab:grossCensus} and \subref{tab:twoGrossCensus} we count elements in the various sets of data and check qubits sorted by their degree (i.e.~number of connections). 
Included also in the calculation of total LPU size, $v_{\text{Bell}}$ is a single check that we decide to measure using an entangled pair of check qubits as in \cref{fig:bellCheck} so that the maximum degree of the connectivity graph does not exceed seven. 
Each side of the Bell pair has degree five.
When measuring a logical operator in the group $\langle\bar X_1,\bar X_7,\bar Z_1,\bar Z_7\rangle$, only certain subsets of the LPU are used, as explained in \app{lpu_meas_procedure}. Several measurements employ all the LPU's qubits and checks, but only $\bar Y_1\bar Y_7$ uses all edges as well.}
\label{fig:LPU_summary}
\end{figure}
}%

We briefly summarize the details of the LPU pictured in \fig{LPU_summary}.
The LPU is similar to the construction for in-module measurements of the gross code in Ref.~\cite{CHRY24}. Sets of qubits and checks on the left side -- $\mathcal{V}_l$, $\mathcal{E}_l$, $\mathcal{U}_l$, and $v_{\text{Bell}}$ -- serve to fault-tolerantly measure $\langle\bar X_1,\bar Z_7\rangle$ and similar sets on the right -- $\mathcal{V}_r$, $\mathcal{E}_r$, $\mathcal{U}_r$, and $v_{\text{Bell}}$ -- serve to measure $\langle\bar X_7,\bar Z_1\rangle$. 
(Note that the labels $l$ and $r$ are not related to the labeling $L$ and $R$ for qubits of the BB code.)
Besides sharing the check $v_{\text{Bell}}$, which is operated as a single check as in \cref{fig:bellCheck}, the left and right sides are connected via a set of bridge qubits $B$ and bridge checks $\mathcal{U}_B$, which enable the fault-tolerant measurement of the rest of the group $\langle\bar X_1,\bar X_7,\bar Z_1,\bar Z_7\rangle$. 
The construction here advances beyond Ref.~\cite{CHRY24} by considering both the gross and two-gross codes, finding lower weight logical operators for logical qubits 1 and 7, and incorporating ideas from Refs.~\cite{WY24,swaroop2024universal} to simplify the constructions.

For more detail about the LPU constructions, refer to the appendices. \app{lpu_BB} presents the details of the constructions for the gross code and the two-gross code. 
\app{lpu_meas_procedure} describes the code deformation process for each different logical measurement. \app{lpu_circuit_scheduling} delves into the circuit scheduling details. \sec{instruction-validation} and \app{cplex} discuss how fault-distances are calculated or bounded for the constructions.

\paragraph{In-module instruction parameters} In \cref{tab:qubit-counts}, the qubit count, $u$, of the LPU is determined by summing up the total numbers of data and check qubits from the tables in \fig{LPU_summary} and including the Bell check $v_{\text{Bell}}$ as two additional check qubits. 
The duration of an in-module operation depends on the circuits we describe in \app{lpu_circuit_scheduling}, but generally this is calculated by performing syndrome measurements in the deformed code for $C$ cycles followed by one cycle of syndrome measurement of the original BB code. This last cycle in the original BB code returns the information to the original code space and readies it for the next operation.

\subsection{Inter-module measurements}\label{sec:inter-module}

To perform quantum operations between code modules and factory modules, we must be able to generate entanglement between them. 
To that end, we assume we have an inter-module device, a \emph{Bell-coupler}, that is able to create a high-fidelity Bell state $(\ket{00}+\ket{11})/\sqrt{2}$ between two modules in time less than the time of a syndrome cycle. Operationally, we use these Bell states to measure checks on sets of physical qubits split between modules, and the ultimate logical operation is the fault-tolerant measurement of a joint logical Pauli operator spanning the modules. The inter-module connectivity is a low-degree graph with modules as nodes and edges representing bundles of $O(d)$ Bell-couplers to connect two adjacent modules. The idea of a universal adapter \cite{swaroop2024universal} dictates how these Bell-couplers should be connected and used. Still, specific constructions can differ in Bell-coupler count, qubit count, and the set of joint logical measurements that can be performed.

\begin{figure}[t]
    \centering
    \input{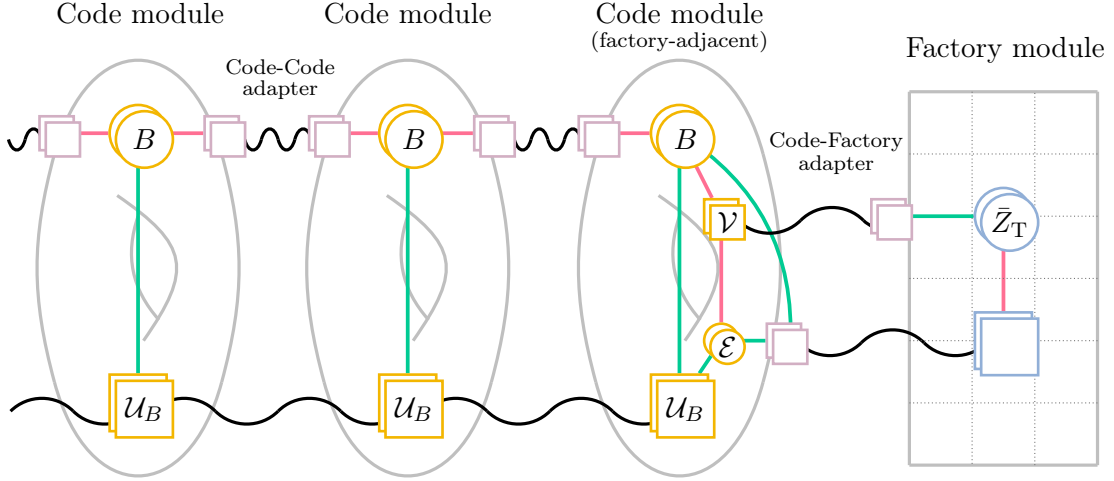}%
    \caption{Operations between two neighboring BB codes are enabled by entangling the bridge qubits and checks in their LPUs via inter-module Bell state preparation (wavy lines). Any of the 36 inter-module Paulis $(\langle\bar X_1,\bar Z_7\rangle\cup\langle\bar X_7,\bar Z_1\rangle)^{\otimes2}$ can be measured this way. The degree of qubits and checks in the LPU bridge is at most four (see \fig{LPU_summary}), so even if we were to connect a code module to three other modules this way, the maximum degree connectivity required remains seven.
    Additionally, magic states produced in a magic state factory can be consumed for non-Clifford gate injection by entangling the vertex checks $\mathcal{V}=\mathcal{V}_l\cup\mathcal{V}_r\cup v_{\text{Bell}}$, edge qubits $\mathcal{E}=\mathcal{E}_l\cup\mathcal{E}_r$, and bridge $B$ of a code LPU with the qubits in the support of a logical $Z$ operator in the factory $\bar Z_{\text{T}}$ and its adjacent $X$ checks. The precise connections in the code-factory adapter depend on the factory, but for one of our BB codes connected to a single surface code factory, the maximum degree connectivity is seven.
    }
    \label{fig:intercode_surgery}
\end{figure}

While there are several justifiable options, we choose to connect two code modules using a coupler and qubit efficient method, illustrated in \fig{intercode_surgery}.
Our construction uses Bell-couplers to connect the corresponding bridge qubits $B$ and bridge checks $\mathcal{U}_B$ from the LPUs in adjacent modules.
Between code modules, this code-code adapter allows the fault-tolerant measurement of any of the 36 inter-module Pauli products $(\langle\bar X_1,\bar Z_7\rangle\cup\langle\bar X_7,\bar Z_1\rangle)^{\otimes2}$. For these inter-module logical measurements, the deformed code distance is limited to $d-1$, the number of bridge qubits,
instead of the full code distance $d$. 
We could add one more bridge qubit, used only for inter-module measurements, to increase the deformed code distance to $d$, but choose not to. In addition to being simpler, this choice is further justifiable:
The minimum number of errors (in a certain phenomenological noise model introduced in \sec{instruction-validation}) that can cause an optimal decoder to fail remains the same for in-module operations and inter-module operations.
Our BB codes and deformed codes for in-module measurements have even code distance, so with a bridge of size $d-1$
they still require $6$ errors for the gross code and $9$ for the two-gross code.

The connection between code modules is limited to measuring $(\langle\bar X_1,\bar Z_7\rangle\cup\langle\bar X_7,\bar Z_1\rangle)^{\otimes2}$, as opposed to $\langle\bar X_1,\bar X_7,\bar Z_1,\bar Z_7\rangle^{\otimes2}$ for inter-module measurements because it reuses the bridge qubits, $B$, and bridge checks, $\mathcal{U}_B$.
The bridge checks are also essential for fault-tolerantly measuring the in-module operators that require both left and right halves of the LPU, such as $\bar Y_1$ for example.
However, the limited set of inter-module measurements is not an obstacle to universality,
so we prefer its simplicity for inter-module operations between code modules.

One exception is that reusing the bridge does become significantly more inconvenient when we seek to couple a code module to a factory module.
Using the previous construction, only the measurements $(\langle\bar X_1,\bar Z_7\rangle\cup\langle\bar X_7,\bar Z_1\rangle)\otimes\bar Z_{\text{T}}$ are accessible, where $\bar Z_{\text{T}}$ is the logical Pauli $Z$ operator on the magic state. We want more capability, particularly the $\bar Y_1\otimes\bar Z_{\text{T}}$ measurement, for our compilation strategies in \sec{small-angle-rotations}.
To do this, we can take advantage of the flexibility of quantum LDPC surgery and the universal adapter constructions~\cite{swaroop2024universal} to introduce another set of Bell-couplers to code modules adjacent to factory modules. We refer to this construction as a \emph{code-factory adapter} (see \cref{fig:intercode_surgery}). A code-factory adapter consists of an additional $O(d)$ Bell-couplers and it enables the measurement of any element of $\langle\bar X_1,\bar X_7,\bar Z_1,\bar Z_7\rangle\otimes\bar Z_{\text{T}}$.
For instance, the code-factory adapter to a surface code factory  uses $2d-1$ Bell-Couplers.
See \app{lpu_meas_procedure} and \fig{code_factory_interface} for more details.

\paragraph{Inter-module instruction parameters} 
Both types of inter-module measurements, code-code and code-factory, require additional check qubits to which Bell-couplers are connected. These check qubits, which constitute the adapter components of our architecture, are counted in \tab{qubit-counts} as $a$ for code-code and $a'$ for code-factory adapters. \fig{intercode_surgery} depicts these additional qubits as pink squares. We see $a=2|B|=2(d-1)$, or 22 and 34 for gross and two-gross codes, respectively. Similarly, there are $a'=2d_\text{factory}-1$ qubits in the code-factory adapter, assuming the factory produces a T state in a distance $d_\text{factory}$ surface code.

\subsection{T injections}\label{sec:T-factories}

Here we briefly describe how to implement the T injection bicycle instruction by preparing a reliable T state (not in a BB code) in a T factory and injecting it into a connected code module. In principle, quantum LDPC surgery, and in particular universal adapters \cite{swaroop2024universal}, enable us to use any magic state factory, but for the present purposes we opt for well-benchmarked examples from the available literature.
We consider four T factories: two using cultivation~\cite{GSJ2024cultivation} (for gross and two-gross at $p=10^{-3}$), and two using distillation (for gross and two-gross at $p=10^{-4}$). 

For the cultivation factories, we use the recently optimized approach from Ref.~\cite{GSJ2024cultivation} because it uses relatively few qubits while being compatible with assumptions about superconducting hardware connectivity. 
Cultivation is a protocol for preparing magic states, checking them with fault-tolerant Clifford measurements, and judiciously increasing the fault distance of the protocol until reaching the desired logical error rate. 
This approach builds on earlier work using cat states and flags~\cite{chamberland2019fault} to make high-weight Clifford measurements fault-tolerant~\cite{shor1996fault}. 
Cultivation does not scale to large code distances but is highly effective at small sizes. 
The small size also make heavy post selection a viable approach to filter out low quality states. 
We extract the parameters needed to perform overall resource estimates for this approach in \tab{Tfactories} using the methodology described in \app{cultivation-cost}.

\begin{table}[h]
\centering
\begin{tabular}{llrrrrr}\toprule
name & approach & $d_\text{factory}$ & $a'$ & $f$ & $P_\text{factory}$  & $\bar{\tau}_\text{factory}$ \\ \midrule
$10^{-3}$ gross factory& cultivation & 15 & 29 & 454 & $3 \times 10^{-6}$  & 351 \\
$10^{-3}$ two-gross factory& cultivation & 15 & 29 & 463 & $2 \times 10^{-8}$  & 2167 \\
$10^{-4}$ gross factory & distillation & 7 & 13 & 810 & $4 \times 10^{-8}$  & 73 \\
$10^{-4}$ two-gross factory& distillation & 25 & 49 & 18\,600 & $6 \times 10^{-25}$  & 407 \\ \bottomrule
\end{tabular}
\caption{T factory properties. The factories are named after the scenario, code and physical error rate, in which they are used.
A mean number of timesteps $\bar{\tau}_\text{factory}$ to produce a T state encoded in a distance $d_\text{factory}$ surface code, with logical error rate $P_\text{factory}$.
The factory and code-factory adapter contain $f$ and $a'$ physical qubits respectively.
\label{tab:Tfactories}}
\end{table}

Numerical analysis of cultivation in Ref.~\cite{GSJ2024cultivation} is performed for a range of physical error rates, but due to low-error-rate sampling limitations, end-to-end parameter estimates are not provided for $p=10^{-4}$, and we therefore only use cultivation for $p=10^{-3}$.
For the distillation factories in \cref{tab:Tfactories}, we assume the protocols on line 1 and line 5 of \cite[Table~1]{litinski2019magic}, both of which use the 15-to-1 protocol~\cite{bravyi2005universal} implemented in surface codes, with the former using two distillation rounds. 
To obtain the number of timesteps, we multiply the number of cycles reported in \cite[Table~1]{litinski2019magic} by 6 since there are 6 timesteps per code cycle for the surface code.
To obtain the distance of the final code, we take the first subscript of the last distillation round reported in \cite[Table~1]{litinski2019magic}.
The second distillation protocol results in a very conservative estimate since it outputs a T state in a distance-25 surface code with logical error rate $P_\text{factory}\sim10^{-24}$, much below our needs in this work, and the qubit requirements for this distillation factory are far larger than those of the cultivation factories. 
Future improvements to cultivation approaches may result in distillation being unnecessary.

\paragraph{T injection instruction parameters} 
The T injection bicycle instruction involves producing a T state in a factory module, then performing an inter-module $\bar{Z}_\text{T}\otimes \bar{P}_1$ measurement between the factory and the pivot qubit in the code module, followed by a destructive $\bar{X}_\text{T}$ measurement on the factory module.  
The number of qubits required in the adapter that connects the factory and code module is $2d_\text{factory} - 1$, as described in \sec{inter-module}.
The logical error rate (to first order) and duration estimates of the T injection instruction are then 
\begin{align}
P_T &= P_\text{factory} + P_C, \nonumber\\
\tau_T &= \bar{\tau}_\text{factory} + \tau_C + 1. \nonumber
\end{align}

\subsection{Benchmarking bicycle instructions}
\label{sec:instruction-validation}

Here we describe our approach to validate the fault-tolerance of each bicycle instruction (and the idle instruction), and to benchmark its performance across a range of regimes.

\paragraph{Circuit noise}
We consider each bicycle instruction for each of the gross code and the two-gross code as a fully-specified quantum circuit, composed of one- and two-qubit quantum operations that each last one timestep (which are specified in the previous subsections, and in the appendices). 
We make the standard modeling assumption that the system begins in an error-free code state prior to the logical operation, and that a final cycle of noise-free stabilizer measurements are performed at the end.
In between, we adopt the standard stochastic circuit-level noise model with error rate $p$, linearized so that all faults are discrete events which occur independently. 
State preparations and measurements each fail with probability $p$, by preparing an orthogonal state and flipping the outcome respectively. 
Idle gates fail independently as $X$, $Y$, or $Z$ errors, each with probability $p/3$. 
CNOT gates fail independently as one of the 15 nontrivial two-qubit Pauli errors, each with probability $p/15$. 
All connections between qubits are assumed to be equally faulty.
We call a set of $w$ faults of that occur during the circuit a weight-$w$ fault configuration.

\paragraph{Decoding}
The circuit implementing each bicycle instruction outputs a syndrome $\sigma$, which is fed to a classical decoding algorithm to identify a correction.
If no faults occur at all, a trivial syndrome will be observed. Consequently, a non-trivial syndrome necessarily implies that at least one fault occurred.
The net effect of the noise and the correction either acts trivially (decoder success) or non-trivially (decoder failure) on the logical action of the circuit.
In this work, we employ the newly proposed `relay-BP' decoder~\cite{RelayPaper}. Relay-BP is designed for accurate real-time decoding with a light-weight gateware implementation.

\paragraph{Circuit distance}
Given a circuit and this noise model, we say a circuit-logical is a fault configuration with trivial syndrome but non-trivial logical action on the circuit. We refer to the minimum weight of any circuit-logical as the circuit distance $d_\text{circ}$.
Since the syndrome is trivial (which is what the decoder receives when no faults occur) any reasonable decoder will fail for a circuit-logical, and as such the circuit distance is a decoder-independent statement.
Moreover, if fewer than $d_\text{circ}/2$ faults occur, then it is possible to correct (although some decoders may not achieve this in practice).
It is therefore common to use the circuit distance $d_\text{circ}$ as a quantitative measure of fault-tolerance.

\paragraph{Phenomenological noise and distance}
Because the circuit distance can be difficult to calculate exactly, we also consider a simpler phenomenological noise model in which all data qubits and check measurements may suffer errors independently, but otherwise the syndrome circuits are perfect. This noise model is considered purely for the purpose of validating LPU constructions and not for noisy simulations.
The phenomenological fault-distance $d_{\text{phenom}}$ is the minimum number of these phenomenological errors that can cause a non-trivial logical action. 
The faults in phenomenological noise forms a (much smaller) subset of the faults in circuit noise, implying $d_{\text{phenom}} \geq d_{\text{circ}}$ and is easier to compute.
For a quantum LDPC surgery operation, the phenomenological fault-distance is the minimum of the distances of the deformed code and a subsystem code describing the code-switching \cite{vuillot2019code,CHRY24,WY24}.
\begin{figure}[tb]
    \centering
    \includegraphics[width=0.8\textwidth]{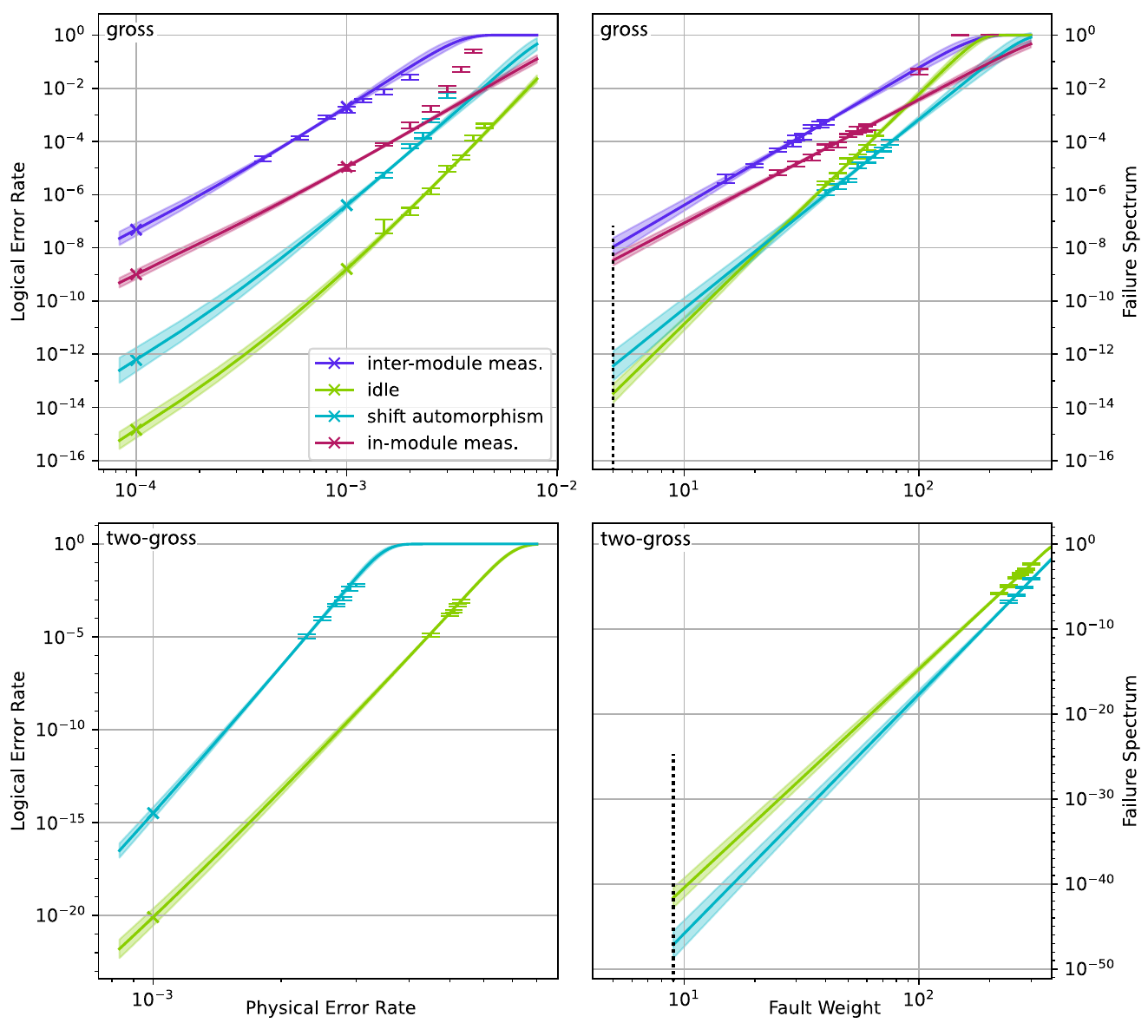}
    \caption{
        Circuit noise simulations for four logical instructions using the gross (top) and two-gross (bottom) codes.
        For each circuit, noise is sampled in two ways and the fraction of fails computed: as a function of physical error rate $p$, estimating the logical error rate, $P(p)$ (left panels); and as a function of the number of faults, $w$, estimating the failure spectrum, $f(w)$ (right panels).
        We fit a two-parameter ansatz (lines) $f_\text{ansatz}(w)$ to the failure fraction data, where $f_\text{ansatz}(w)$ is set to $0$ for $w < d^*_\text{circ}/2$, where $d^*_\text{circ}$ is our estimate of the circuit distance
        ($10$ for gross and $18$ for two-gross) with uncertainties (shaded) computed by bootstrap data resampling.
        We also show the inferred $P_\text{ansatz}(p)$ (computed from a weighted sum over $f_\text{ansatz}(w)$), which agree with the independently obtained data estimating $P(p)$.
        More details of our approach are provided in \cref{app:logical-op-simulations}.
        Logical error rates from fits at $p = 10^{-3}$ and $10^{-4}$ (crosses) are reported in \tab{logical_ops}.
    }
    \label{fig:data-logical-error-rates}
\end{figure}

\begin{table}[tb]
    \centering
    \begin{tabular}{lrrrrrr}\toprule%
         & \multicolumn{3}{c}{gross code} & \multicolumn{3}{c}{two-gross code} \\ \cmidrule(r){2-4} \cmidrule(l){5-7}
         & idle/aut. & in-mod. & inter-mod. & idle/aut. & in-mod. & inter-mod.\\ \midrule
         $d_{\text{phenom}}$ & 12 & 12 & $\le11$ & 18 & 18 & $\le17$ \\ 
         $d_{\text{circ}}$ & $\le10$ & $\le10$ & $\le10$ & $\le18$ & $\le18$ & $\le17$ \\ \bottomrule
    \end{tabular}
    \caption{Known and upper bounded phenomenological and circuit distances. Cases that are known exactly are calculated using integer programming. See \app{cplex} for more details. For problem sizes that are too large for exact solution, upper bounds can be obtained by running a decoder \cite{bravyi2024high}.}
    \label{tab:distances}
\end{table}

We validate the bicycle instructions in multiple ways. 
We have computed the phenomenological fault-distances for many of the bicycle instructions for both the gross and two-gross codes using mixed-integer programs solved with CPLEX \cite{cplex2022v22} and found them to either be equal to the original code distances or one less (as expected for our inter-module construction), see \tab{distances}. 
We have also attempted to numerically find low-weight circuit-logicals. While this only results in upper bounds on the circuit distance (also provided in \tab{distances}), not finding any smaller circuit logicals after considerable searching is at least an encouraging indication of having a high circuit distance.
See \app{cplex} for more details on these approaches.
Lastly, we perform full circuit-level simulations of the bicycle instructions and observe the behavior of the logical error rates at low physical error rates, see \fig{data-logical-error-rates}.
This data is used to extract the logical error rates of each bicycle instruction presented in \tab{logical_ops}, and we provide more details on our simulation approach in \app{logical-op-simulations}.

\paragraph{Failure-spectrum ansatz fitting} 
To estimate logical failure rates beyond the reach of standard Monte Carlo methods for low logical error rates, we employ a technique introduced in Ref.~\cite{LowLogicalErrorPaper}, inspired by importance sampling. 
This technique is most easily applied using a noise model in which all faults occur independently with equal probability, but the strength-$p$ circuit noise model we wish to model has faults which occur with various multiples of $p/15$. 
To this end, we approximate the circuit noise model by another model where each fault occurs with probability $q = p/15$, which is constructed from the original model by including additional copies of some faults.
For example, a fault with probability $3p/15$ in the original model is replaced by three independent faults, each occurring with probability $q$.
This yields a corresponding logical error rate, $P_\text{iid}(q)$, that matches the logical error rate, $P(p)$, up to $O(p^2)$ corrections in the probabilities of individual faults.
Let $N$ be the number of faults in the resulting noise model.
We then make use of the relation $P_\text{iid}(q) = \sum_{w=0}^{N} f(w) \binom{N}{w} q^w (1 - q)^{N - w}$, where $f(w)$ is the \emph{failure spectrum}, defined as the fraction of weight-$w$ fault configurations that lead to logical failure. 

In \cref{fig:data-logical-error-rates}, we fit the data estimating $f(w)$ to a simple ansatz $f_\text{ansatz}(w)$ that vanishes for $w < w_0 = \lceil d_\text{circ} / 2 \rceil$ and depends on two fit parameters. 
This ansatz enforces the correct low-$p$ scaling of $P(p)$ and has been observed to accurately capture its behavior at intermediate $p$ across a wide range of QEC systems~\cite{LowLogicalErrorPaper}.
For these fits, we take the circuit distance $d_\text{circ}$ to be 10 for gross and 18 for two-gross, consistent with the upper bounds in \cref{tab:distances}. 
For the gross code, in-module and inter-module measurements are estimated using $\bar{Y}_1$ and $\bar{X}_1 \otimes \bar{X}_1$ as representatives, with circuits that initialize the LPU, and run $C$ syndrome cycles in the merged code before disentangling the LPU.
We observe some deviation from the ansatz $f_\text{ansatz}(w)$ at higher $w>80$ for $\bar{Y}_1$, and we make the conservative choice to exclude those points from the fit (including them would result in lower logical error estimates).
Further details are given in Ref.~\cite{LowLogicalErrorPaper} and \cref{app:logical-op-simulations}.

We remark that in simulations of the inter-module measurement we choose $C$, the number of cycles performed in the deformed code, to be equal to the corresponding upper bound on circuit distance provided in \tab{distances}. Specifically, $C=10$ for the gross code in-module and inter-module operations, and $C=18$ for the two-gross in-module operations. For measurements, this ensures that errors that flip the logical measurement result are suppressed equivalently to errors on the unmeasured logical qubits, see \app{cplex}. 

The extrapolated logical error rate for the two-gross idle instruction in \cref{fig:data-logical-error-rates} appears strikingly low compared to that of a surface code of similar distance. 
At $p = 0.001$, we estimate a logical error rate of about $10^{-20}$ for the two-gross code, versus about $10^{-10}$ for a rotated $d=17$ surface code (using the approximate formula $0.1 \cdot (p/0.01)^{d/2}$ for the surface code per-cycle logical error rate~\cite{fowler2012surface,wang2011surface}). 
While we caution against over-interpreting this extrapolation in isolation, the result may reflect a genuine advantage: the two-gross code has significantly fewer low-weight logical operators. For instance, it has just 336 weight-18 $X$-logicals, see \app{basis}, \fig{logical_counts}, versus $875\,178$ in the $d=18$ rotated toric code~\cite{beverland2019role}, which we expect to be comparable to the surface code.
A greater number of low-weight logical operators generically results in an increase in logical error rate by the introduction of an entropic contribution.

\subsection{Future directions}
\label{sec:future-directions-ft}

Here we present a number of future research directions that relate to the bicycle architecture presented in this section. Throughout our design and simulation of the bicycle architecture in this first iteration, we made several choices often balancing simplicity and efficiency. 
Our main goal was to get a baseline for the capabilities of fault-tolerant architectures based on quantum LDPC codes and surgery, directly comparable to baseline surface code architectures \cite{litinski2019game}, but there is great potential for improvements:

\begin{itemize}
\item \textbf{Modifying the bicycle instruction set.}
Recent work points to the potential of larger LPUs to measure several operators in parallel \cite{cowtan2025} or enable a larger set of measurements, even the whole logical Pauli group, without relying on code automorphisms~\cite{he2025extractors}.
Could there be a practical version of these constructions applicable to BB codes? Notably, an affirmative answer would drastically reduce the time overhead incurred when mapping a logical circuit to the bicycle architecture, which is studied in detail in the next section. Furthermore, while we used some strategies from Ref.~\cite{swaroop2024universal} to enable a greater variety of measurements between just code and factory modules, we could also include additional adapters to enable a larger set of code-code module measurements
that could lead to savings during compilation.

\item \textbf{Connectivity.}
We paid attention to the connectivity in our architecture design, particularly the maximum qubit degree. 
However, it is known that the qubit degree can be lowered by time-dynamics \cite{mcewen2023relaxing} and code morphing \cite{shaw2025lowering}. 
Could these techniques also be applicable to the complete bicycle architecture? Concatenation also may be used to reduce connectivity constraints \cite{pattison2023hierarchical} or enhance computational capacity \cite{gidney2023yoked,berthusen2025adaptive}.

\item \textbf{Achieving the code distance.}
The circuit distance for the gross code is known to be at most 10, which is less than the code distance of 12, and while the circuit distance of the two-gross code is not known to be any less than its code distance of 18, it would not be surprising if this was actually the case. See \tab{distances}. Thus, it would be useful to find syndrome measurement schedules with larger circuit distance for these codes, while simultaneously achieving logical error rates that are no worse than the current schedules at practical physical error rates.

\item \textbf{Simulations of realistic noise.}
Improved understanding of the bicycle architecture can also come from more sophisticated and comprehensive simulations. 
As one example, we should relax the assumption that all circuit components fail with the same rate. 
This is especially significant for measurements and Bell-couplers, which may be expected to be as much as an order of magnitude worse than gates. 
While this will undoubtedly lead to worse estimates, how much so is not clear, and there is even cause for optimism from similar studies of 2-dimensional codes \cite{carroll2024subsystem,ramette2024fault,jacinto2025network}.

\item \textbf{Improved inter-module operations.}
Our present simulations suggest the logical error rates of inter-module operations to be particularly limiting to the overall computational capability of the architecture. Though it certainly warrants further study, we do not believe this to be a fundamental problem. There are a few potential reasons the logical error rate of inter-module operations is currently subpar. First, there is the structure of the code-code adapter itself, which introduces a large number of weight-11 or weight-17 (depending on the code) logical operators. Other constructions, such as we illustrate in \cref{app:lpu_meas_procedure}, \cref{fig:code_code_2_2}, can reduce the number of these logical operators and hopefully also improve logical error rates. Second, the Relay decoder could be performing sub-optimally for a couple reasons. Faced with weight-2 checks in the code-code adapter (see \cref{fig:code_code_1_1}), a belief-propagation decoder like the relay-BP decoder may more often fail to converge as it struggles to decide on which qubit in the weight-2 check the error occurred, though the two choices are equivalent. This could possibly be remedied by adjusting error priors or combining two separate decoding strategies, one for the LPU and one for the BB code, as in Ref.~\cite{CHRY24}. Alternatively, though we did tweak them slightly compared to those for idle and in-module operations, the relay-BP decoder parameters are unlikely to be fully optimized for the inter-module measurements. This optimization is known, for idle operations at least, to make order-of-magnitude differences in logical error rates \cite{RelayPaper}.

\item \textbf{Probe low-logical error regime.} The data extrapolations we have presented here show very low logical error rates for the two-gross code, considerably below that estimated for a surface codes of similar distance. 
It would be valuable to identify how accurate this extrapolation is. 
\end{itemize}

In some places, we made assumptions that we would like to either justify or relax in future work. 
For example, when a logical measurement requires $C$ cycles to perform fault tolerantly, we have assumed that another logical measurement can begin right after the $C^{\text{th}}$ cycle has completed as quickly as the LPU qubits can be measured and reinitialized. Similarly, we do not separately consider a BB code module initialization step at the start of a computation, and instead assume that the logical state of each module is fault-tolerantly initialized from a physical product state by the first $C$-cycle logical measurement it undergoes. Both assumptions have precedent in surface code lattice surgery \cite{litinski2019game}.

We also refer the reader to a number of additional opportunities for improvement in \cref{sec:major-challenges}.

\section{Compiling to the bicycle architecture}\label{sec:compiling}

Having outlined the components and instructions of the bicycle architecture, in this section we describe a protocol to compile a universal quantum circuit into a \emph{bicycle circuit}, which is a circuit with bicycle instructions, on a linear connectivity (see \cref{fig:gross-architecture})\footnote{We discuss in \cref{sec:compilation-future-directions} how the compilation algorithm we propose can be adapted to any module connectivity with minor adaptions.}.
Our approach intends to strike a balance between simplicity and efficiency, and aims to achieve reasonable performance while also defining a starting point for future optimizations.

\begin{figure}[htb]
    \centering
    \begin{subcaptiongroup}
	\includegraphics{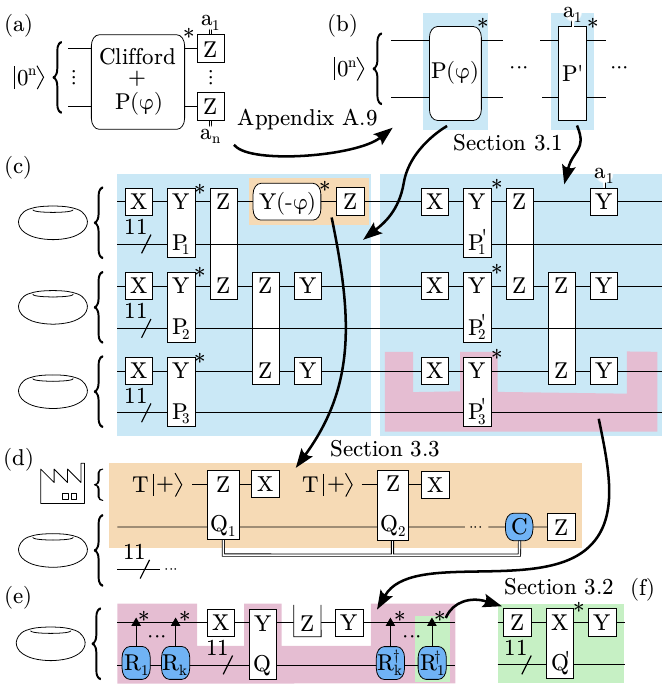}
        \phantomcaption\label{fig:compilerInput}
        \phantomcaption\label{fig:pbc}
        \phantomcaption\label{fig:distributedCircuit}
        \phantomcaption\label{fig:smallAngleSynthesis}
        \phantomcaption\label{fig:measurementSynthesis}
        \phantomcaption\label{fig:nativeRotation}
    \end{subcaptiongroup}
    \captionsetup{subrefformat=parens}
    \caption{%
        \label{fig:compilerSummaryFigure} An overview of the compilation process detailed in \cref{sec:compiling}.
        We use rectangular boxes for measurements and rounded for unitaries, see \cref{eq:diagrammar}.
        Register wires drawn through an operation are unaffected.
        We mark operations that are not bicycle instructions with an asterisk and expand each further.
        \subref{fig:compilerInput} The input of the compiler is a Clifford+$P(\varphi)$ circuit
        that assumes a state prepared in $\ket{0^n}$ and at the end measures every qubit measured in $Z$.
        \subref{fig:pbc} The input is translated to Pauli-generated rotations and Pauli measurements.
        \subref{fig:distributedCircuit} Decomposing gates across multiple code blocks.
        We draw tori to emphasize qubits in the same code module.
        \subref{fig:smallAngleSynthesis} Implementing $Y(-\varphi) = e^{-i\frac{\varphi}{2} Y}$ using a T factory.
        The Paulis $Q_2, \ldots, Q_k$ depend on previous measurement results.
        The resulting Clifford correction, $C$, can be absorbed into the subsequent measurement.
        \subref{fig:measurementSynthesis} Implementing arbitrary Pauli measurements using conjugation by Clifford rotations, $R_i(\pi/2)$. 
        \subref{fig:nativeRotation}
        This can be applied with an $X \otimes Q'$ in-module bicycle measurement and automorphisms, requiring an ancilla (marked by a triangle).
    }
    \label{fig:pbcGrossArchitecture}
\end{figure}

We assume the input circuit acts on $n$ qubits, initially prepared in the $\ket{0^n}$ state, and consists of a sequence of Clifford unitaries, Pauli-generated rotations $P(\varphi) \coloneqq e^{i \frac{\varphi}{2} P}$ for $P \in \mathcal{P}_{n}$ (where $\mathcal{P}_{n}$ is $n$-qubit Pauli group), and final computational basis measurements\footnote{Note that throughout this section, we drop the bar over logical operations for convenience, since here all operations are at the logical level.}.
A summary of the compilation algorithm is presented in \fig{pbcGrossArchitecture}
and consists of the following steps:
\begin{enumerate} 
    \item Compile the input circuit into one composed entirely of Pauli-generated rotations and multi-qubit Pauli measurements (also known as Pauli-Based Computation)~\cite{bravyi2016trading,litinski2019game}, which we review in \app{PBC};
    \item Distribute Pauli-generated rotations and measurements across code modules. See \sec{entanglement}; 
    \item Synthesize arbitrary Pauli measurements on a code module using bicycle instructions. See \sec{local-Paulis};
    \item Synthesize a small-angle $X(\varphi)$ gate on the pivot qubit of the code module adjacent to the factory module via T state injections.
    See \sec{small-angle-rotations}.
\end{enumerate}
Finally, in \sec{compilation-future-directions}, we discuss opportunities for improving the compilation strategy proposed here.

Different from the unitary circuits common in the quantum computing literature,
the bicycle instructions are mostly measurement-based
and require additional consideration of measurement outcomes and Pauli corrections.
To facilitate clear analysis of these circuits, we introduce a diagrammatic circuit language.
For any unitary $V$ and anti-commuting Pauli operators $P$ and $Q$, we define the following graphical representations:
\begin{align}\label{eq:diagrammar}
    V &\coloneqq \vcenter{\hbox{\includegraphics{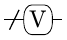}}},
    & e^{i \frac{\pi}{4} P} &\coloneqq \vcenter{\hbox{\includegraphics{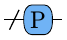}}},
    & \frac{\idty + {(-1)}^a P}{2} &\coloneqq \vcenter{\hbox{\includegraphics{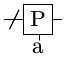}}},
    & & \vcenter{\hbox{\includegraphics{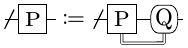}}}.
\end{align}
From left to right, these represent general unitary gates (including Pauli-generated rotations), Clifford gates generated by a Pauli, projective measurements with outcome $a \in \set{0, 1}$, and \emph{measurement projections}.
Measurement projection diagrams simplify circuit diagrams by omitting explicit Pauli corrections by assuming they are implicitly applied.
For a measurement projection of $P$, a Pauli correction $Q$ is conditionally applied when the measurement outcome is `1' to ensure that the state afterwards is stabilized by $P$.
This is achieved by applying any Pauli $Q$ that anti-commutes with $P$ and stabilized the state just before the measurement of $P$.
Pauli corrections for a circuit containing measurement projection diagrams can therefore be systematically determined if needed.

\subsection{Distributing operations across data modules}
\label{sec:entanglement}

To translate $n$-qubit Pauli measurements and Pauli-generated rotations (see \fig{pbc}) into bicycle instructions, we first address the challenge of distributing these operations across modules, each hosting 12 logical qubits with limited inter-module connectivity.
We assign the $n$ logical qubits to $M = \lceil n/11 \rceil$ modules, using 11 qubits per module for data and reserving one qubit (the pivot) as an ancilla to support both inter-module coordination and in-module operations (see \sec{local-Paulis}).
In this subsection, we compile the circuit into one using only bicycle instructions and in-module Pauli measurements (along with a single-qubit rotation for the Pauli-generated rotation) as depicted in \cref{fig:distributedCircuit}.
The in-module Pauli measurements are compiled into in-module bicycle instructions in later subsections.

Our approach relies on the fact that the basic operations of measuring an $n$-qubit Pauli $P$ and applying the Pauli-generated rotation $P(\varphi)$ can be implemented using an ancilla and a controlled-$P$ gate:
\begin{equation}\label{eq:pauliMeasurement}
\includegraphics{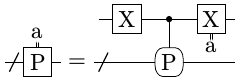}
\end{equation}
for Pauli measurements, and
\begin{equation}\label{eq:pauliRotation}
\includegraphics{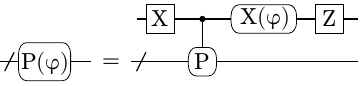}
\end{equation}
for Pauli-generated rotations, requiring the ability to apply $X(\varphi)$ to the ancilla.
We now systematically re-express these circuits until we obtain equivalent circuits using the desired operations.

First, factoring $P$ into $P = P_1 \otimes \dots \otimes P_M$, where each $P_i$ acts on the 11 non-pivot qubits of module $i$,
the controlled-$P$ gate in \cref{eq:pauliMeasurement,eq:pauliRotation} can be implemented as follows (illustrated for $M=3$ modules)\footnote{%
If $P$ does not have support on some code modules then we may optimize out operations on these code modules in some cases.
For rotations, we find the largest range of modules starting from the factory module that have nontrivial $P_i$.
For measurements, we consider the largest range of modules that have nontrivial $P_i$.
From here on, we assume that we have computed this range for the respective Pauli-generated rotation or Pauli measurement and we are only looking at those modules within this range.
}:
\begin{equation}\label{eqn:distribute_controlled_rotations}
    \includegraphics{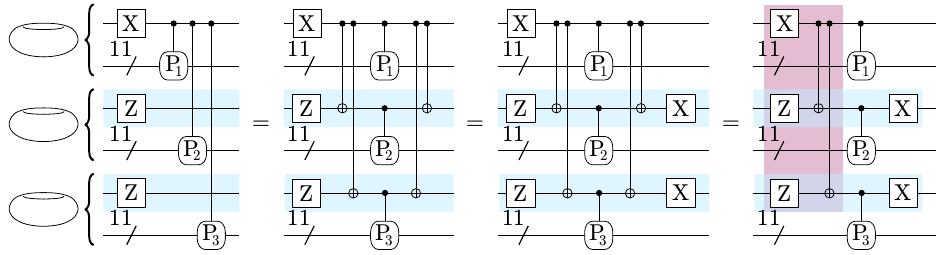}
\end{equation}
The first circuit diagram in \cref{eqn:distribute_controlled_rotations} prepares each pivot (blue) in $\ket{0}$ using a $Z$ measurement projection and leaves it untouched, while the second circuit prepares each pivot in $\ket{0}$, involves it in nontrivial operations, and returns it to $\ket{0}$ at the end.
Applying an $X$ measurement projection to these pivots at the end of the second diagram does not affect the other qubits and yields the third circuit diagram.
To obtain the fourth diagram, simply remove CNOTs targeting qubits measured in $X$, as their action is trivial up to Pauli corrections.

Next, note that the red region in the fourth circuit in \cref{eqn:distribute_controlled_rotations} prepares a GHZ state, which can equivalently be prepared as
\begin{equation}\label{eq:ghz_prep}
\includegraphics{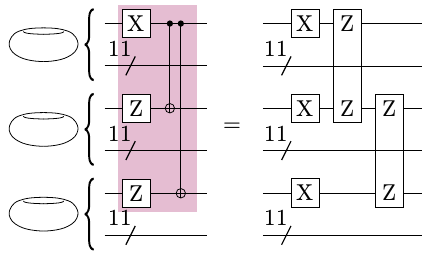},
\end{equation}
where register wires are drawn through the box to indicate qubits unaffected by the operation.
Replacing the CNOT-based GHZ preparation in the fourth circuit of \cref{eqn:distribute_controlled_rotations} yields a circuit for controlled-$P$ that uses only $X$ and $Z$ measurements of pivot qubits, $ZZ$ measurements of pivot qubits on adjacent modules (bicycle instructions), and controlled-$P_i$ unitaries within modules
(which are not bicycle instructions). 
The result is that by using \cref{eqn:distribute_controlled_rotations,eq:ghz_prep} for \cref{eq:pauliMeasurement,eq:pauliRotation},
we find that each module requires one of the following two circuits:
\begin{align}\label{eq:rotation_control}
    &\includegraphics{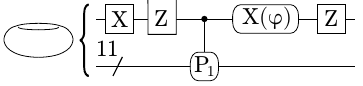} & &\text{and} & &\includegraphics{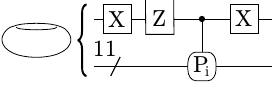}
\end{align}
For simplicity, we depict only one end of one $ZZ$ measurement occurring on a code module.
Specifically, for \cref{eq:pauliMeasurement}, all modules use the circuit on the right-hand side of \cref{eq:rotation_control}, while for \cref{eq:pauliRotation}, the first module uses the circuit on the left-hand side, and all others use the right-hand side.
Hence, we write $P_1$ on the left-hand side.
At this stage, the original operations ($P$ measurement and $P(\varphi)$) have been reduced to inter-module bicycle measurements and in-module operations as shown in \cref{eq:rotation_control}.

We now describe how to further decompose the remaining in-module operations.
If $P_i = \idty$, the controlled-$P_i$ gate is removed, completing the implementation for this module.
Otherwise, we proceed with the following circuit transformations.

We now show how to implement it using Pauli measurements. 
This is achieved by first decomposing the controlled-$P_i$ gates in terms of Clifford rotations about $Z\otimes P_i$, $Z$, and $P_i$ separately as~\cite{litinski2019game}
\begin{equation}\label{eq:controlledRotationIdentity}
    \includegraphics{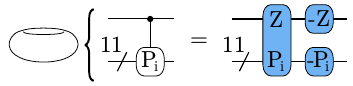}.
\end{equation}
Applying \cref{eq:controlledRotationIdentity} to the left circuit of \cref{eq:rotation_control}, we obtain
\begin{align}
    \vcenter{\hbox{\includegraphics{rotation_control_commutation}}} &= \vcenter{\hbox{\includegraphics{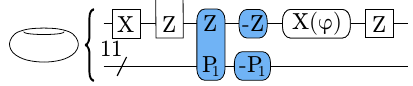}}} \\
    &= \vcenter{\hbox{\includegraphics{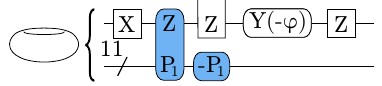}}}, \label{eq:removeZRotation}
\end{align}
where we have commuted the $Z\otimes P_1$ rotation forward, and the $-Z$ rotation to the end of the circuit and absorbed it into the final $Z$ measurement projection.
Similarly, we apply \cref{eq:controlledRotationIdentity} to the right circuit of \cref{eq:rotation_control}.

We use the fact~\cite[Prop.~2.2]{kliuchnikov2023stabilizer} that, for any two anti-commuting Paulis $Q$ and $R$,
\begin{equation}\label{eq:antiCommutingMeasurement}
    \includegraphics{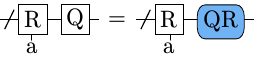}
\end{equation}
to turn the $Z\otimes P_i$ Clifford rotation into a $Y \otimes P_i$ measurement.
By using \cref{eq:antiCommutingMeasurement} on \cref{eq:removeZRotation}, we obtain
\begin{align}
    \vcenter{\hbox{\includegraphics{remove_z_rotation}}} &= \vcenter{\hbox{\includegraphics{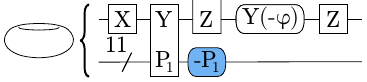}}} \label{eq:grossRotation} \\
    \intertext{and similarly, for measurements, the right side of \cref{eq:rotation_control} equals}
    \vcenter{\hbox{\includegraphics{rotation_control_no_rotation}}} &= \vcenter{\hbox{\includegraphics{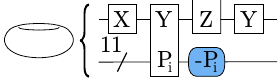}}}. \label{eq:grossMeasurement}
\end{align}
These forms are expressed entirely in terms of Pauli measurements and known Clifford corrections (and a single-qubit rotation $Y(-\varphi)$ to be handled in \cref{sec:small-angle-rotations}).
The remaining $-P_i$ Clifford rotations could be implemented using the techniques from \sec{local-Paulis}.
However, since these are Clifford rotations that affect only a single module, they can instead be commuted to the end of the circuit.
The process of commuting Clifford rotations will change the basis of future Pauli-generated rotations and measurements,
which may change the cost of implementing them.
Assuming we commuted the $-P_i$ Clifford rotations to the end,
we obtain the circuit in \cref{fig:distributedCircuit} for implementing Pauli-generated rotations and measurements across modules.

\subsection{In-module Pauli measurement synthesis}%
\label{sec:local-Paulis}

This subsection focuses on handling the in-module $Y \otimes P_i$ measurements for arbitrary $P_i \in \mathcal{P}_{11}$ in \cref{eq:grossRotation,eq:grossMeasurement} (highlighted in the compilation outline \cref{fig:measurementSynthesis}).
First observe that it suffices to implement any of the measurements $X \otimes P_i$, $Y \otimes P_i$, or $Z \otimes P_i$, since \cref{eq:grossRotation} and \cref{eq:grossMeasurement} each admit three equivalent forms related by basis changes on the pivot qubit.
For example, for \cref{eq:grossRotation}:
\begin{equation}\label{eq:rotationBases}
\vcenter{\hbox{\includegraphics{rotation_y_gross}}} 
= \vcenter{\hbox{\includegraphics{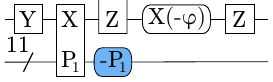}}} 
= \vcenter{\hbox{\includegraphics{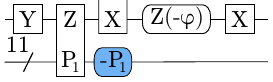}}}.
\end{equation}
Using all three forms requires implementing a rotation $Y(-\varphi)$, $X(-\varphi)$, or $Z(-\varphi)$, which we address in the next subsection.
All other operations are bicycle instructions, aside from these rotations and the $X \otimes P_i$, $Y \otimes P_i$, and $Z \otimes P_i$ measurements.

In this subsection we therefore aim to measure whichever of $X \otimes P_i$, $Y \otimes P_i$, or $Z \otimes P_i$ is easiest for each $P_i \in \mathcal{P}_{11}$.
Our approach builds on Ref.~\cite{CHRY24}, using bicycle instructions to construct 11-qubit Clifford unitaries that map a small set of easy Pauli measurements to a general set, with differences arising from changes in the LPU design.

Recall that bicycle instructions include a set $\mathcal{M}$ of 15 in-module measurements and generate a set $\mathcal{A}$ of 36 shift automorphism unitaries (see \cref{sec:bicycle-arch} and \cref{app:automorphisms}).
We define a set $\mathcal{N}$ of \emph{native measurements}, given by
\begin{equation}
\mathcal{N} = \{ (U \otimes U) P (U \otimes U)^\dagger ~|~ P \in \mathcal{M}, ~ U \in \mathcal{A} \}.
\end{equation}
We view native measurements $\mathcal{N}$ as having approximately the same cost as bicycle measurements $\mathcal{M}$, since each element of $\mathcal{N}$ is implemented using one element of $\mathcal{M}$ along with a small number of shift automorphisms (which generate $\mathcal{A}$ and are faster and less noisy than bicycle measurements; see \cref{tab:logical_ops}).
We similarly define the set of \emph{native rotations} as
\begin{equation}
\{ R(\pi/2) ~|~ Q \otimes R \in \mathcal{N} \text{ for any } Q \in \{X, Y, Z\} \},
\end{equation}
which can be implemented using native measurements (forming \fig{nativeRotation} in the compilation outline).
For example, the rotation $R(\pi/2)$ can be realized given a native measurement $X \otimes R$, $Y \otimes R$ or $Z \otimes R$ using:\footnote{%
The single-qubit measurements in each circuit can be switched to produce an equivalent circuit.
}
\begin{equation}\label{eq:native_rotation}
\includegraphics{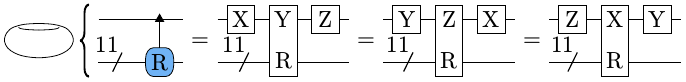},
\end{equation}
which can be proved using \cref{eq:antiCommutingMeasurement}
and that a measurement projection of a register in the same basis as it was rotated removes support of that rotation on the register.
The triangle on the pivot indicates it participates in the implementation as an ancilla and cannot host another state.

We note that $\mathcal{M}$ is the same for the gross and the two-gross codes but $\mathcal{A}$ is different.
Both codes support a set of 540 native measurements and a set of 510 native rotations, but containing slightly different elements.
In both codes, every $X_i$ and $Z_i$ for $i \in 1, 2 \dots, 12$ is a native measurement, and the native rotations generate the Clifford group~\cite{CHRY24}.

\begin{figure}[tbp]
	\centering
	\includegraphics{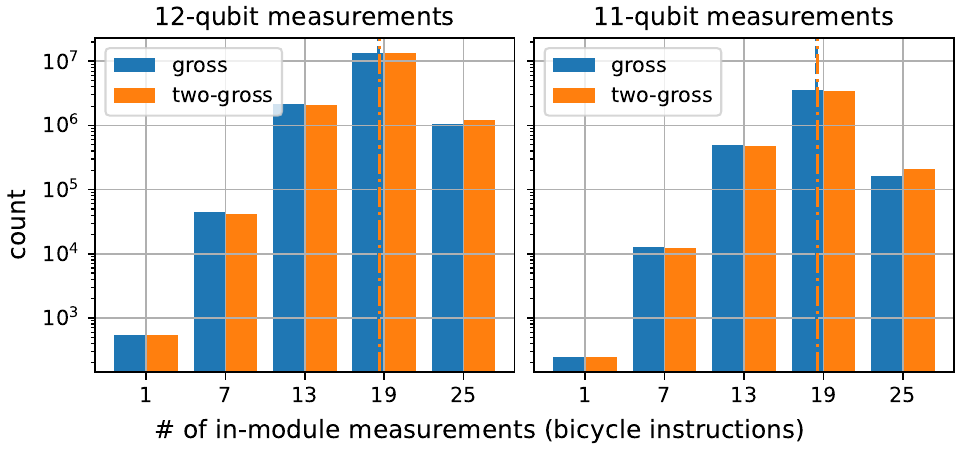}
	\caption{
    Histograms of the number of in-module bicycle measurements to implement Pauli measurements.
    \textbf{Left:} Histogram of $\beta(Q \otimes P_i)$, the number of bicycle measurements to measure $Q \otimes P_i$ for a specific choice of $Q \in \{X,Y,Z \}$, with mean values 18.55 (gross) and 18.67 (two-gross).
    We don't use this in the compilation algorithm since we can choose the pivot basis.
    \textbf{Right:} Histogram of $\beta(P_i) \coloneqq \min\left( \beta(X \otimes P_i), \beta(Y \otimes P_i), \beta(Z \otimes P_i) \right)$, obtained by minimizing over the pivot basis $Q \in \{X, Y, Z\}$, with mean values 18.48 (gross) and 18.58 (two-gross).
	}\label{fig:overhead}
\end{figure}

We now return to the problem of measuring $X \otimes P_i$, $Y \otimes P_i$, and $Z \otimes P_i$ for arbitrary $P_i \in \mathcal{P}_{11}$.
Because the native rotations generate the Clifford group, we can measure $Q\otimes P_i$ for $Q \in \{X,Y,Z \}$ and any $P_i \in \mathcal{P}_{11}$ by conjugating a native measurement $Q\otimes R$ with a sequence of native rotations as
\begin{equation}\label{eq:arbitraryMeasurement}
    \includegraphics{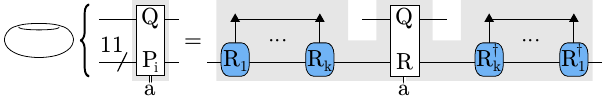},
\end{equation}
where $P_i = (R_1 \dots R_k) R (R_1 \dots R_k)^\dagger$.
Note that the input on the pivot qubit must be inserted halfway through,
outside the gray box,
since the native rotations consume the pivot, forming what is known as a quantum comb~\cite{chiribella2008}.

We compute the number of bicycle measurements in \cref{eq:arbitraryMeasurement} for arbitrary $Q \otimes P_i$,
defined as $\beta(Q \otimes P_i)$.
On the left of \cref{fig:overhead}, we provide a histogram of the number of distinct Paulis $Q \otimes P_i \in \mathcal{P}_{12}$ which result in each value of $\beta$.
In our algorithm, we are free to choose the pivot basis.
So, on the right of \cref{fig:overhead}, we instead compute $\beta(P_i) \coloneqq \min(\beta(X \otimes P_i),\beta(Y \otimes P_i),\beta(Z \otimes P_i))$,
which is thus the quantity of relevance for our algorithm.

We directly apply \cref{eq:arbitraryMeasurement} to implement \cref{eq:grossRotation}, corresponding to \fig{measurementSynthesis} in the compilation outline.
For a given $P_1 \in \mathcal{P}_{11}$, we choose $Q \in \{X, Y, Z\}$ to minimize $\beta(Q \otimes P_1)$ and denote the corresponding sequence of native rotations by $R_1, \dots, R_k$.
If the minimizing choice of $Q$ is $Y$, we obtain
\begin{multline}\label{eq:generalRotation}
    \includegraphics{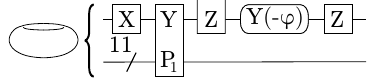}\\
    =  \vcenter{\hbox{\includegraphics{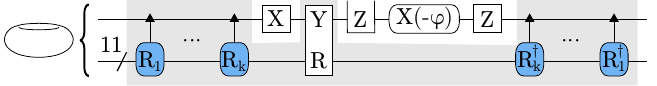}}}
\end{multline}
The same approach is applied to \cref{eq:grossMeasurement}.

\subsection{Small-angle rotations}%
\label{sec:small-angle-rotations}
In this section, we show how a small-angle Pauli-generated rotation $X(\varphi)$, $Y(\varphi)$ or $Z(\varphi)$ for $\varphi \in \mathbb R$
followed by an in-module measurement $Q \in \set{X_1,Y_1,Z_1}$
can be approximated by bicycle instructions and a classically conditioned Pauli measurement of the pivot,
as depicted in \cref{fig:smallAngleSynthesis}.

The first step is approximating $Z(\varphi)$ by gates in $\mathcal T \coloneqq \set{T, X(\frac{\pi}{4}), Y(\frac{\pi}{4})}$
followed by a Clifford circuit.
The rotations $X(\varphi)$ and $Y(\varphi)$ follow as a corollary.
\textcite{selinger2014optimal} give an algorithm that synthesizes non-Clifford
$Z(\varphi)$ up to some accuracy $\varepsilon \in (0, 1)$ into a sequence of $T$, $S$, and $H$ operators followed by a Clifford circuit, $C_0$,
\begin{equation}\label{eq:selingerApprox}
Z(\varphi) \approx H_k T H_{k-1} \ldots H_1 T C_0
\end{equation}
for positive integer $k$, $H_k \in \set{\idty, H, SH}$, and $H_1, \dots, H_{k-1} \in \set{H, SH}$.

We now commute Clifford gates to the end.
The commutation relations of the Hadamard gate are
\begin{align}\label{eq:HCommutation}
    HT &= X\paren*{\frac{\pi}{4}} H, & H X\paren*{\frac{\pi}{4}} &= TH, & \text{and } H Y\paren*{\frac{\pi}{4}} &= Y\paren*{-\frac{\pi}{4}} H = Y\paren*{\frac{\pi}{4}} Y \paren*{\frac{\pi}{2}}H
\end{align}
and the commutation relations of the Clifford Pauli-generated rotations with angle $\frac{\pi}{2}$ are described by
\begin{equation}\label{eq:SCommutation}
    \forall\set{\sigma_1,\sigma_2,\sigma_3} = \set{X,Y,Z} : \sigma_1 \paren*{\frac{\pi}{2}} \sigma_2 \paren*{\frac{\pi}{4}} = \sigma_3 \paren*{\frac{\pi}{4}} \sigma_1 \paren*{\frac{\pi}{2}}.
\end{equation}
Applying \cref{eq:HCommutation,eq:SCommutation} to \cref{eq:selingerApprox}, we obtain
\begin{equation}\label{eq:pi4String}
    H_k T H_{k-1} \ldots H_1 T C_0 = T_k T_{k-1} \dots T_1 C_1
\end{equation}
where $T_1, \dots, T_k \in \mathcal T$,
and where $C_1$ is some Clifford circuit.

For other bases of rotations, we use $X(\varphi) = HZ(\varphi)H$ and $Y(\varphi) = X(\frac{\pi}{2}) Z(\varphi) X(\frac{\pi}{2})$.
By applying \cref{eq:HCommutation,eq:SCommutation}, we obtain a circuit of the form in \cref{eq:pi4String} as well.

The T injection bicycle instruction can implement any gate in $\mathcal T$ with a classically-controlled Clifford correction.
More generally, for any Pauli $P$, it holds that~\cite{litinski2019magic}
\begin{equation}\label{eq:TInjection}
    P \left(\frac{\pi}{4}\right) = \vcenter{\hbox{\includegraphics{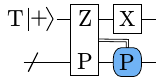}}}
\end{equation}
by using a T state, $T\ket{+}$.
The T injection bicycle instruction enables $P$ to be $X_1$, $Y_1$, or $Z_1$
but requires a $P(\frac{\pi}{2})$ correction for half of the measurement outcomes.
Using \cref{eq:SCommutation}, these conditional corrections can be commuted to the end of the circuit
by classically conditioning the basis of each $T_j$ gate, for $j \in \set{2,\ldots,k}$, on prior results.
Let the Clifford circuit at the end, a function of the measurement outcomes, be $C$,
then the result is depicted in \cref{fig:smallAngleSynthesis} for a final measurement in $Z$ (can also be $X$, see \cref{eq:rotationBases}).
The circuit $C$ can be absorbed into the following measurement to become a conditional $X/Y/Z$-basis measurement.

\paragraph{Clifford+T} As an important example 
consider the special case where $\varphi = \frac{\pi}{4}$,
which occurs in Clifford+T circuits.
Note that \cref{eq:TInjection} is an exact implementation ($\epsilon = 0$) of the desired rotation.
Suppose $Q \in \set{X,Y,Z}$, then we observe \cref{eq:TInjection} implies
\begin{equation}\label{eq:tRotationConcrete}
    \vcenter{\hbox{\includegraphics{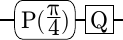}}} = \vcenter{\hbox{\includegraphics{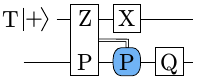}}}.
\end{equation}
We can substitute \cref{eq:tRotationConcrete} at the end of \cref{eq:generalRotation}
to give an exact implementation Pauli-generated rotations of angle $\frac{\pi}{4}$.
It is easy to see that the conditional Clifford, $P(\frac{\pi}{2})$,
can be absorbed by the final measurement to become a conditional measurement in some Pauli basis.

\subsection{Future directions}
\label{sec:compilation-future-directions}

We proposed a compilation scheme (\fig{pbcGrossArchitecture}) for a universal gate set targeting the bicycle architecture.
The only type of entanglement being consumed are GHZ states that connect the code modules.
As long as the code modules form a connected graph,
we can always prepare a GHZ state by measuring along a spanning tree.
For architectures containing a Hamiltonian path, the circuit for preparing the GHZ state is two rounds of inter-module measurements.
Moreover, we can synthesize arbitrary-angle rotations without incurring additional global entanglement overhead,
which increases the rate at which T states can be consumed by the circuit and reduces overall noise
since most of the system is well-protected during idling.
Now let us discuss some open problems:
\begin{itemize}
    \item \textbf{Parallelism} The generality of our compilation scheme may also limit it from producing more efficient circuits.
    While the translation in \cref{fig:pbc} removes all Clifford gates,
    the resulting Pauli-generated rotations may have large support.
    This means that we are entangling the entire quantum architecture for every rotation or measurement
    at the cost of increased noise cost.
    It also forces all Pauli-generated rotations to be performed serially.
    Maintaining the parallelism of Pauli-generated rotations of disjoint support could improve both speed and noise cost~\cite{litinski2019game,chamberland2022universal,beverland2022edpc,beverland2022assessing}.
    The additional overhead of explicit implementation of Clifford gates~\cite{Pllaha_2021,Bravyi_2021}
    rather than simply eliminating them as we have done is to be balanced by the additional support gained by going from $P_i$ to $P_i'$, which increases implementation cost.
    In particular, Clifford gates that cross module boundaries will need to consider
    the available architecture connectivity to be implemented through inter-module measurements.
    
    \item \textbf{Reducing time overhead} Synthesis of arbitrary Pauli measurements incurs significant overhead (\fig{overhead}).
    Especially for sequences of Pauli-generated rotations with a lot of structure,
    it is possible to apply similar Clifford rotations between subsequent rotations, reducing the overhead.
    
    \item \textbf{Toffoli synthesis} The compilation scheme is well-suited for small-angle rotations, but many important quantum algorithms feature large numbers of Toffoli gates,
    which do not benefit from small-angle synthesis.
    Improving this architecture by adding tailor-made schemes for Toffoli gates or other types of non-Clifford gates could also increase consumption rate of T states.
    Moreover, it is possible to synthesize $\text{CCZ}\ket{+}^{\otimes 3}$ states directly,
    each of which can be used to synthesize a Toffoli gate
    instead of seven (four in some cases) $T\ket{+}$ states~\cite{Gidney2019efficientmagicstate,beverland2020lower}.
    
    \item \textbf{Small-angle synthesis} There exist probabilistic small-angle synthesis routines that on average use fewer T states.
    These routines require the use of ancillas and a measurement-and-feedforward circuit~\cite{Kliuchnikov2023shorterquantum}.
    With this moderate space overhead, the leading constant factor in average number of T states needed to inject a small-angle rotation is reduced by about $5.4\times$ when compared to \cite{selinger2014optimal,pygridsynth}.
    It could be useful to allocate some additional space to incorporate these synthesis routines in the architecture.
    Moreover, there are brute-force methods for finding the minimal T count to approximating a small-angle rotation up to a precision of about $10^{-17}$~\cite{kliuchnikov2013fast},
    which is sufficient for many large scale applications of quantum computing.
\end{itemize}

We also refer the reader to a number of additional opportunities for improvement in \cref{sec:major-challenges}.

\section{End-to-end resource estimates}
\label{sec:results}

Quantifying resources such as qubit count and runtime is essential for benchmarking and identifying bottlenecks across algorithm design~\cite{roetteler2017quantum,berry2019qubitization,gidney2021factor,lee2021even,vonburg2021catalysis}, compilation~\cite{litinski2019game,beverland2022edpc}, and fault-tolerant architectures~\cite{litinski2019game,litinski2022active}. 
Existing general-purpose resource estimation frameworks~\cite{azureresource2023,beverland2022assessing,qualtran2023,harrigan2024expressing} focus primarily on architectures implemented using two-dimensional local codes such as the surface code. 
In this section, we extend these methods to evaluate the bicycle architecture and compilation strategy introduced in this work and compare its performance to that of the conventional surface code architecture as described in Ref.~\cite{beverland2022assessing}.
At the end of this section we briefly comment on recent work considering less conventional surface code architectures incorporating code concatenation \cite{gidney2023yoked,gidney2025factor} or transversal gates \cite{zhou2024algorithmic,zhou2025resource}.

We consider two illustrative circuit families. 
The first examines random circuits to characterize the architecture’s performance on unstructured, demanding workloads (results are shown in \cref{fig:logicalCapabilities}). 
The second analyzes time evolution of a two-dimensional transverse-field Ising model (TFIM) simulation as previously considered in Ref.~\cite{beverland2022assessing}, providing a contrasting example of a structured algorithm with scientific relevance. 
A summary of our approach and results follows, with additional details in \cref{app:logical-capability}.

\paragraph{Approach} We assume modules are laid out in 1D nearest-neighbor connectivity, as illustrated in \cref{fig:linearConnectivity},
with $M$ code modules (gross or two-gross) and one factory module (\cref{tab:Tfactories}).
The logical error rate $P_i$ and duration $\tau_i$ of each idle and bicycle instruction can be found in \cref{tab:logical_ops}.
The physical qubits, $q$, and number of code modules, $M$, are related by
\begin{equation}\label{eq:qubit-count}
    q = M (c+u+a) - a + a' + f,
\end{equation}
where the other parameters are given in \cref{tab:qubit-counts}.
The number of logical data qubits available for computation is $n=11M$.

We consider a circuit of the same structure between our two examples, consisting of a sequence of Pauli-generated rotations.
We implement the compilation algorithm from \cref{sec:compiling}
to compile the input circuit on $n$ logical qubits to a bicycle circuit.
For simplicity, we ignore the state preparation and final measurement steps.
Then, the input circuit $\mathcal C = \prod_j Q_j(\varphi_j)$,
with $j$ ranging over positive integers, $Q_j(\varphi_j) \coloneqq \exp(i \frac{\varphi_j}{2} Q_j)$ and $Q_j \in \mathcal{P}_{n}$ and $\varphi_j \in \mathbb{R}$.
We also neglect any Clifford commutation steps that need to happen during the compilation algorithm.
The uniform distribution over Paulis is invariant under Clifford conjugation so our capability estimates are not affected,
but the resource estimates for TFIM simulation require propagating in-module Clifford corrections that could change the results.
We then optimize $\mathcal{C}_1$ by removing subsequent identical in-module measurements to produce an optimized circuit $\mathcal{C}_2$.
The resulting circuit $\mathcal{C}_2$ specifies a partial ordering of bicycle instructions but does not include explicit timing.
To estimate runtime, we augment $\mathcal{C}_2$ with the number of timesteps of each idle or bicycle instruction from \cref{tab:logical_ops} to produce $\tilde{\mathcal{C}_2}$.
We can thus read off the total runtime from $\tilde{\mathcal{C}_2}$.

We assume identical logical errors and consider the computation a failure if any logical error occurs.
The \emph{circuit failure probability} is obtained from $\tilde{\mathcal C_2}$ and \cref{tab:logical_ops}, to leading order, by
\begin{equation}
\label{eq:error-budget}
    P = \sum_{j \in \mathcal I \cup \set{I}} N_j P_j,
\end{equation}
where $j$ ranges over the idle and bicycle instructions
and $N_j$ is the number of times instruction $j$ appears in $\tilde{\mathcal{C}_2}$.
Note that, for the purposes of calculating $P$,
an idle instruction is inserted whenever a module needs to wait for its next operation,
thus rounding up the idling time to the nearest integer number of idles.

\paragraph{Capability estimation: random circuits}

We estimate the capability of the bicycle architecture
by estimating the largest random circuit $\mathcal C$ that can run with circuit failure probability at most $\frac{1}{3}$ for several differently sized systems.
We fix the number of physical qubits, $q$, and physical error rate, $p$,
and investigate the capabilities of the bicycle architecture in five cases of interest:
Ranging from smaller devices with a higher noise of $p=\num{e-3}$,
to the improved capabilities of devices with lower physical noise of $p=\num{e-4}$ scaling to very large devices (up to $q=500k$).

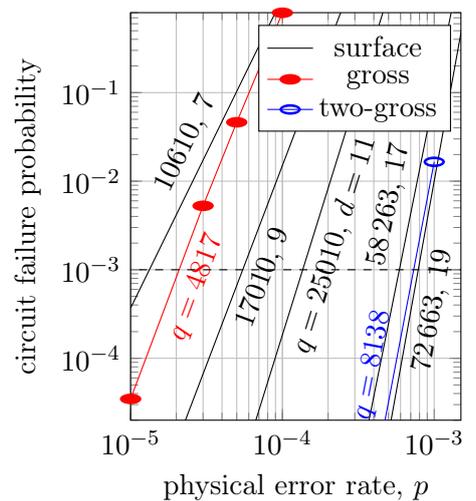
\begin{figure}
{
    \sisetup{
        exponent-mode=scientific,
        table-format = 1.1e2,
        table-auto-round,
    }
    \centering
    \subcaptionbox{Capability estimates\label{tab:failure}}{%
        \centering
        \begin{tabular}{r@{\hskip 1pt}rlrSS[exponent-mode=input,table-format=4.0]}\toprule
            $p$ & $q$ & code & $n$ & $N_T$ & $\frac{\text{timesteps}}{\text{T gate}}$ \\ \midrule
            $10^{-3}$ & 5k & gross & 121 & 13.59580266807207 & 3367.153846 \\
            & & surface ($d=5$) & 45 & 7 & 381 \\
            & 50k & two-gross & 704 & 3471247.948411036 & 7057.337650 \\
            & & surface ($d=17$) & 42 & 2.5e6 & 2269 \\
            $10^{-4}$ & 5k & gross & 110 & 222089.62968351436 & 3090.551150 \\
            & & surface ($d=7$) & 21 & 377928 & 115 \\
            & 50k & gross & 1342 & 37033.2338363515 & 3135.156775 \\
            & & surface ($d=7$) & 250 & 31746 & 115 \\
            & & two-gross & 440 & 7015434087082015.0 & 5435.873010 \\
            & & surface ($d=17$) & 27 & 1.2e15 &  509 \\
            & 500k & two-gross & 6886 & 448297625597020.8  & 5436.914760 \\
            & & surface ($d=17$) & 416 & 7.9e13 & 509 \\
            \bottomrule
        \end{tabular}
    }%
}
    \hfill%
    \subcaptionbox{TFIM requirements\label{fig:TFIMRequirements}}{%
        \centering
        \begin{tikzpicture}
            \begin{loglogaxis}[
                xlabel={physical error rate, $p$},
                ylabel={circuit failure probability},
                ymajorgrids=true,
                xminorgrids=true,
                xmajorgrids=true,
                xmin=1e-5,
                xmax=1.5e-3,
                ymax=8e-1,
                ymin=2e-5,
                width=0.35\linewidth,
                height=2.75in,
                legend pos=north east,
            ]
                \addplot+ [no markers, black, domain=1e-5:1e-4] {0.03 * (x / 0.01)^(7/2) * 7 * (2 * 184000 * 100 + 19036800 - 184000)} node[sloped,above,pos=0.5] {$10610$, $7$};
                \addlegendentry{surface}
                \addplot+ [no markers, black, domain=1e-5:3e-4, forget plot] {0.03 * (x / 0.01)^(9/2) * 9 * (2 * 184000 * 100 + 19036800 - 184000)} node[sloped,below,pos=0.5] {$17010$, $9$};
                \addplot+ [no markers, black, domain=5e-5:5e-4, forget plot] {0.03 * (x / 0.01)^(11/2) * 11 * (2 * 184000 * 100 + 19036800 - 184000)} node[sloped,below,pos=0.5] {$q=25010$, $d=11$};
                \addplot+ [no markers, black, domain=2e-4:2e-3, forget plot] {0.03 * (x / 0.01)^(17/2) * 17 * (2 * 184000 * 100 + 19036800 - 184000)} node[sloped,above,pos=0.55] {$58\,263$, $17$};
                \addplot+ [no markers, black, domain=2e-4:2e-3, forget plot] {0.03 * (x / 0.01)^(19/2) * 19 * (2 * 184000 * 100 + 19036800 - 184000)} node[sloped,below,pos=0.55] {$72\,663$, $19$};
                \addplot [red, mark=donut*] coordinates {
                    (1e-5, 3.4766673348943354e-05)
                    (3e-5, 0.005268591620036828)
                    (5e-5, 0.04624877775057292)
                    (1e-4, 0.7974506221710217)
                    (1e-3, 33258.93337461165)
                } node [sloped, below, pos=0.15] {$q=4817$};
                \addlegendentry{gross};

                \addplot [blue, mark=donut] coordinates {
                    (1e-4, 1.6531588280000003e-11)
                    (1e-3, 0.016531603535021058)
                } node[sloped, above, pos=0.74] {$q=8138$};
                \addlegendentry{two-gross}
        
        
                \addplot[dashed, black, samples=2, domain=1e-6:1e-2] {1e-3};
                
            \end{loglogaxis}
        \end{tikzpicture}
    }
    \captionsetup{subrefformat=parens}
    \caption{%
        \subref{tab:failure} The mean T count, $N_T$, equals the length of a sequence of random Pauli-generated rotations $e^{i \frac{\pi}{8} Q_j}$
        that have full support on $n$ logical qubits with circuit failure rate less than $\frac{1}{3}$ (8 trials).
        We compile a sequence of up to $10^5$ random Pauli-generated rotations to the bicycle architecture and estimate the noise rate of the circuit
        using a linear fit (see \cref{fig:errorRate}).
        All digits for the estimated number of T gates are significant.
        We then supplement these numbers with estimates surface code architectures that have similar capabilities as explained in \cref{app:logical-capability}
        and compute the average time of performing a rotation in timesteps,
        assuming one T factory.
        \subref{fig:TFIMRequirements} Additive estimates of the circuit failure probability when performing transverse-field Ising model simulation.
        We add simplified estimates for some surface code architectures with varying code distances, $d$, at the same T count.
        For each architecture, we denote the required number of physical qubits, $q$, using one $10^{-4}$ gross factory ($f=810$) for the gross code and surface code architectures at $d\le 11$ or one $10^{-3}$ two-gross factory ($f=463$) for the two-gross code and surface code architectures at $d>11$.
        We also mark our accuracy target of $\epsilon = 10^{-3}$ (dashed).
    }%
    \label{fig:resourceEstimates}
\end{figure}

We construct $\mathcal C$ by randomly selecting the Clifford unitaries, which corresponds to randomly selecting non-trivial Paulis $Q_j \in \mathcal P_{n}$ and setting $\varphi_j = \frac{\pi}{4}$ for all $j$,
so that one T injection is sufficient for each rotation.
We find the maximum $k \in \mathbb N$ such that the circuit failure probability when implementing $\mathcal C$ on the architecture is at most $\frac{1}{3}$,
which is suitable for quick convergence on decision problems by a majority vote over multiple trials.
Using $k$, we find the surface code architecture with one T factory that has the closest capabilities (on a logarithmic scale)
and plot selected results in \cref{fig:logicalCapabilities}.
The closest surface code architecture always has fewer logical qubits.
From the compiled circuit, we compute the total runtime and normalize by the number of rotations
to estimate the time to implement one Pauli-generated rotation gate to the bicycle architecture showing the results in \cref{tab:failure}.

Note that the runtime for the bicycle architecture is longer (per T) than for the surface code.
The runtime of both architectures could be improved, sometimes at the cost of additional space.
The main limitation for the bicycle architecture seems to be the overhead in compiling to bicycle instructions.
To reduce this overhead, we discuss increasing the LPU size and synthesis improvements in \cref{sec:future-directions-ft,sec:compilation-future-directions}.
Surface code architectures seem to be mainly limited by the distillation rate of the single T factory.
At the cost of additional qubits for additional T factories,
the surface code architecture can get closer to its theoretical limit of $6d$ timesteps per rotation.

\paragraph{Requirements for simulating the transverse-field Ising model}
We estimate the physical noise rate, $p$, that is required for the gross and two-gross code to successfully perform a 2D $10 \times 10$ TFIM simulation to about $10^{-3}$ error.
The TFIM is a scientific toy model and analyzing its time dynamics has been highlighted as among the easiest problems believed to be classically intractable that could be analyzed on a quantum computer~\cite{beverland2022assessing}.
Relatively small system sizes of around one hundred logical qubits are just beyond current classical computation capabilities making it a likely candidate for early quantum computing insights about otherwise intractable questions.

{
\sisetup{
    exponent-mode=scientific,
    round-mode=figures,
    round-precision=2, 
}
We choose the same parameters for the TFIM Hamiltonian as \cite{beverland2022assessing}
and obtain a circuit with $\num{184000}$ Pauli-generated rotations (of some small angle).
We neglect the error contribution of the T factory, only allocating its physical qubits, $f$ (see \cref{tab:Tfactories}).
We find that $q=4817$ qubits would be sufficient for the bicycle architecture with gross codes
by using one $10^{-4}$ gross factory.
This is considerably fewer qubits than the $17\,010$ qubits we estimate would be needed for the $d=9$ surface code as shown in \cref{fig:TFIMRequirements}.
However, to successfully simulate the TFIM would require a very low physical error rate of $p \le \num{2.0855E-05} \approx 10^{-4.68}$ for the gross code, whereas the $d=9$ surface code could achieve it with a higher physical error rate of $p \le \num{5.4E-05} \approx 10^{-4.26}$.

To suppress the error further in large computations, we construct a bicycle architecture with two-gross codes
using a $10^{-3}$ two-gross factory.
This architecture needs needs $q=8138$ physical qubits
and a physical noise rate $p \le \num{7.3220E-04} \approx 10^{-3.14}$
to successfully simulate a TFIM.
We use the $10^{-3}$ two-gross factory because the logical error rate of T injection roughly needs to be on the order of $10^{-10}$ since we need $N_T = \num{19036800}$ T injections
and improvements in the physical noise rate from $10^{-3}$ to $10^{-3.14}$ could plausibly improve logical error rate of the T factory from $2 \times 10^{-8}$ to $10^{-10}$.
A comparable surface code architecture requires a physical noise rate $p \le \num{5.8E-04}$ at distance $d=17$ (requiring $q=58\,263$ physical qubits) or $p \le \num{7.8E-04}$ at distance $d=19$ (requiring $72\,663$ physical qubits).

\paragraph{Further remarks on resource estimates}
The bicycle architecture shows about an order of magnitude improvement in the number of logical data qubits, $n$, over surface code architectures
when fixing the T count, $N_T$.
When compared to our application regions in \cref{fig:logicalCapabilities} (details in \cref{app:logical-capability}),
the bicycle architecture at 5k physical qubits with $p=10^{-3}$ requires perhaps an order of magnitude improvement in its T count capability before entering the ``unexplored classically'' regime.
It seems plausible that such an order of magnitude improvement could be achieved by about one order of magnitude improvement in logical error rates on inter-module measurements.
In general, we expect that further research and development will end up increasing the capabilities, both in logical T count and logical qubit count,
whereas algorithmic improvements will lower the capabilities required.

For scientific toy problems, like simulating the TFIM, a bicycle architecture with two-gross codes seems more suitable than using gross codes
and is possible at a physical noise rate of about $10^{-3.14}$,
yet requires much fewer physical qubits than a comparable surface code architecture.
However, improvements in- and inter-module measurements, could make a bicycle architecture with gross codes also feasible for simulating TFIM at $p=10^{-4}$.
See \cref{sec:future-directions-ft} for more discussion on improving the logical error rate of inter-module measurements.
}

\paragraph{Advances in surface code architectures}
Concurrent work has pushed surface code architectures and associated resource estimates in new directions by including concatenation with an outer code (``yoking'')~\cite{gidney2023yoked,gidney2025factor} or going from superconducting hardware to a neutral atom platform to incorporate transversal gates and correlated decoding~\cite{zhou2024algorithmic,zhou2025resource} into the design.
These are significant changes to the conventional surface code architecture \cite{beverland2022assessing}, which we compare against in this section and in \cref{fig:logicalCapabilities}.
We consider it likely that improvements to the bicycle architecture similar to these recent results will be developed with further research. 

\section{Conclusions and opportunities}
\label{sec:conclusion}

We introduced the bicycle architecture as a fault-tolerant quantum computing platform based on quantum LDPC codes. By employing the gross and two-gross bivariate bicycle codes, we constructed a modular architecture satisfying all six criteria for a scalable fault-tolerant architecture. 
The bicycle architecture is \textit{fault-tolerant} with respect to circuit noise, \textit{addressable} via logical processing units that support selective Pauli measurements, \textit{universal} through a combination of Clifford operations and T injection, \textit{adaptive} with real-time decoding and feedforward logic, \textit{modular} through self-contained code blocks connected via long-range Bell pairs, and \textit{efficient} by offering significantly lower qubit overhead than conventional surface code architectures.

A central contribution of this work is the design of custom fault-tolerant instructions tailored to the properties of bivariate bicycle codes. These bicycle instructions include shift automorphisms, in-module and inter-module Pauli measurements, and T injection operations.
We showed how high-level quantum algorithms can be compiled to the bicycle instructions
and avoid the need to directly implement Clifford gates.
Moreover, we show how to synthesize small-angle Pauli-generated rotations via local T state injection and synthesis.

Together, the compilation algorithm and the bicycle instructions define a coherent programming and execution model for the bicycle architecture, enabling end-to-end resource estimation. To demonstrate this we considered two problems: Random quantum circuits and a transverse-field Ising model on a $10\times10$ 2D grid. Our results show that the bicycle architecture can support significantly larger logical circuits (about an order of magnitude) within fixed physical qubit budgets compared to the conventional surface code architecture, albeit with increased runtime due to compilation into the bicycle instructions.

A key advantage of a large-scale fault-tolerant quantum computer based on the bicycle architecture is that we believe it is more practical to build than the conventional surface code architecture. While bicycle codes require more and longer-range couplers within a module, the problem of engineering the appropriate connectivity is not insurmountable, even in superconducting hardware~\cite{bravyi_future_2022}, with demonstrations already appearing in the literature~\cite{wang2025demonstration}. And, provided this connectivity, the advantage of the bicycle architecture with respect to fabrication yield is particularly significant.

For instance, consider superconducting qubit technologies like those used by IBM and Google, where each qubit requires 1--2 junctions and each coupler 2--3 junctions. Assuming the larger junction counts and with an optimistic junction yield of 99.98\%, the probability of successfully fabricating a gross code module (memory and LPU) for the bicycle architecture is about 2 in 5, and for the two-gross code, about 1 in 5. 
In stark contrast, yielding a fully functioning monolithic surface code chip with $5\,000$ physical qubits happens just 1 time in about 3,000 samples, and perfect chips with $50\,000$ or $500\,000$ qubits are astronomically unlikely. Instead of waiting for a miracle, defects have to be tolerated in any realistic monolithic architecture. For example, encoding around defects is possible \cite{stace2009thresholds,auger2017fault,lin2024codesign,debroy2024luci,leroux2024snakes} but typically results in lower effective code distance and thus lower computational capacity.

Alternatively, but still unsatisfactorily, surface codes can be made modular. For example, consider surface code modules with 11 computational logical qubits and a single ancilla logical qubit for Clifford gates.\footnote{The surface code module we consider for this argument is optimistically idealized. In reality with just 2D-local connectivity, more ancilla logical qubits per surface code module are necessary for Clifford computation \cite{litinski2019game}.} Matching the logical capability of a two-gross code module on random circuits requires surface codes with distance 17, see \fig{resourceEstimates}. However, for distance 17, one expects a fully operational surface code module from just 1 in $66\,000$ samples. Using instead smaller modules, each with one surface code logical qubit, would improve fabrication yield of individual modules, but at the expense of requiring 10 times more inter-module couplers, each with its own risk of connection failure.

In summary, building a system that is modular and 10 times smaller has an outsized effect on decreasing the engineering challenge, placing much looser demands on yield and fabrication tolerances.

\subsection*{Opportunities}
\label{sec:major-challenges}

Here we discuss directions of research that we expect could yield very significant improvements for the bicycle architecture.
\begin{enumerate}
    \item \textbf{Reducing time overhead}. 
    While \fig{logical-circuit-capabilities} shows that our bicycle architecture requires an order of magnitude fewer physical qubits than the surface code to implement a given logical circuit, it would take more than an order of magnitude longer to run (see \cref{tab:failure}). We expect there to be compilation improvements, particularly ones unique to the bicycle architecture, to reduce this time overhead. For instance, a tradeoff we have cursorily considered is allocating more logical qubits per block to be pivot qubits, which can sometimes yield modest improvements. However, it is also possible that the strategy of compiling any logical measurement using a sequence of Clifford conjugations, \cref{eq:arbitraryMeasurement}, could itself be replaced to improve the time overhead. Similarly, circuits can be optimized to rely primarily on Pauli matrices with smaller synthesis overheads.
    
    \item \textbf{Decoding with speed and accuracy}
    To meet the fault-tolerant architecture criterion regarding adaptivity, we require real-time decoding of each of our bicycle instructions. For some instructions, such as idles and automorphisms which involve just a single code, this is expected to be possible given recent work \cite{RelayPaper}. For others, particularly the largest instructions for doing inter-module measurements, accurate real-time decoding is less clearly possible with current techniques. Perhaps careful choice of decoder parameters or new ways of breaking up the problem~\cite{CHRY24} could be used to solve these larger decoding problems successfully.
    
    \item \textbf{Increasing code and circuit distances}.
    At some point, reducing the physical error rates becomes untenable and we will also have to consider codes of larger distance.
    Luckily, the BB code family contains some examples \cite{bravyi2024high}. Also, it is reasonable to expect multiple ``gross code families” to exist based on descriptions of BB codes as coupled copies of the toric code \cite{liang2025generalized}. As an example, if we select integer $r\in\mathbb{Z}$ and bit $b\in\{0,1\}$, we can define a BB code with $\ell = 6(r+b)$, $m=6r$ and the same polynomials that we used for the gross and two-gross codes: $A=1+y+x^{3}y^{-1}$, $B=1+x+x^{-1}y^{-3}$. We find $n=72r(r+b)$, $k=12$ and conjecture $d=6(2r+b-1)$.
    With fixed $A$, $B$ polynomials independent of $r$, the code family is actually 2D-local, and therefore must satisfy the Bravyi-Poulin-Terhal bound \cite{bravyi2010tradeoffs} on code parameters $kd^2=O(n)$. However, 2D-local BB codes achieve much better ratios $kd^2/n$ than the surface code. Assuming the conjectured distance, as $r$ grows this ratio approaches $24$, compared with $1$ for the surface code. For the gross code ($r=1,b=1$) and the two-gross code ($r=2,b=0$), this ratio $kd^2/n$ is already reasonably large, 12 and 13.5, respectively. Alternatively, given the capability of long-range range coupling, we expect to able to implement a wide variety of quantum LDPC codes \cite{breuckmann2021quantum}, including perhaps those with constant rate \cite{tillich2014quantum,gottesman2014fault}, good code parameters \cite{panteleev2022asymptotically}, or concatenation \cite{pattison2023hierarchical,berthusen2025adaptive}. Also in the concatenation category, yoking LDPC codes might similarly be expected to improve the space overhead like for surface codes \cite{gidney2023yoked}.

    \item \textbf{Modifying the bicycle instructions}.
    Other opportunities lie in improving and extending the bicycle instructions we have presented. For instance, there are likely simpler circuits, perhaps employing circuit morphing techniques \cite{shaw2025lowering} and time-dynamics \cite{mcewen2023relaxing}, to improve the automorphism, in-module, and inter-module logical operations. Some of our more complicated and logical error-prone instructions, such as the inter-module measurements, could especially benefit from simplifying improvements. Further, our time overheads in general could be improved by in some way reducing the time cost of surgery operations to less than $d$ syndrome cycles, or alternatively employing techniques for parallelization of logical measurements \cite{cowtan2025}, though both strategies are expected to increase the qubit overhead. In another direction, one might reduce the time overhead by enlarging the set of fault-tolerant logical operations beyond our bicycle instructions, \fig{bicycleInstructions}. For example, we might include more shift automorphisms, perhaps implemented physically using additional qubit connectivity. Additionally, in-module and inter-module operations may be extended to include more logical Pauli measurements \cite{he2025extractors}, again with the cost of more auxiliary qubits and connectivity.

    \item \textbf{Utilizing long-range connectivity for the T factory}.
    Currently, we assume a T factory based on the surface code, which needs only 2D-local connectivity. Can we do better with the higher connectivity we assume is available for the bicycle architecture? There are potentially many ways to achieve this, and we remark on three. First, one could consider performing state distillation or even some form of cultivation directly within a BB code module. This could make use of the LPU we proposed here or another one tailored to the operations required \cite{zhang2025constant}. Second, one could implement other LDPC codes with properties suitable for mass magic production~\cite{zhu2023non,golowich2024quantum,zhu2025topological}. Whether there are sufficiently small and practical versions of these constructions is an important open problem. Finally, one could consider scaling cultivation to create the states required to do truly massive computations without ever relying on distillation, a task that might be simplified with the addition of longer range connectivity~\cite{davydova2025universal,bauer2025planar,huang2025generatinglogicalmagicstates,guanyu2025universal}. Interestingly, each of the existing approaches could be capable of producing magic states for non-Clifford gates other than T, such as CCZ, which could in turn facilitate the compilation of particular algorithms.

\end{enumerate}

\subsection*{Acknowledgements}
We thank
\begin{itemize*}[label={}, afterlabel={}, itemjoin={{, }}, itemjoin*={{, and }}]
    \item Thomas Alexander
    \item Ewout van den Berg
    \item Sebastian Brandhofer
    \item Lev Bishop
    \item Oliver Dial
    \item Bence Het\'{e}nyi
    \item Zhiyang (Sunny) He
    \item Ali Javadi-Abhari
    \item Anirudh Krishna
    \item Tristan M\"{u}ller
    \item Diego Rist\`{e}
    \item Vikesh Siddhu
    \item Shraddha Singh
    \item Barbara Terhal
    \item Kento Ueda
    \item Drew Vandeth
    \item Helena Zhang
    \item Guanyu Zhu
\end{itemize*}
for helpful discussions. The authors acknowledge the IBM Research CCC Service for providing resources that have contributed to the research results reported within this paper.

\nocite{gnuparallel}

\printbibliography%

\appendix

\clearpage
\section{Appendices}

\subsection{Logical code bases}\label{app:basis}

Here, we present logical Pauli bases for the gross and two-gross codes. These are used to build logical processing units (LPUs) using the techniques of qLDPC surgery \cite{CKBB22,CHRY24,WY24,ide2024fault,swaroop2024universal}. These LPUs, along with shift automorphisms of the codes arising from their translational symmetry, enable the measurement of any logical qubit in either the $X$ or $Z$ basis and also enable the full 11-qubit Clifford group on any 11 of 12 logical qubits. Later in this section, we present certain requirements on logical bases that lead to this computational capability and also simpler connectivity within the LPU.

First, we review the polynomial formalism to describe BB codes \cite{panteleev2021degenerate,bravyi2024high}. 
Suppose we label the unit cell at integer coordinate $(i,j)$ by $x^iy^j$, where one can think of $x,y$ as formal variables. We are working on the surface of the $\ell\times m$ torus, so unit cells $(i+h\ell,j+km)$ are all the same for any integers $h,k$. We can express this redundancy in our labels by requiring $x^\ell=1$ and $y^m=1$. The qubits and checks in cell $x^iy^j$ are also all called $x^iy^j$, but we can distinguish the four cellmates by saying $L$, $R$, $X$, or $Z$ as well.

Define polynomials
\begin{equation}\label{eq:toric_layout_BB}
A=1+y+x^{a_{\rightarrow}}y^{a_{\uparrow}},\quad B=1+x+x^{b_{\rightarrow}}y^{b_{\uparrow}}.
\end{equation}
If we consider an arbitrary $X$ check with label $x^iy^j$, it connects to exactly the three $L$ qubits labeled by the three terms in the polynomial $x^iy^jA$. It also connects to the three $R$ qubits labeled by the terms in $x^iy^jB$.

To describe $Z$ checks satisfactorily with polynomials, define the transpose of monomial $x^iy^j$ to be $(x^iy^j)^\top=x^{\ell-i}y^{m-j}=x^{-i}y^{-j}$. We take the transpose of a polynomial by taking the transpose of each of its monomial terms and adding them together. For instance,
\begin{equation}
A^\top=1+y^{-1}+x^{-a_{\rightarrow}}y^{-a_{\uparrow}},\quad B^\top=1+x^{-1}+x^{-b_{\rightarrow}}y^{-b_{\uparrow}}.
\end{equation}
Then, a $Z$ check labeled $x^iy^j$ is connected to three $L$ qubits $x^iy^jB^\top$ and three $R$ qubits $x^iy^jA^\top$. The total weight of every check (in the complete basis of $X$ and $Z$ checks we have described) is six, the number of terms in $A$ plus the number in $B$. 

\begin{figure}[!ht]
    \centering
    \input{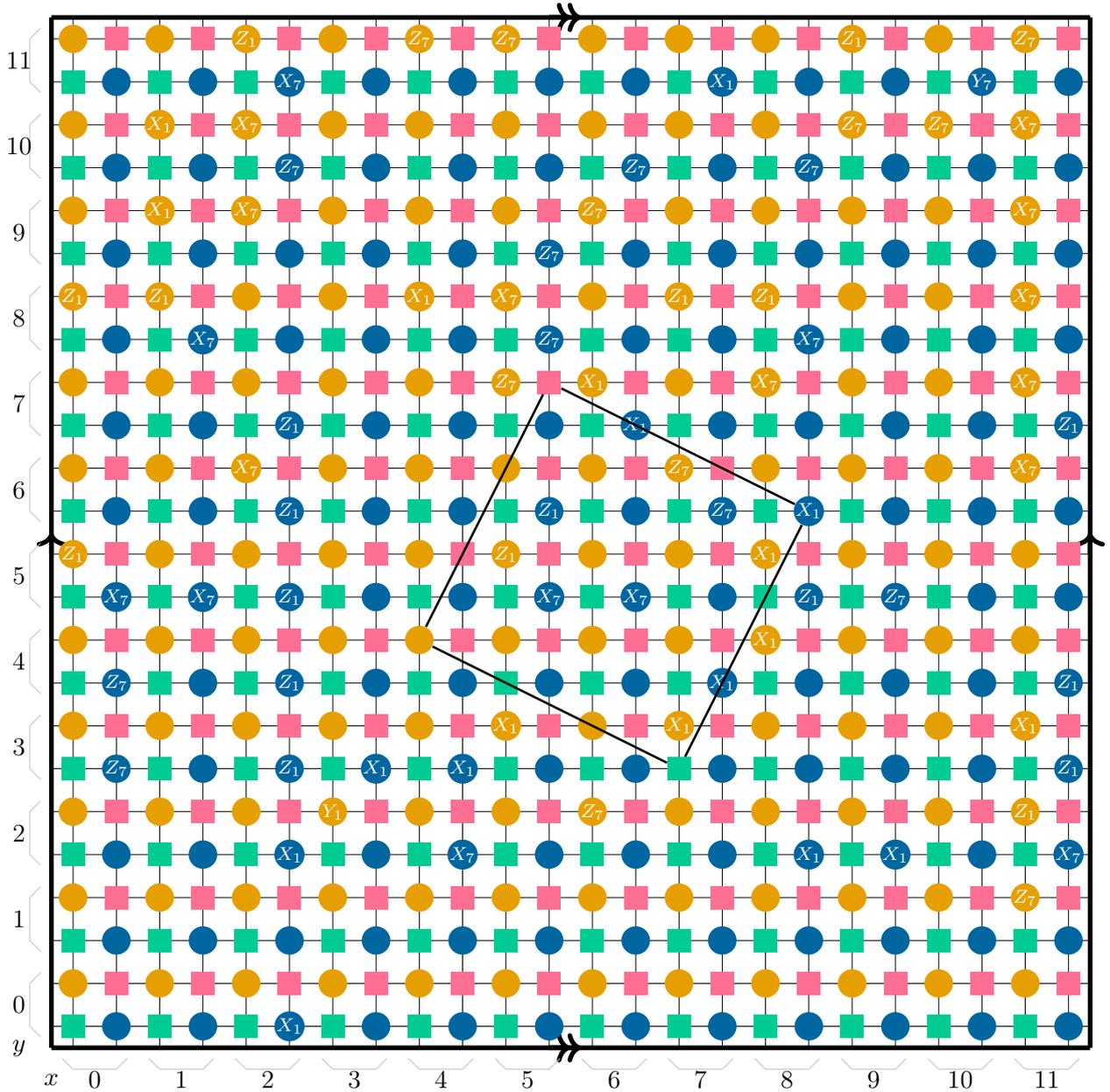}%
    \caption{The $[[288,12,18]]$ two-gross code is assembled from a $12\times 12$ grid of unit cells (see \cref{fig:unit_cell}). Only four example long-range connections are drawn, but these can be extrapolated to all unit cells by translational symmetry. Qubits in logical operators $\bar X_1$, $\bar X_7$, $\bar Z_1$, and $\bar Z_7$ are labeled $X_1$, $X_7$, $Z_1$, and $Z_7$, respectively. Where $\bar X_1$ and $\bar Z_1$ (resp.~$\bar X_7$ and $\bar Z_7$) overlap, the qubit is labeled $Y_1$ (resp.~$Y_7$). 
    }
    \label{fig:two_gross_code}
\end{figure}

The two BB codes we consider in this paper are depicted in \cref{fig:gross_code,fig:two_gross_code}. The gross code is defined with $\ell=12$, $m=6$, $a_{\rightarrow}=-b_{\uparrow}=3$ and $a_{\uparrow}=b_{\rightarrow}=-1$ and its code parameters are $[[144,12,12]]$. The two-gross 
code is defined with $\ell=12$, $m=12$, and the same long-range connections as the gross code, $a_{\rightarrow}=-b_{\uparrow}=3$ and $a_{\uparrow}=b_{\rightarrow}=-1$. The two-gross code has parameters $[[288,12,18]]$ and is depicted in \cref{fig:two_gross_code}.

We employ qLDPC surgery to perform logical Pauli measurements on the gross and two-gross codes. In both cases, our constructions also enable the full group of Clifford gates on 11 out of the 12 logical qubits, sacrificing one logical qubit for gate synthesis in the vein of Pauli-based computation \cite{bravyi2016trading,litinski2019game}. While we save the detailed explanations of the qLDPC surgery constructions to the next section, they rely on convenient bases of logical Pauli operators, which we describe now.

We denote a Pauli $X$ (resp.~$Z$) operator acting on the $L$ qubits in polynomial $p$ and the $R$ qubits in polynomial $q$ by writing $X(p,q)$ (resp.~$Z(p,q)$). Then, one can check that $X(p_x,q_x)$ and $Z(p_z,q_z)$ commute if and only if $1\not\in p_xp_z^\top+q_xq_z^\top$, where the notation $\alpha\in p$ means monomial $\alpha$ is a term in polynomial $p$, and the polynomial arithmetic is performed over $\mathbb{F}_2[x,y]/\langle x^\ell-1,y^m-1\rangle$ (polynomials in two variables $x,y$ with coefficients reduced modulo two and $x^\ell=1,y^m=1$). So $X(p,q)$ is a logical operator if and only if it commutes with all $Z$ checks, or $Z(\alpha B^\top,\alpha A^\top)$ for all monomials $\alpha$. This happens if and only if $1\not\in \alpha pB+\alpha qA$ for all $\alpha$, and this means $pB+qA$ must be the zero polynomial. We summarize this reasoning and the similar reasoning for $Z$-type logical operators as follows:
\begin{align}\label{eq:x_logical}
X(p,q)\text{ is a logical operator} &\Longleftrightarrow pB+qA=0,\\\label{eq:z_logical}
Z(r,s)\text{ is a logical operator} &\Longleftrightarrow rA^\top+sB^\top=0.
\end{align}

There are two properties of logical operators we point out. First, if $X(p,q)$ and $Z(r,s)$ are logical, then so are $X(\alpha p,\alpha q)$ and $Z(\alpha r,\alpha s)$ for any monomial $\alpha$. We say $X(p,q)$ and $X(\alpha p,\alpha q)$ are related by a shift automorphism, or that they are shifts of of one another. Geometrically, this property of logical operators simply expresses the translational symmetry of the BB code on the torus.

The second property relates $X$ and $Z$ operators: $X(p,q)$ is a logical operator if and only if $Z(q^\top,p^\top)$ is a logical operator. This follows immediately from Eqs.~\eqref{eq:x_logical} and \eqref{eq:z_logical}, and is indicative of a ZX-duality \cite{breuckmann2024fold,bravyi2024high,eberhardt2024logical}.

To use qLDPC surgery efficiently on BB codes, we follow the strategy of Ref.~\cite{CHRY24} in selecting a logical basis. Suppose we identify and fix polynomials $p,q,r,s$ and monomials $\mu,\nu$ such that
\begin{equation}\label{eq:pqrs_basis}
\begin{array}{ll}
\bar X_1=X(p,q),&\bar X_7=X(r,s),\\
\bar Z_1=Z(\nu s^\top,\nu r^\top),&\bar Z_7=Z(\mu q^\top,\mu p^\top)
\end{array}
\end{equation}
are all logical operators. The construction of these operators via ZX-dualities and shifts implies that the bipartite graph connecting qubits in $\bar X_1$ with their adjacent $Z$ checks is isomorphic to the same graph for qubits of $\bar Z_7$ and their adjacent $X$ checks. The same is true for $\bar X_7$ and $\bar Z_1$. This property is extremely convenient for our LPU construction, as it means we can reuse sets of fixed-connectivity auxiliary qubits and checks to measure two different logical operators, effectively cutting the size of the LPU in half.

It is also helpful to the LPU construction to require other properties of the logical basis. 
\begin{enumerate}
\item $\bar X_1$ and $\bar Z_1$ anti-commute and $\bar X_7$ and $\bar Z_7$ anticommute. Automatically from the construction in \cref{eq:pqrs_basis}, all other pairs of Paulis in that equation commute.
\item The operators in \cref{eq:pqrs_basis} and their shifts generate the entire 12-qubit logical Pauli group.
\item If two different operators in \cref{eq:pqrs_basis} commute, they do not overlap. If two operators anticommute, they overlap on exactly one qubit.
\item The set of $Z$-checks that overlap with $\bar X_1$ is disjoint from the set of $Z$-checks that overlap with $\bar X_7$. Likewise, the set of $X$-checks that overlap with $\bar Z_1$ is disjoint from the set of $X$-checks that overlap with $\bar Z_7$.
\end{enumerate}
These properties ensure that the logical processing unit (LPU) that we create with qLDPC surgery and attach to the orignal BB code is both sufficient for Clifford computation (properties 1 and 2) and does not increase the degree of qubit and checks in the original code by more than one (properties 3 and 4).

\begin{figure}[!ht]
    \centering
    \includegraphics[width=0.8\linewidth]{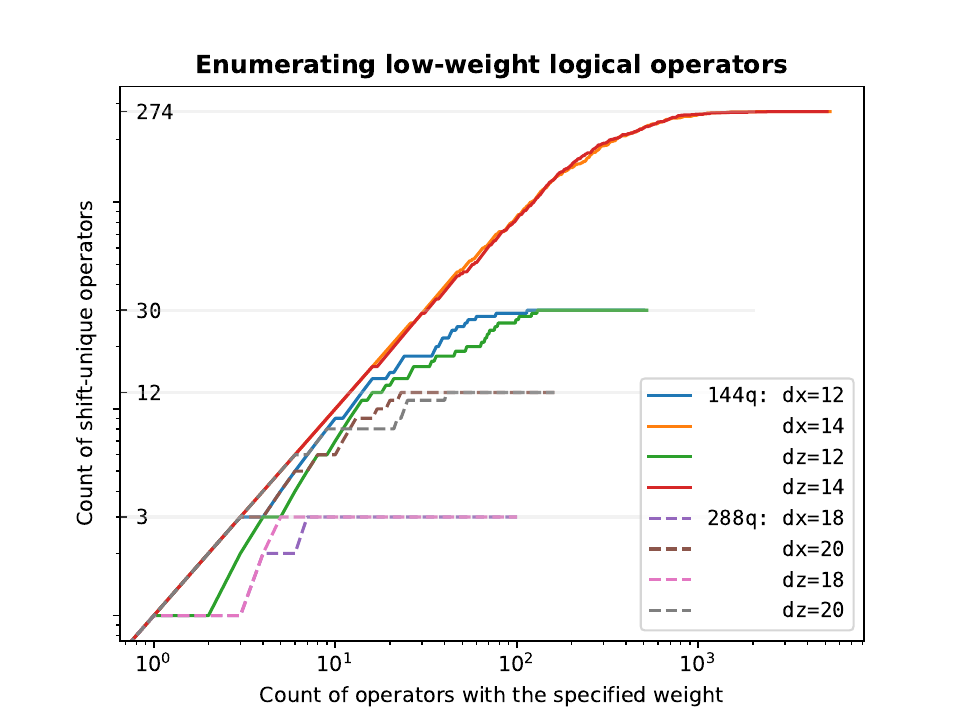}
    \caption{We can use the BP+OSD decoder \cite{panteleev2021degenerate} to find low-weight logical operators as described in Ref.~\cite{bravyi2024high} by solving a decoding problem that demands commutation with all stabilizers and anti-commutation with a random non-trivial logical operator. Usually the decoder will output a minimum weight logical, weight $d=12$ and $d=18$ for the 144 qubit gross and 288 qubit two-gross codes, respectively. However, we can introduce random priors to the BP+OSD decoder to coax it more often into finding higher weight logicals as well. Every decoding run, we pull each qubit prior error probability from the uniform distribution on $[0.01,0.99)$. For the gross code, we set the maximum BP iterations to 20 and the OSD order to 0. For the two-gross code, we use maximum BP iterations 40 and OSD order 7 (specifically, combination sweep \cite{roffe2020}). We enumerate $X$ and $Z$ type logical operators with weights $d$ and $d+2$ by running the decoder many times. In the figure, we plot the count of shift-unique logicals (those that do not differ by just a shift automorphism) versus the count of logicals found with the specified weight. For each curve, to build confidence that we found everything, we run decoder trials until we have obtained at least 10 times the shift-unique count more operators of the specified weight without finding a new shift-unique operator. Including all operator shifts, we find in total $1884$ weight-$12$ and $19728$ weight-$14$ logical $X$ operators for the gross code, $336$ weight-$18$ and $1728$ weight-$20$ logical $X$ operators for the two-gross code, and the same counts for logical $Z$ operators which are always related to the $X$ operators via the $ZX$-duality transform described in the text. In particular, our enumeration of weight $d=12$ operators for the gross code matches the exhaustive enumeration in Ref.~\cite{ott2025decision} via a different method. Though we did not rigorously optimize decoder parameters and priors, enumerating all the logicals for this plot takes less than 25 minutes on a laptop with the BP+OSD implementation in Ref.~\cite{Roffe_LDPC_Python_tools_2022}.}
    \label{fig:logical_counts}
\end{figure}

We find bases in the form of \cref{eq:pqrs_basis} for the gross and two-gross code by enumerating all weight $d$ and $d+2$ operators for the two codes, \cref{fig:logical_counts}. To satisfy the four properties above, there are generally many choices, but we can choose among them to minimize the qubit overhead of the LPU, the technical details of which will be described later. For the gross code, we select $\mu=\nu=xy$ and
\begin{equation}\label{eq:pqrs_gross}
\begin{array}{ll}
p = x^4+x^5+x^6y+x^4y^2+x^5y^4+x^6y^5,& q = x^3+x^4+x^3y+x^3y^2+x^4y^2+x^3y^5,\\
r = 1+x^8+xy+x^9y+x^3y^4+x^{11}y^4,& s = x+x^9+x^4y^4+x^8y^4+y^5+x^8y^5.
\end{array}
\end{equation}
For the two-gross code, we select $\mu=\nu=xy$ and
\begin{equation}\label{eq:pqrs_twogross}
\begin{array}{l}
p = x^2+(x^2+x^8+x^9)y^2+(x^3+x^4)y^3+x^7y^4+x^8y^6+x^6y^7+x^7y^{11},\\ q = x^3y^2+(x^5+x^7+x^{11})y^3+x^8y^4+x^8y^5+x^6y^7+x^4y^8+xy^9+xy^{10},\\
r = (x^4+x^{11})y^2+(1+x+x^5+x^6)y^5+(x+x^8)y^8+(x^2+x^{10})y^{11},\\ s = (x^2+x^{11})(y^6+y^9+y^{10})+(x^8+x^{11})y^7+(x^5+x^{11})y^8.
\end{array}
\end{equation}
We comment that while $\bar X_1$, $\bar X_7$, $\bar Z_1$, and $\bar Z_7$ are all minimal weight 12 for the gross code, for the two-gross code they are each weight 20, two more than the code distance 18. For the two-gross code, it turns out that any choice of only weight 18 logical operators fails to satisfy property 2, and choosing some to be weight 18 and others weight 20 satisfies only 0-3.

We also note these logical basis polynomials are written to satisfy \cref{eq:x_logical,eq:z_logical} for the $A$ and $B$ polynomials in \cref{eq:toric_layout_BB}, which differ from the $A$ and $B$ polynomials originally introduced in Ref.~\cite{bravyi2024high} for these two codes. We made this change, which just amounts to a relabeling of qubits and checks of the codes, to better connect to the toric layouts depicted in \cref{fig:gross_code} and \cref{fig:two_gross_code}. 

It happens for these two codes and the choices of bases above, that a stronger version of property 2 holds true. A logical basis for all 12 operators can be defined in terms of shifts of those in \cref{eq:pqrs_basis}. Specifically, for all $i=1,2,3,4,5,6$, there are monomials $\alpha_i$ and $\beta_j$ such that
\begin{equation}
\begin{array}{ll}\label{eq:basisdefn}
\bar X_i=\alpha_i\bar X_1=X(\alpha_ip,\alpha_iq),& \bar X_{i+6}=\beta_i^\top\bar X_7=X(\beta_i^\top r,\beta^\top_is),\\
\bar Z_i=\beta_i\bar Z_1=Z(\beta_i\nu s^\top,\beta_i\nu r^\top),& \bar Z_{i+6}=\alpha^\top_i\bar Z_7=Z(\alpha^\top_i\mu q^\top,\alpha^\top_i\mu p^\top).
\end{array}
\end{equation}
In fact, there are several such choices of $\alpha_i$ and $\beta_i$ for both codes. 
Here is one choice for the gross code
\begin{equation}\label{eq:grossbasis}
\alpha=(1,\; x^3y^5,\; x^{11}y^5,\; x^{10}y,\; x^5y^4,\; x^4y^2),\quad\beta=(1,\; x^2y^4,\; xy^2,\; x^2y^5,\; xy,\; x^3y).
\end{equation}
and one choice for the two-gross code
\begin{equation}\label{eq:twogrossbasis}
\alpha=(1,\;y^3,\;x^3y^7,\;x^{11}y^{11},\;x^2y^9,\;x^7y^4),\quad\beta=(1,\;xy,\;x^4,\;x^5y^4,\;x^4y^3,\;x^3y^5).
\end{equation}

\subsection{Shift automorphisms}\label{app:automorphisms}

In \cref{sec:shift_automorphisms}, we described basic shift automorphisms of BB codes that can be performed by first swapping qubits with adjacent checks, and then swapping them from there into new qubit positions, with both these swaps happening along edges of the Tanner graph of the BB code. Here, we analyze the shift automorphism group generated by these basic shifts using the polynomial description of a BB code in \cref{eq:toric_layout_BB}. We first consider a more general case where both polynomials $A=A_0+A_1+A_2$ and $B=B_0+B_1+B_2$ are the sum of three monomial terms.

A shift automorphism is defined by a monomial $\delta$ and implements a global qubit permutation $(L,\alpha)\rightarrow(L,\delta\alpha)$ and $(R,\alpha)\rightarrow(R,\delta\alpha)$ for all $\alpha$. For the gross and two-gross codes, it can be shown that shifts by $\delta\in\langle x^6,y^6\rangle$ act as logical identity. Thus, each has only 35 logically nontrivial shift automorphisms.

Within the existing connectivity of the gross and two-gross codes, a generating set of permutations $\delta$ can be implemented using fault-tolerant quantum circuits. Such a circuit operates as follows: the data on the $L$ qubits is moved to the $X$ check qubits along edges corresponding to a term in $A$, and is moved back onto the $L$ qubits using a different term. Simultaneously the data on the $R$ qubits are moved to the $Z$ checks and back. This implements a permutation $\delta\in\{A_iA_j^T:i,j\in\{0,1,2\},i\neq j\}$. Similarly, if $L$ qubits travel through $Z$ checks (and $R$ qubits through $X$ checks),  a permutation $\delta\in\{B_iB_j^T:i,j\in\{0,1,2\},i\neq j\}$ is implemented. Hence, there are 6 fault tolerant circuits for implementing automorphisms and their inverses, for a total of 12. These constitute the unitary logical operations $\bar U_1,\ldots, \bar U_{6}$ in \cref{sec:bicycle-arch}. For both codes the basic shift automorphisms are $\{x,y,x^3y^{-1},xy^3,x^3y^{-2},x^2y^3\}$ and their inverses. It is easy to verify by enumeration that products of at most two basic shifts suffice to generate all nontrivial shift automorphisms.

We note that generally the full automorphism group of a BB code is larger than just the group of shift automorphisms. See for instance Refs.~\cite{eberhardt2024logical,sayginel2024fault}. This is true also for the gross and two-gross codes. However, these additional automorphisms are expected to be harder to implement using only the qubit connectivity required for syndrome measurements.

Permutations of physical qubits in a CSS code result in logical CNOT circuits. In general, a CNOT circuit can be described by an invertible matrix $M$ over $\mathbb{F}_2$ acting on computational basis states as $\ket{x} \to \ket{Ax}$.
In the logical basis described by \cref{eq:basisdefn}, \cref{eq:grossbasis}, and \cref{eq:twogrossbasis}, it furthermore holds that this matrix is of the form $M = A \otimes A$ for $A \in \mathbb{F}_2^{6 \times 6}$.
This can be seen by the fact that the logical operators of qubits 1 through 6 all take the form $\alpha \bar X_1$, $\beta \bar Z_1$, and that qubits 7 through 12 all take the form $\alpha \bar X_7$, $\beta \bar Z_7$. For any $\alpha,\beta$, even those other than the ones in \cref{eq:grossbasis} and \cref{eq:twogrossbasis}, $\alpha \bar X_1$, $\beta \bar Z_1$ commute with $\beta \bar Z_7$, $\alpha \bar X_7$ respectively. Hence, if we apply an automorphism to one of $\bar X_1$ through $\bar X_6$ and obtain an operator of the form $\alpha \bar X_1$, this operator must be expressible as a linear combination of $\bar X_1$ through $\bar X_6$ only (modulo stabilizers). Similarly, applying an automorphism to one of $\bar X_7$ through $\bar X_{12}$ results in an operator $\alpha \bar X_7$ that can be expressed as a linear combination of $\bar X_7$ through $\bar X_{12}$ only (modulo stabilizers).

For the particular choice of basis in \cref{eq:grossbasis}, and \cref{eq:twogrossbasis}, the matrices $A$ corresponding to the shifts $\delta = x$ and $\delta = y$ are
{\allowdisplaybreaks %
\begin{align}
    A_x^\text{gross} &= \begin{bmatrix}
0 & 1 & 0 & 1 & 0 & 0 \\
0 & 1 & 0 & 0 & 0 & 1 \\
0 & 0 & 1 & 1 & 0 & 0 \\
1 & 1 & 0 & 1 & 1 & 0 \\
0 & 1 & 0 & 0 & 1 & 0 \\
1 & 1 & 1 & 1 & 0 & 1
    \end{bmatrix},
    &
    A_y^\text{gross} &= \begin{bmatrix}
1 & 0 & 0 & 0 & 0 & 1 \\
1 & 1 & 1 & 0 & 0 & 1 \\
0 & 0 & 0 & 0 & 1 & 0 \\
0 & 1 & 0 & 0 & 0 & 0 \\
0 & 1 & 1 & 0 & 0 & 1 \\
0 & 0 & 1 & 1 & 0 & 1 
    \end{bmatrix}, \\
    A_x^\text{two-gross} &= \begin{bmatrix}
0 & 1 & 1 & 1 & 0 & 1 \\
1 & 0 & 1 & 0 & 1 & 1 \\
1 & 0 & 1 & 0 & 1 & 0 \\
1 & 0 & 1 & 1 & 1 & 1 \\
0 & 1 & 1 & 1 & 1 & 1 \\
1 & 0 & 0 & 1 & 1 & 0
    \end{bmatrix},
    &
    A_y^\text{two-gross} &= \begin{bmatrix}
1 & 1 & 1 & 1 & 1 & 0 \\
1 & 1 & 0 & 1 & 1 & 1 \\
0 & 1 & 1 & 0 & 0 & 0 \\
1 & 0 & 0 & 0 & 1 & 0 \\
1 & 0 & 0 & 1 & 1 & 1 \\
1 & 0 & 0 & 0 & 0 & 1
    \end{bmatrix}.
\end{align}}%
We note that all these matrices have order 6.  Since any shift can by achieved by a combination of $x$ and $y$ shifts,
the matrices above suffice to specify the group of logical gates spanned by the shift automorphisms. Let the unitary representation on 6 qubits of $A_i^C$ be $u_i^C$. Then the unitary representation of a shift automorphism (on 12 qubits) is $U_i \in \left\langle u_x^C \otimes u_x^C, u_y^C \otimes u_i^C \right\rangle$.

\subsection{Logical processing units for logical measurements}
\label{app:lpu_BB}

\subsubsection*{An improved logical processing unit for the gross code}
We construct an ancilla system capable of measuring any element of $\langle\bar X_1,\bar X_7,\bar Z_1,\bar Z_7\rangle$ with phenomenological fault-distance 12. The construction uses just 90 ancilla qubits to perform these in-module measurements. Moreover, the LPUs belonging to different BB codes can be connected to measure joint logical Paulis as we describe in \cref{sec:inter-module}.

We draw the connectivity of the ancilla system and gross code together in \cref{fig:LPU_summary}. The ancilla system alone can be described using an auxiliary graph as in Ref.~\cite{WY24}. In this view, there is a graph $\mathcal{G}=(\mathcal{V},\mathcal{E})$ and cycle basis $\mathcal{U}$ with qubits identified to edges $\mathcal{E}$ and checks of opposite types (e.g.~$X$ and $Z$) associated to vertices $\mathcal{V}$ and cycles $\mathcal{U}$.\footnote{This association of checks to graphical structures is similar to Kitaev's description of the toric code \cite{kitaev2003fault} which has $X$ checks on vertices and $Z$ checks on faces of a square lattice. A notable difference however is that the LPU includes a $Z$ check for every cycle (in a cycle basis) of the graph, not just homologically trivial cycles. Thus, viewed as a CSS code, the code defined on this graph alone encodes no logical qubits.} This graph is also connected to the gross code, so that all vertex checks $\mathcal{V}$ pick up support on certain qubits of the gross code, and certain checks in the gross code pick up support on certain edge qubits in $\mathcal{E}$. We will work through this construction in more detail in this section.

The construction of the entire auxiliary graph $\mathcal{G}$ proceeds in stages. First, we construct an auxiliary graph $\mathcal{G}_l$ to measure $\bar X_1$ (and by symmetry $\bar Z_7$) and an auxiliary graph $\mathcal{G}_r$ to measure $\bar X_7$ (and by symmetry $\bar Z_1$). Then the graphs $\mathcal{G}_1$ and $\mathcal{G}_7$ are joined by a vertex identification and the addition of ``bridge'' edges \cite{CHRY24,swaroop2024universal} to obtain $\mathcal{G}$.

To construct $\mathcal{G}_l$, create a set of vertices in 1-1 correspondence with the qubits supporting $\bar X_1=X(p,q)$. These qubits are labeled by the 12 monomial terms of $pL+qR$,\footnote{Note we include dummy variables $L$ and $R$ in monomials and polynomials in this section to conveniently distinguish $L$ and $R$ qubits.} and we label the vertices with the same monomials. The vertex labeled with a monomial $\gamma\in pL+qR$ is physically connected to qubit $\gamma$ and, to ensure we can also measure $\bar Z_7=Z(xyq^\top,xyp^\top)$, to qubit $xy\gamma^\top$ as well, exploiting the ZX-duality relation between $\bar X_1$ and $\bar Z_7$ in \cref{eq:pqrs_basis}. We note that for monomials like $\gamma=\alpha T$ that include the dummy variable $T=L,R$, we define the transposes as $(\alpha L)^\top=\alpha^\top R$ and $(\alpha R)^\top=\alpha^\top L$.

Edges are now added to $\mathcal{G}_l$ in such a way that checks of the original gross code do not gain support on more than one qubit when deforming the code for a logical measurement of $\bar X_1$. This necessitates adding an edge between vertices $\gamma,\delta\in pL+qR$ if qubits $\gamma$ and $\delta$ participate together in a $Z$ check. 
The qubit on the edge $(\gamma,\delta)$ of $\mathcal{G}_l$ is physically connected to that $Z$ check. There are 18 such edges in $\mathcal{G}_l$. These same edges serve, by symmetry, to guarantee $X$ checks adjacent to qubits in $\bar Z_7$ do not gain more than one qubit in support, and the edge qubits are thus also each connected to one $X$ check. 

In the general case (and for the two-gross code), a second set of edges would now be added to $\mathcal{G}_l$ to increase the graph's edge expansion, or, more to the point, to ensure that the deformed codes for measuring $\bar X_1$ and $\bar Z_7$ have code distances $12$ \cite{WY24,ide2024fault}. However, for the choice of gross code basis in \cref{eq:pqrs_gross}, no such additional expanding edges are necessary; the deformed codes already have phenomenological fault-distances of $12$, see \cref{app:cplex}.

Now, $\mathcal{G}_l$ is a graph with $12$ vertices and $18$ edges. A cycle basis of minimum size therefore has $18-12+1=7$ cycles. However, some basis cycles can be eliminated based on check redundancies in the original code \cite{WY24}. Which cycles can be eliminated depends on the operator being measured and, since cycles of $\mathcal{G}_l$ are also cycles of the complete auxiliary graph $\mathcal{G}$ (which we describe in detail later), we need to also keep cycles if they are necessary for logical measurements using $\mathcal{G}$. Measuring $\bar X_1\bar Z_1\bar X_7\bar Z_7$ (using the complete graph $\mathcal{G}$) is a limiting case, which requires 5 of the 7 basis cycles of $\mathcal{G}_l$. We choose to eliminate the longest cycles we can, and the remaining basis cycles can be written as sequences of monomially-labeled vertices:
\begin{align}\nonumber
&x^4L\rightarrow x^3y^2R\rightarrow x^3yR\rightarrow x^5L\rightarrow x^4L,\\\nonumber
&x^3y^5R\rightarrow x^3R\rightarrow x^6y^5L\rightarrow x^4y^2L\rightarrow x^3y^5R,\\\nonumber
&x^3y^5R\rightarrow x^3R\rightarrow x^6y^5L\rightarrow x^4R\rightarrow x^5y^4L\rightarrow x^3y^5R,\\\nonumber
&x^4L\rightarrow x^3y^2R\rightarrow x^6yL\rightarrow x^5y^4L\rightarrow x^4R\rightarrow x^4L,\\\nonumber
&x^3yR\rightarrow x^3y^2R\rightarrow x^6yL\rightarrow x^4y^2R\rightarrow x^5L\rightarrow x^3yR.
\end{align}
This cycle basis is denoted $\mathcal{U}_l$. This completes the description of $\mathcal{G}_l$ and its cycle basis.

Similarly, $\mathcal{G}_r$ can be constructed to measure $\bar X_7=X(r,s)$ and $\bar Z_1=Z(xys^\top,xyr^\top)$. Because this construction follows the same procedure as $\mathcal{G}_l$ and also requires no additional expanding edges, we skip the discussion and just write down the reduced cycle basis, which in this case consists of just three cycles:
\begin{align}\nonumber
&x^9yL\rightarrow x^8y^4R\rightarrow x^8y^5R\rightarrow x^{11}y^4L\rightarrow x^9yL,\\\nonumber
&xyL\rightarrow xR\rightarrow x^4y^4R\rightarrow x^3y^4L\rightarrow xyL,\\\nonumber
&1L\rightarrow y^5R\rightarrow x^9R\rightarrow x^9yL\rightarrow x^8y^4R\rightarrow 1L.
\end{align}
This cycle basis is denoted $\mathcal{U}_r$.

We now build a graph $\mathcal{G}$ starting from graphs $\mathcal{G}_l$ and $\mathcal{G}_r$. To start, we identify vertex $x^4R$ in $\mathcal{G}_l$ with vertex $x^9yL$ in $\mathcal{G}_r$. Intuitively, this is done because $\bar X_1$ (resp.~$\bar Z_7$), measured by the $\mathcal{G}_l$ system, and $\bar Z_1$ (resp.~$\bar X_7$), measured by the $\mathcal{G}_r$ system, overlap on a single qubit $x^4R$ (resp.~$x^9yL$) of the gross code, and we would like to connect this qubit to just one vertex check in the auxiliary graph so that, including its 6 gross code checks, it is involved in 7 checks and no more. We note that vertex identification is done also in the gross code construction of Ref.~\cite{CHRY24}. We should also note that identifying vertices this way and connecting the resulting vertex to both gross code qubits $x^4R$ and $x^9yL$ leads to a vertex check of weight 8. To avoid this, the check can be implemented by a Bell pair, as in \cref{fig:LPU_summary}. Finally, all cycles of $\mathcal{G}_l$ and $\mathcal{G}_r$ are still cycles in $\mathcal{G}$, provided that vertices $x^4R$ and $x^9yL$ are identified.

In addition to the vertex identification, we add 11 edges to complete the construction of $\mathcal{G}$. Ultimately, the reason these edges are added is to increase the edge expansion \cite{WY24,ide2024fault} or, more accurately, the relative expansion \cite{swaroop2024universal} of $\mathcal{G}$. However, following the perspective of \cite{CHRY24}, we call these bridge edges $B$. These 11 bridge edges are depicted in this diagram.
\begin{equation}
\arraycolsep=1.25pt
\begin{array}{ccccccccccccccccccccccc}\vspace{5pt}
\mathcal{G}_l:& x^6y^5L & \smdash & x^4y^2L & \smdash & x^3y^5R & \smdash & x^5y^4L & \smdash & x^6yL & \smdash & x^4y^2R & \smdash & x^5L & \smdash & x^4L & \smdash & x^3y^2R & \smdash & x^3yR & \smdash & x^3R \\\vspace{5pt}
&|&&|&&|&&|&&|&&|&&|&&|&&|&&|&&| \\
\mathcal{G}_r:& x^{11}y^4L & \smdash & x^8y^5R & \smdash & x^8y^4R & \smdash & 1L & \smdash & xR & \smdash & x^4y^4R & \smdash & x^8L & \smdash & x^9R & \smdash & y^5R & \smdash & x^3y^4L & \smdash & xyL
\end{array}
\end{equation}
Here, the top vertices form a path in $\mathcal{G}_l$ (i.e.~each is connected to the next) avoiding vertex $x^4R$, the bottom vertices form a path in $\mathcal{G}_r$ avoiding $x^9yL$, and corresponding pairs are connected with bridge edges. This naturally adds 10 length-4 cycles (the squares in the diagram) to the cycle basis of $\mathcal{G}$. There is an additional cycle that goes through the identified vertex and a single bridge edge, which can be chosen as:
\begin{equation}
x^9yL=x^4R\rightarrow x^6y^5L\rightarrow x^{11}y^4L\rightarrow x^9yL=x^4R.
\end{equation}
These additional 11 cycles are denoted $\mathcal{U}_B$. This completes the description of $\mathcal{G}$ and its cycle basis. See \cref{fig:LPU_summary} for a depiction of the ancilla system connectivity.

\subsubsection*{A logical processing unit for the two-gross code}

The construction of the two-gross code’s LPU proceeds along the same lines as the construction for the gross code, so we simply specify the relevant information without much elaboration. Again, it enables the measurement of any element of $\langle\bar X_1,\bar X_7,\bar Z_1,\bar Z_7\rangle$. The only qualitative difference is that for the two-gross code, we need additional edge qubits to achieve phenomenological fault-distance 18, which we can check with CPLEX as described in \cref{app:cplex} for all of the 15 logical measurements. The additional edges are randomly sampled until a set is found to ensure the fault-distance. We attempt to minimize the number of additional edges by sampling sets of edges of increasing number (i.e.~first try various sets of one edge, then sets of two, etc.) until a satisfying choice is found.

The graph $\mathcal{G}_l=(\mathcal{V}_l,\mathcal{E}_l)$ has a vertex for each qubit of $\bar X_1$, and an edge for each of the 30 adjacent $Z$ checks. Two additional edges are added to increase the fault-distance:
\begin{equation}
x^3y^3L\leftrightarrow x^7y^{11}L, \quad xy^9R\leftrightarrow x^8y^5R.
\end{equation}
Now $\mathcal{G}_l$ has a total of 20 vertices and 32 edges. After eliminating redundant cycles, we find a basis of 11 cycles:
\begin{align}
&x^{2}L\rightarrow x^{3}y^{3}L\rightarrow x^{4}y^{3}L\rightarrow x^{2}L,\\
&x^{7}y^{11}L\rightarrow x^{8}y^{2}L\rightarrow x^{9}y^{2}L\rightarrow x^{7}y^{11}L,\\
&xy^{10}R\rightarrow xy^{9}R\rightarrow x^{4}y^{8}R\rightarrow xy^{10}R,\\
&x^{8}y^{5}R\rightarrow x^{8}y^{4}R\rightarrow x^{11}y^{3}R\rightarrow x^{8}y^{5}R,\\
&x^{6}y^{7}L\rightarrow x^{5}y^{3}R\rightarrow x^{8}y^{2}L\rightarrow x^{7}y^{11}L\rightarrow x^{6}y^{7}R\rightarrow x^{6}y^{7}L,\\
&x^{3}y^{3}L\rightarrow x^{4}y^{3}L\rightarrow x^{5}y^{3}R\rightarrow x^{8}y^{2}L\rightarrow x^{7}y^{11}L\rightarrow x^{3}y^{3}L,\\
&x^{7}y^{11}L\rightarrow x^{6}y^{7}R\rightarrow x^{8}y^{6}L\rightarrow x^{7}y^{3}R\rightarrow x^{9}y^{2}L\rightarrow x^{7}y^{11}L,\\
&x^{2}y^{2}L\rightarrow xy^{10}R\rightarrow xy^{9}R\rightarrow x^{8}y^{5}R\rightarrow x^{11}y^{3}R\rightarrow x^{2}y^{2}L,\\
&x^{7}y^{4}L\rightarrow x^{7}y^{3}R\rightarrow x^{8}y^{6}L\rightarrow x^{8}y^{5}R\rightarrow x^{8}y^{4}R\rightarrow x^{7}y^{4}L,\\
&x^{2}y^{2}L\rightarrow x^{11}y^{3}R\rightarrow x^{8}y^{4}R\rightarrow x^{7}y^{4}L\rightarrow x^{3}y^{2}R\rightarrow x^{2}y^{2}L,\\
&x^{3}y^{3}L\rightarrow x^{3}y^{2}R\rightarrow x^{7}y^{4}L\rightarrow x^{7}y^{3}R\rightarrow x^{9}y^{2}L\rightarrow x^{7}y^{11}L\rightarrow x^{3}y^{3}L.
\end{align}
We denote this cycle basis $\mathcal{U}_l$. This completes the description of $\mathcal{G}_l$.

Similarly, the graph $\mathcal{G}_r=(\mathcal{V}_r,\mathcal{E}_r)$ has a vertex for each qubit of $\bar X_7$ and edges corresponding to the 30 adjacent $Z$ checks. It also has two additional edges,
\begin{equation}
x^5y^5L\leftrightarrow x^{11}y^{10}R, \quad x^{11}y^9R\leftrightarrow x^2y^6R
\end{equation}
and a cycle basis $\mathcal{U}_r$ consisting of 9 cycles after redundant cycle checks are removed:
\begin{align}
&x^{11}y^{2}L\rightarrow xy^{5}L\rightarrow y^{5}L\rightarrow x^{11}y^{2}L,\\
&x^{4}y^{2}L\rightarrow x^{5}y^{5}L\rightarrow x^{6}y^{5}L\rightarrow x^{4}y^{2}L,\\
&x^{11}y^{8}R\rightarrow x^{2}y^{6}R\rightarrow x^{11}y^{9}R\rightarrow x^{11}y^{8}R,\\
&x^{11}y^{8}R\rightarrow x^{2}y^{6}R\rightarrow x^{11}y^{7}R\rightarrow x^{11}y^{8}R,\\
&x^{2}y^{10}R\rightarrow x^{2}y^{9}R\rightarrow x^{5}y^{8}R\rightarrow x^{2}y^{10}R,\\
&x^{5}y^{5}L\rightarrow x^{2}y^{6}R\rightarrow x^{11}y^{9}R\rightarrow x^{11}y^{10}R\rightarrow x^{5}y^{5}L,\\
&x^{5}y^{5}L\rightarrow x^{6}y^{5}L\rightarrow x^{8}y^{8}L\rightarrow x^{10}y^{11}L\rightarrow x^{11}y^{10}R\rightarrow x^{5}y^{5}L,\\
&x^{11}y^{2}L\rightarrow xy^{5}L\rightarrow x^{11}y^{6}R\rightarrow x^{8}y^{7}R\rightarrow x^{8}y^{8}L\rightarrow x^{10}y^{11}L\rightarrow x^{11}y^{2}L,\\
&y^{5}L\rightarrow xy^{8}L\rightarrow x^{11}y^{9}R\rightarrow x^{11}y^{10}R\rightarrow x^{10}y^{11}L\rightarrow x^{11}y^{2}L\rightarrow y^{5}L.
\end{align}
This completes the description of $\mathcal{G}_r$, which has a total of 20 vertices and 32 edges.

We merge graphs $\mathcal{G}_l$ and $\mathcal{G}_r$ by identifying the vertex $x^3y^2R$ from $\mathcal{G}_l$ with the vertex $x^{10}y^{11}L$ from $\mathcal{G}_r$. An additional 17 bridge edges are added in the following pattern
\begin{align}
\arraycolsep=1.25pt
\begin{array}{ccccccccccccccccccc}\vspace{5pt}
\mathcal{G}_l:&x^{7}y^{4}L & \smdash & x^{7}y^{3}R & \smdash & x^{8}y^{6}L & \smdash & x^{6}y^{7}R & \smdash & x^{7}y^{11}L & \smdash & x^{3}y^{3}L & \smdash & x^{2}L & \smdash & x^{4}y^{3}L & \smdash & x^{5}y^{3}R & \smdash\\\vspace{5pt}
&|&&|&&|&&|&&|&&|&&|&&|&&|&\dots\\\vspace{20pt}
\mathcal{G}_r:&x^{8}y^{8}L & \smdash & x^{6}y^{5}L & \smdash & x^{4}y^{2}L & \smdash & x^{5}y^{5}L & \smdash & x^{2}y^{6}R & \smdash & x^{11}y^{9}R & \smdash & x^{11}y^{10}R & \smdash & x^{2}y^{9}R & \smdash & x^{5}y^{8}R & \smdash\\\vspace{5pt}
&&\smdash & x^{6}y^{7}L & \smdash & x^{4}y^{8}R & \smdash & xy^{9}R & \smdash & xy^{10}R & \smdash & x^{2}y^{2}L & \smdash & x^{11}y^{3}R & \smdash & x^{8}y^{5}R & \smdash & x^{8}y^{4}R&\\\vspace{5pt}
&&\dots&|&&|&&|&&|&&|&&|&&|&&|&\\
&&\smdash & x^{2}y^{10}R & \smdash & x^{2}y^{11}L & \smdash & xy^{8}L & \smdash & y^{5}L & \smdash & x^{11}y^{2}L & \smdash & xy^{5}L & \smdash & x^{11}y^{6}R & \smdash & x^{8}y^{7}R&
\end{array}
\end{align}
In addition to the 16 length-4 cycles indicated by squares in the diagram, another length-3 cycle completes the cycle basis of the full graph:
\begin{equation}
x^{10}y^{11}L=x^3y^2R\rightarrow x^7y^4L\rightarrow x^8y^8L\rightarrow x^{10}y^{11}L=x^3y^2R.
\end{equation}
The full graph has 39 vertices, 81 edges, and a basis of 37 cycles. 

\subsection{Logical measurement protocols}
\label{app:lpu_meas_procedure}

We noted in \app{lpu_BB} that the LPU can be described by a graph $\mathcal{G}=(\mathcal{V},\mathcal{E})$ with cycle basis $\mathcal{U}$ \cite{WY24,ide2024fault},\footnote{A graph with cycle basis is also a 3-term chain complex $\mathcal{V}\rightarrow\mathcal{E}\rightarrow\mathcal{U}$.} in which qubits are associated to edges $\mathcal{E}$, checks acting as $X$ on edge qubits are associated to vertices $\mathcal{V}$, and $Z$ checks are associated to cycles $\mathcal{U}$. Vertex checks and edge qubits are also connected to qubits and checks, respectively, in the original BB code, see \cref{fig:LPU_summary}. These connections to the original code are not actually part of the graph $\mathcal{G}$, and are described separately (e.g.~by port and matching matrices \cite{swaroop2024universal}).

Logical measurement is done according to the principles of gauging measurement \cite{WY24}. Suppose we are measuring logical operator $L$ and $L_q\in\{I,X,Y,Z\}$ denotes the support of $L$ on qubit $q$. There are four steps:
\begin{enumerate}
\item {\bf Initialize} all edge qubits in $\ket{0}$.
\item {\bf Measure} all checks of the LPU. By the stabilizer update rules \cite{gottesman1997stabilizer}, this deforms the original code so that some of its checks gain support on edge qubits. Measure all the checks of this deformed code as well. The logical measurement result is $\bar m = \prod_{v\in\mathcal{V}} m_v$ where $m_v=\pm1$ are the results of measuring the vertex checks. Repeat this step a total of $C$ times to ensure fault-tolerance of the logical measurement result in the presence of measurement errors.
\item {\bf Return} to the original code by measuring the checks of the original code and measuring all edge qubits in the $Z$ basis. Let $m_e=\pm1$ be the result of measuring edge qubit $e$.
\item {\bf Correct} the original code by a Pauli operator. To describe the correction, fix an arbitrary vertex $v_0\in\mathcal{V}$. For any other vertex $v$, let $\mu_v$ be an arbitrary path from $v_0$ to $v$. If $\prod_{e\in\mu_v}m_e=-1$ apply a correction of $L_q$ to each qubit $q$ in the original code on which check $v$ had support.
\end{enumerate}
After these steps, it can be shown that the logical projector $(I+\bar m L)/2$ has been applied to the logical state \cite{WY24}. As is typical in error-correction, the Pauli correction in step 4 can simply be tracked in software and does not therefore need to be applied. We remark that, if we say the Pauli correction for one choice of $v_0$ is $Q$, then making another choice of $v_0$ leads to either the same correction by $Q$ or a correction by $QL$, which is equivalent to $Q$ since the logical state was projected into an eigenstate of $L$. The choices of paths in step 4 are also arbitrary because, after errors are corrected, all cycles $\gamma$ should satisfy $\prod_{e\in \gamma}m_e=+1$. 

The logical measurement result is vulnerable to $C$ measurement errors, so we choose $C=d_{\text{circ}}$ where $d_{\text{circ}}$ is an upper bound on the circuit distance of the original code, 10 for the gross code and 18 for the two-gross code.

We use the same LPU system with fixed-connectivity to measure a variety of different logical operators. We do this by carefully choosing connected subgraphs of $\mathcal{G}$ for each case \cite{CHRY24}. In several cases, the chosen subgraph is strictly smaller than the original graph. For the duration of the particular measurement, only the qubits and checks in the subgraph are involved in the measurement procedure described above. We elaborate on these subgraphs for the different use-cases of the LPU. 

The LPU enables in-module measurement of any logical Pauli in the group $\langle\bar X_1,\bar X_7,\bar Z_1,\bar Z_7\rangle$. With notation from \cref{fig:LPU_summary}, using just the qubits in $\mathcal{E}_l$ and checks in $\mathcal{V}_l$, $\mathcal{U}_l$ and $v_{\text{Bell}}$, the three non-trivial operators in $\langle\bar X_1,\bar Z_7\rangle$ can be measured. We refer to this collection of qubits and checks as half-LPU $l$. Note that only the appropriate subset of connections from vertex to code and from edge to code should be used so that the product of all active vertex checks equals the desired logical operator. Similarly, half-LPU $r$, consisting of the qubits in $\mathcal{E}_r$ and checks in $\mathcal{V}_r$, $\mathcal{U}_r$ and $v_{\text{Bell}}$, can be used to measure the non-trivial elements of $\langle\bar X_7,\bar Z_1\rangle$. For any of the other 9 non-trivial elements of $\langle\bar X_1,\bar X_7,\bar Z_1,\bar Z_7\rangle$ not in these two cases, we measure it using the entire LPU.

\begin{figure}[t]
\centering
\begin{subfigure}{0.75\textwidth}
\centering
\includegraphics[width=0.75\textwidth]{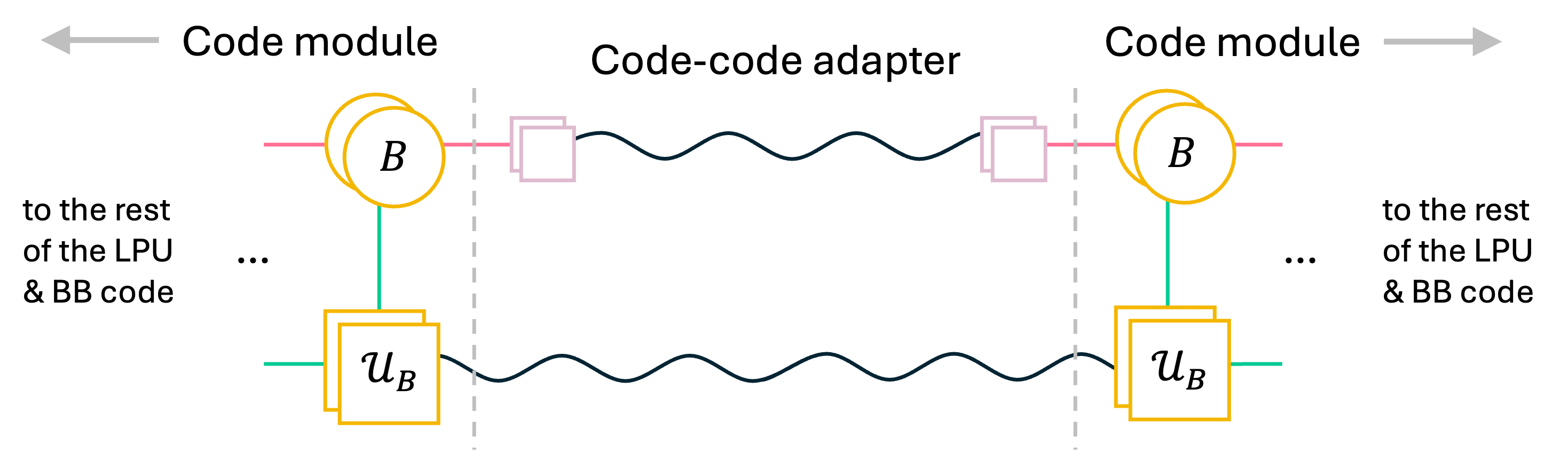}
\caption{Connectivity of the code-code adapter in general}\label{fig:code_code}
\end{subfigure}
\hfill
\begin{subfigure}{0.49\textwidth}
\centering
\vspace{15pt}
\includegraphics[width=\textwidth]{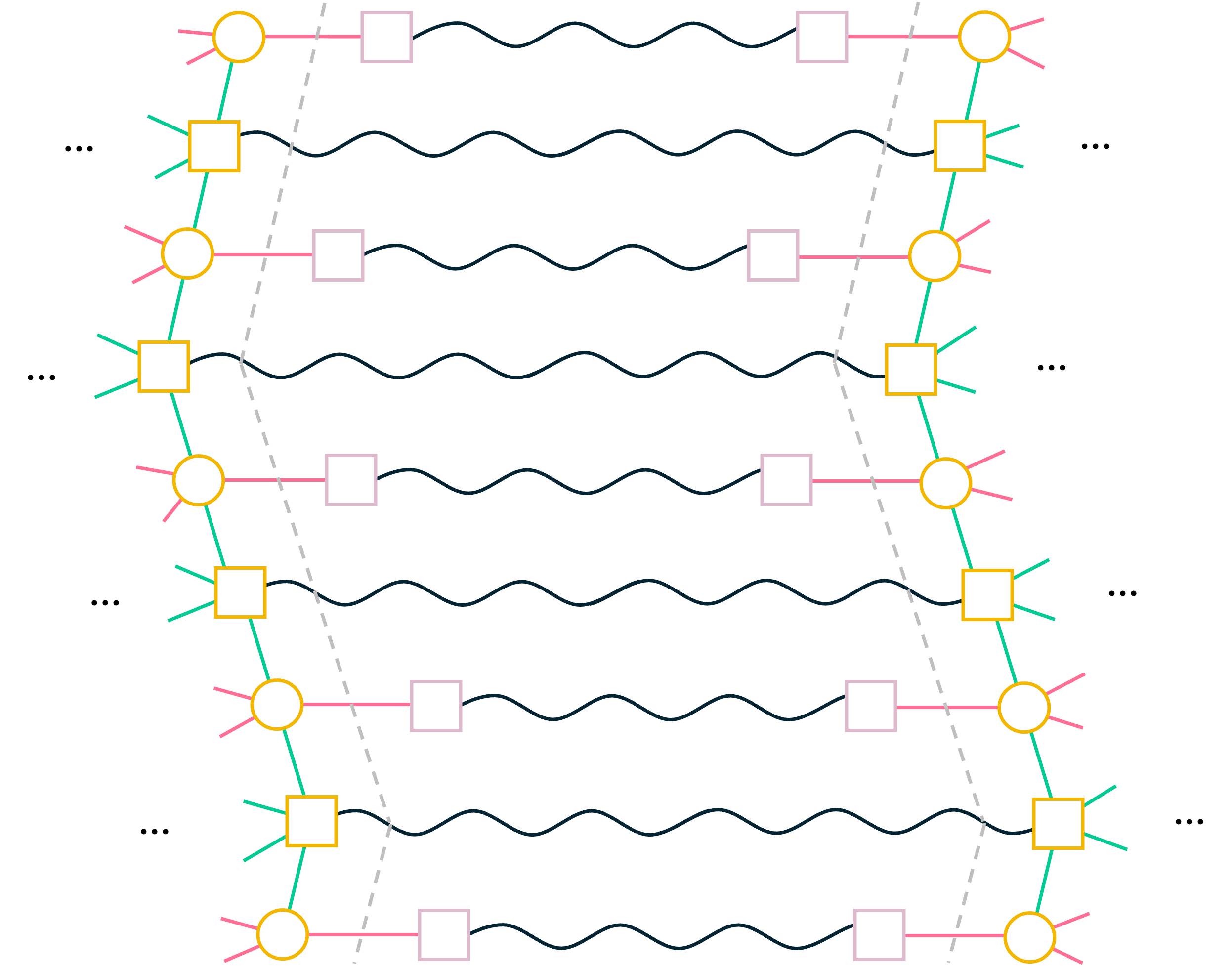}
\caption{Example code-code adapter: 1-1 Bell checks}\label{fig:code_code_1_1}
\end{subfigure}
\hfill
\begin{subfigure}{0.49\textwidth}
\centering
\vspace{15pt}
\includegraphics[width=\textwidth]{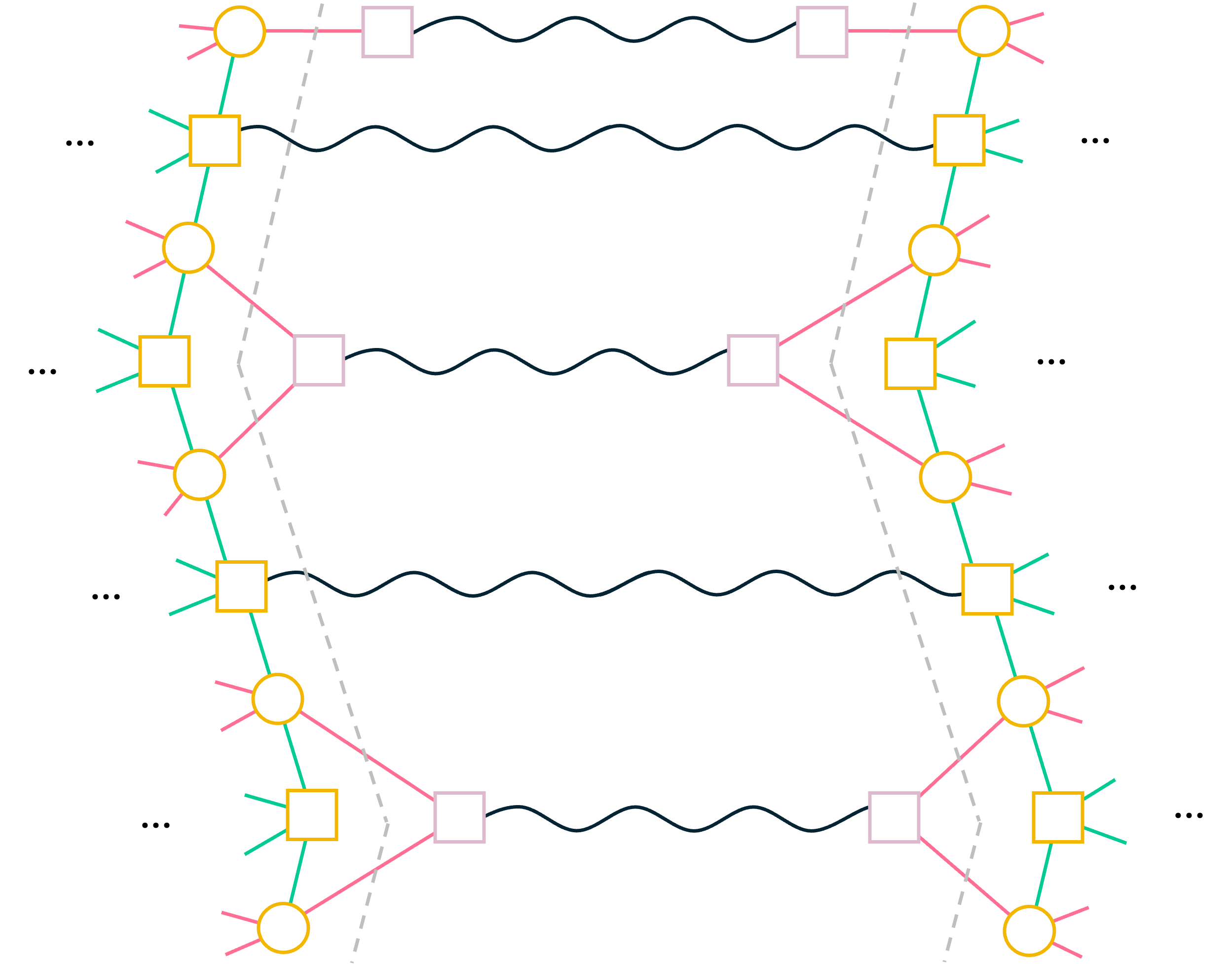}
\caption{Example code-code adapter: 2-1 Bell checks}\label{fig:code_code_2_2}
\end{subfigure}
\caption{Elaborating on the code-code adapter from \cref{fig:intercode_surgery}. (a) A redrawing of the general structure of the code-code adapter from \cref{fig:intercode_surgery}. The details of the connectivity are made clearer with the next two examples. (b) The only implementation of the code-code adapter we use and simulate in this work (though for simplicity this example is smaller than for either of our codes). For each bridge qubit (yellow circle) of the two codes, an additional check qubit (pink square) is connected, and these check qubits in turn are connected by Bell-couplers (wavy lines) to implement Bell checks as in \cref{fig:check_circuits}c. Similarly, Bell-couplers are used to turn the bridge checks (yellow squares) into Bell checks. (c) Alternative code-code adapter constructions exist. For instance, the new check qubits can connect many-to-one with bridge qubits, which saves on Bell-couplers but increases check weights. We leave exploration of this tradeoff and its potential to improve logical error rates of inter-module instructions to future work.}
\label{fig:code_code_interface}
\end{figure}

For inter-module measurements between BB code modules, the LPUs, connected by Bell-couplers as in \cref{fig:intercode_surgery}, enable the measurement of any element of $(\langle\bar X_1,\bar Z_7\rangle\cup\langle\bar X_7,\bar Z_1\rangle)^{\otimes2}$. The total graph in these cases consists of the appropriate half-LPUs, $l$ or $r$, from both BB code modules, the bridge qubits $B$ from both modules, Bell checks introduced via Bell-couplers that serve to identify the two sets of bridge qubits, and the weight-4 bridge checks $\mathcal{U}_B$ from one of the modules prepared in Bell states with the bridge checks of the other module via Bell-couplers.\footnote{The weight-3 bridge checks, one from each code module, are not used for inter-module measurements. Thinking graph theoretically, connecting two disconnected graphs via $w$ edges increases the cycle space rank by $w-1$, not $w$.} \cref{fig:code_code_interface} makes the schematic from \cref{fig:intercode_surgery} and this discussion more explicit with a couple examples. 

\begin{figure}[t]
\centering
\includegraphics[width=0.7\textwidth]{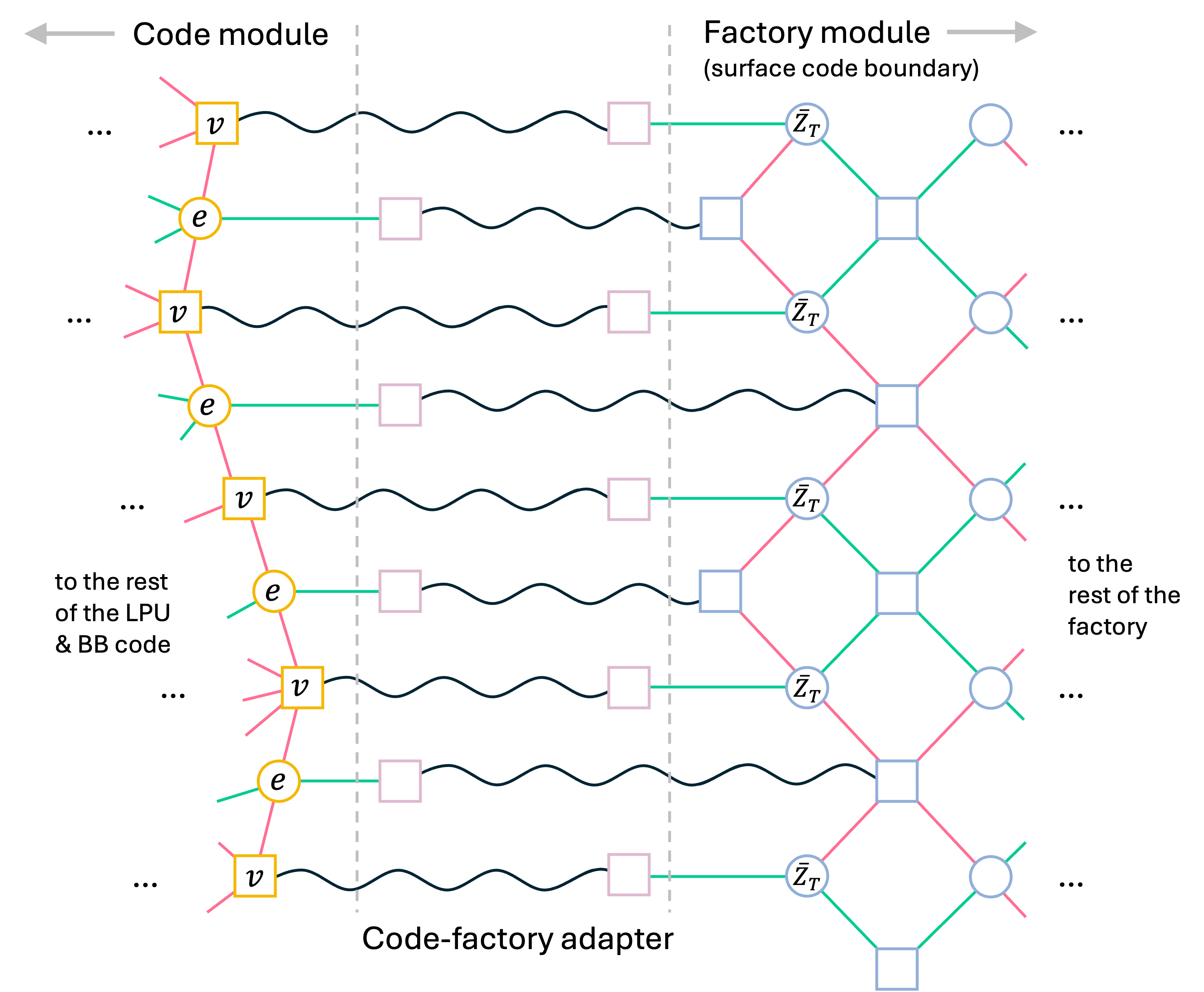}
\caption{Elaborating on the code-factory adapter from \cref{fig:intercode_surgery}. We draw an example connectivity between the LPU of a code module (left, yellow) and a magic state factory (right, blue) based on the surface code. We identify a path in the LPU of the code module with a number of vertices (labeled $v$) equal to the weight of the logical $\bar Z_{\text{T}}$ operator in the factory. These vertices are connected to the qubits in the support of $\bar Z_{\text{T}}$ by introducing new check qubits (pink) and Bell couplers (wavy lines). Some checks of the surface code necessarily gain support on the edge qubits (labeled $e$) in the path, and this extra support can be incorporated by again introducing more qubits and Bell couplers. If instead the surface code has distance larger than the longest path in the LPU (as is the case for the gross code), one option is to allow some of the check qubits in the adapter on the surface code side to connect to two consecutive qubits in the support of $\bar Z_{\text{T}}$, similar to \cref{fig:code_code_2_2}.}
\label{fig:code_factory_interface}
\end{figure}

A major reason we opt for connecting the bridge qubits of adjacent code blocks together (rather than for instance, directly connecting the bridge qubits of one block to the vertex checks of the other) is to keep the maximum degree connectivity under control. Since the bridge qubits have few connections, just 3 or 4, within a code module, connecting a code module to even three other modules through the bridge qubits does not increase the connectivity required beyond degree 7.

However, this arithmetic changes for code-factory adapters. In this case we opt to connect vertex checks in a BB code LPU via Bell-couplers to the qubits of a logical operator $\bar Z_{\text{T}}$ of the magic state factory, see \cref{fig:intercode_surgery}. This is the same idea used to perform joint measurements with the toric code in Ref.~\cite{swaroop2024universal}. For the case of connecting either of our two codes to a surface code magic state factory, we can ensure the maximum qubit or check degree remains 7 by finding in the graph defining the LPU (including bridge edges) a path through $d$ vertices that does not traverse vertices or edges that are already degree 7. Connect qubits of $\bar Z_{\text{T}}$ to the vertices of this path, and surface code $X$ checks adjacent to $\bar Z_{\text{T}}$ to the edges. See \cref{fig:code_factory_interface}. We note that the effective graph for joint measurements between a BB code and factory is still exactly the graph of the original BB code LPU, although several checks are implemented over Bell couplers and vertices (edges) of the graph connect to qubits (checks) in both the BB code and the factory. This means we can measure any operator in $\langle\bar X_1,\bar X_7,\bar Z_1,\bar Z_7\rangle\otimes\bar Z_{\text{T}}$, which benefits our compilation methods. We would have had a more limited set of measurements available if we connected to the factory through the LPU bridge qubits. 

We remark that direct connections to the vertex checks should be used sparingly if we want to also preserve a maximum qubit and check degree of 7. Thus, we just do this to connect a code module to at most one factory module. Still, there are similar ideas with a small $O(d)$ qubit overhead that can be used to expand the set of joint measurements between code modules without increasing the maximum qubit and check degrees. We leave study of these options to future work.

\subsection{Circuit scheduling for logical measurements}
\label{app:lpu_circuit_scheduling}

Our main goal in this section is to describe how to schedule syndrome measurement circuits for the deformed BB codes encountered in qLDPC surgery, i.e.~how to implement the {\bf Measure} step described in \app{lpu_meas_procedure}. We start, however, by describing how to schedule syndrome measurement circuits for arbitrary qLDPC codes in which checks are measured either with single check qubits or as Bell checks, see \cref{fig:check_circuits}a and c.

The circuit scheduling problem for a qLDPC code can be approached through graph coloring \cite{delfosse2021bounds} or integer (linear) programming \cite{vittal2024flag}. In this work we just report the results from graph coloring and leave benchmarking the improvement due to integer programming to future work.

The description of the scheduling problem is the same regardless of the method used to solve it. We start with a connectivity graph like that in \cref{fig:LPU_summary}, which is a graph in which nodes are of two types -- check qubits (squares) or data qubits (circles). Edges between check qubits and data qubits are labeled with non-identity single-qubit Paulis, and check qubits may have at most one edge to another check qubit to indicate a pair of Bell check qubits. If two check qubits are connected this way, there should be no data qubit that they are both connected to. This graph represents connectivity between data and check qubits, but it also implicitly represents a set of checks to be measured. Note the distinction between ``check qubit'' and ``check'' in the language in this section. Specifically, if a check $\Gamma$ is not a Bell check, $\Gamma$ just corresponds to a single check qubit in the connectivity graph. If check $\Gamma$ is a member of the set of Bell checks, denoted $\mathcal{C}_{\text{Bell}}$, then it corresponds to two (connected) check qubits $\gamma$ and $\tilde \gamma$ in the connectivity graph. We use $\gamma\sim\delta$ and $\delta\sim\gamma$ to indicate a check qubit $\gamma$ is connected to a data qubit $\delta$ in the connectivity graph.\footnote{Equivalently, one might formulate the scheduling problem by starting with an arbitrary check matrix $H\in\mathbb{F}_2^{r\times2n}$ for a stabilizer code with $r$ checks and $n$ data qubits, and then separately specify which of the checks are to be implemented by Bell checks and how the data qubits of those checks should be connected to the Bell pairs.}

We schedule just one syndrome cycle and then repeat the same schedule for multiple cycles. For every check qubit $\gamma$ connected to data qubit $\delta$ there is a two-qubit gate for which we create a variable $t_{\gamma,\delta}\in\mathbb{Z}$, indicating the timestep in which the gate will be performed. For convenience, we let $t_{\Gamma,\delta}$ for Bell check $\Gamma\in\mathcal{C}_{\text{Bell}}$ mean either $t_{\gamma,\delta}$ or $t_{\tilde\gamma,\delta}$ depending on whether $\gamma$ is connected to $\delta$ or $\tilde\gamma$ is. On the other hand, if  $\Gamma\not\in\mathcal{C}_{\text{Bell}}$, then $\Gamma$ is implemented by a single check qubit $\gamma$, and we define $t_{\Gamma,\delta}=t_{\gamma,\delta}$. Finally, the Bell state on each Bell check $\Gamma$ needs to be prepared via a single CNOT (assumed to be from $\gamma$ to $\tilde\gamma$), which we schedule at timestep $t_\Gamma\in\mathbb{Z}$. These integer variables $t_{\gamma,\delta}$ and $t_\Gamma$ representing two-qubit gate times are all we need to determine. Given integer values for all of them, we schedule initialization and measurement of check qubits as late as possible and as early as possible, respectively. 

We follow Ref.~\cite{vittal2024flag} in expressing the constraints on these variables. First, we must avoid scheduling two gates on the same qubit in the same timestep by enforcing the constraints
\begin{align}\nonumber
t_{\gamma,\delta}&\neq t_{\gamma,\delta'},\quad\text{for all check qubits $\gamma$ and distinct data qubits $\delta\sim\gamma$ and $\delta'\sim\gamma$,}\\\label{eq:unequality_constraints}
t_{\gamma,\delta}&\neq t_{\gamma',\delta},\quad\text{for all data qubits $\delta$ and distinct check qubits $\gamma\sim\delta$ and $\gamma'\sim\delta$.}
\end{align}

Second, we must schedule the Bell preparation CNOTs to avoid all the rest. We assume this preparation is done before any of the other gates from the Bell pair to the data, though this is not strictly necessary. For all Bell checks $\Gamma\in\mathcal{C}_{\text{Bell}}$, let $\Gamma$ be implemented by check qubits $\gamma,\tilde\gamma$ and enforce the constraints
\begin{align}\nonumber
t_\Gamma &< t_{\gamma,\delta},\quad\text{for all data qubits $\delta\sim\gamma$,}\\\label{eq:bellprep_constraints}
t_\Gamma &< t_{\tilde\gamma,\delta},\quad\text{for all data qubits $\delta\sim\tilde\gamma$.}
\end{align}

Third, checks that overlap other checks must be carefully scheduled so that the measurements are equivalent to measurements of the checks sequentially. This is expressed by considering the ``anticommuting overlap'' of checks $\Gamma$ and $\Gamma'$. We denote this by $A_{\Gamma,\Gamma'}$, and it is the set of all data qubits $\delta$ such that the support of check $\Gamma$ on $\delta$ (i.e.~a single-qubit Pauli) anticommutes with the support of check $\Gamma'$ on $\delta$ (i.e.~another single-qubit Pauli). Then, we enforce the constraints
\begin{equation}\label{eq:overlap_constraints}
\prod_{\delta\in A_{\Gamma,\Gamma'}}\left(t_{\Gamma,\delta}-t_{\Gamma',\delta}\right)>0,\quad\text{for all pairs of checks $\Gamma,\Gamma'$ with $A_{\Gamma,\Gamma'}\neq\emptyset$.}
\end{equation}
Note here we have made use of the aforementioned convenient notation $t_{\Gamma,\delta}$ where $\Gamma$ is a check and not a check qubit. We also remark that for the BB codes and the deformed codes encountered during our surgery operations, the size of any anticommuting overlap $A_{\Gamma,\Gamma'}$ is only ever 0 or 2.

Finally, we deviate from Ref.~\cite{vittal2024flag} by introducing an upper bound on the number of timesteps. This is an additional input to the problem, an integer $T_{\text{max}}$. We enforce
\begin{equation}\label{eq:maxtime_constraints}
0<t_{\gamma,\delta}<T_{\text{max}}+1,\quad\text{for all check qubits $\gamma$ and data qubits $\delta\sim\gamma$.}
\end{equation}
For large enough $T_{\text{max}}$, any optimization problem with the above constraints is feasible, because one can just sequentially measure all the checks. On the other hand, for small enough $T_{\text{max}}$ (e.g.~less than the maximum degree of the connectivity graph), any such optimization problem becomes infeasible. In the integer programming solution to scheduling, we do not directly optimize $T_{\text{max}}$ for reasons explained later, so having it as separate knob can be useful.

We can simplify the problem in our case of interest: scheduling circuits for the deformed BB codes encountered in qLDPC surgery. Specifically, we already know a schedule for the BB code, which both of our scheduling methods continue to use. This is formalized by setting the variables $t_{\gamma,\delta}$ for edges $\gamma\sim\delta$ lying within the BB code to their values from \cref{fig:check_circuits}b plus some choice of constant integer increment $\Delta_{\text{BB}}\ge0$ independent of $\gamma$ and $\delta$. The remaining gates to schedule lie on edges within the LPU or between the LPU and BB code.\footnote{It is useful to have $\Delta_{\text{BB}}$ in general because it may be optimal to schedule some of the gates connecting the LPU and BB code to happen before the start of the standard BB schedule, which setting $\Delta_{\text{BB}}>0$ allows.} The complete schedule must only satisfy the constraints in \cref{eq:unequality_constraints}-\cref{eq:maxtime_constraints}) to be valid, though it is nice if it also is constructed to minimize time or spacetime volume in some way. The specific optimization functions differ for our two methods.

\paragraph{Graph coloring} A simple but sub-optimal method for finding a valid solution involves graph coloring of bipartite graphs. While finding optimal edge-colorings of general graphs is NP-hard, edge-coloring bipartite graphs is possible in polynomial time \cite{konig1916grafok}. A similar method was used in \cite{CHRY24}. Edges $\delta \sim \gamma$ are scheduled in the following order:

\begin{enumerate}
    \item Bell states are initialized on $(\gamma,\tilde\gamma) \in \mathcal{C}_\text{Bell}$.
    \item Edges with $\gamma$ in the LPU and $\delta$ in the BB code are performed.
    \item Edges contained entirely in the BB code are scheduled according to the standard syndrome cycle presented in \cite{bravyi2024high}.
    \item Edges contained entirely in the LPU are scheduled in two phases:
    \begin{enumerate}
        \item We consider the subgraph of edges $\gamma\sim \delta$ where $\gamma$ acts as an $X$-check. We find an optimal edge-coloring of this bipartite graph using the `gcol' python package. We then associate edges with the same color to layers in the circuit.
        \item We do the same for edges with $\gamma$ acting as a $Z$-check.
    \end{enumerate}
    \item Edges with $\delta$ in the LPU and $\gamma$ in the BB code are performed.
\end{enumerate}

Once the order of the gates has been determined as above, the gates are scheduled as early as possible while meeting the constraint \cref{eq:overlap_constraints}. We find that the subcircuit obtained from the graph coloring of the LPU edges dominates the circuit depth, and also removes the opportunity for subsequent cycles to overlap. For $C_\text{mem}$ syndrome cycles the number of timesteps required is $12C_\text{mem}$.

\paragraph{Integer programming} To define the function to be optimized, we start by observing that the time needed for $C$ syndrome cycles is not always $C$ times the time needed for one syndrome cycle. For example, once a check qubit has been measured, it may be possible to immediately reset it and continue with the next cycle without waiting for all the measurements on other check qubits to finish.

To capture this observation, we define the lifetime of a check or data qubit to be the number of timesteps used to perform all the two-qubit gates on that qubit for one cycle. That is, for check qubit $\gamma$ and data qubit $\delta$, the lifetimes are
\begin{align}
l_\gamma &= \max_\delta t_{\gamma,\delta}-\min_\delta t_{\gamma,\delta}+1.\\
l_\delta &= \max_\gamma t_{\gamma,\delta}-\min_\gamma t_{\gamma,\delta}+1,
\end{align}
For a data qubit $\delta$, after $l_\delta$ timesteps spent in one syndrome cycle, it becomes available to begin the next cycle. A check qubit $\gamma$ needs two more timesteps for initialization and measurement and so becomes available to begin the next cycle after $l_\gamma+2$ timesteps. Therefore,
\begin{equation}\label{eq:Rcycle_time}
\text{time to complete $C$ cycles} = \max(\max_\gamma(l_\gamma+2),\max_\delta l_\delta)C+b,
\end{equation}
where the constant $b$ captures the delay between the first check qubits measured and the last check qubits measured. 

Notice that $l_\gamma$ must be at least the degree of $\gamma$ in the connectivity graph, and likewise $l_\delta$ must be at least the degree of $\delta$. Recall from \cref{sec:gross-review} that the time for $C_\text{mem}$ syndrome cycles of the BB code memory is $8C_\text{mem}+1$. We see now the coefficient of 8 is optimal because the degrees of all check and data qubits are 6.\footnote{Also, the constant of 1 is optimal for translationally invariant circuits~\cite{bravyi2024high}.} 

We choose to minimize the coefficient in \cref{eq:Rcycle_time}, $F=\max(\max_\gamma(l_\gamma+2),\max_\delta l_\delta)$, subject to constraints in \cref{eq:unequality_constraints}-\cref{eq:maxtime_constraints}. Though not obviously a linear integer programming problem, there are standard tricks that introduce auxiliary variables to make it so. The linearization of \cref{eq:overlap_constraints} and reformulation of the unequality constraints in \cref{eq:unequality_constraints} as inequalities is done in Ref.~\cite{vittal2024flag}. Furthermore, the minimization of a $\text{max}_{v\in V}v$ function over a set of variables $V$ can be rephrased by minimizing a new variable $v_{\text{max}}$ subject to the additional constraints $v_{\text{max}}\ge v$ for all $v\in V$. The max and min functions in the definitions of the qubit lifetimes can be similarly captured by introducing auxiliary variables $t^{\text{max/min}}_{\gamma/\delta}$. We end up with the following integer linear program, where $V_{\text{all}}$ is the set of all original and auxiliary integer variables.
\begin{equation}
\label{eq:optimize_schedule}
\begin{aligned}
\min_{V_{\text{all}}} \quad & F\\
\textrm{s.t.} \quad & F\ge t^{\text{max}}_\gamma-t^{\text{min}}_\gamma+1+2,\forall\text{ checks }\gamma\\
  &F\ge t^{\text{max}}_\delta-t^{\text{min}}_\delta+1,\forall\text{ data }\delta\\
  \quad & t^{\text{max}}_\gamma\ge t_{\gamma,\delta},\forall\text{ checks }\gamma\text{ and $\forall$ data }\delta\sim\gamma\\
  \quad & t^{\text{min}}_\gamma\le t_{\gamma,\delta},\forall\text{ checks }\gamma\text{ and $\forall$ data }\delta\sim\gamma\\
  \quad & t^{\text{max}}_\delta\ge t_{\gamma,\delta},\forall\text{ data }\delta\text{ and $\forall$ checks }\gamma\sim\delta\\
  \quad & t^{\text{min}}_\delta\le t_{\gamma,\delta},\forall\text{ data }\delta\text{ and $\forall$ checks }\gamma\sim\delta\\
  \quad & \text{\cref{eq:unequality_constraints}-\cref{eq:maxtime_constraints} linearized \cite{vittal2024flag}}
\end{aligned}
\end{equation}
We solve this problem using CPLEX \cite{cplex2022v22}.

We find for the gross code and all measurement modes of the LPU that $C$ syndrome cycles in the deformed code requires $9C+b$ timesteps, where $b=1,2$ depending on the mode. The coefficient of 9 is optimal because the maximum degree of check qubits and data qubits in the deformed codes is always $7$. The $b=1,2$ values are concluded to be optimal also (assuming you have frozen the BB code schedule as we did), because when setting $T_{\text{max}}$ smaller than 10 or 11 respectively, the corresponding problems become infeasible. Solving \cref{eq:optimize_schedule} takes roughly a second for the gross code cases.

We can take the optimization a step further. Because we just minimize $F$, CPLEX is not incentivized to optimize any particular check qubit lifetime below the maximum check qubit lifetime 7, even if the check qubit has degree less than 7. This could mean some check qubits are being kept alive longer than they need to be, which is in turn detrimental to the logical error rate. Therefore, after the initial optimization of $F$, having found some set of solutions $V^*$, we iteratively revisit each check qubit $\gamma$, and create a new optimization problem with the objective to minimize $l_\gamma$ and in which the only free variables are $t_{\gamma,\delta}$ for all $\delta$ connected to $\gamma$ and also $t_\Gamma$ if $\gamma$ is part of a Bell check $\Gamma$. All other variables are set to their solutions from $V^*$. All constraints from \cref{eq:unequality_constraints}-\cref{eq:maxtime_constraints} are again enforced. The solution of this problem for $\gamma$ serves to update the values of $t_{\gamma,\delta}$ in $V^*$ before moving onto the next problem for a different check. We proceed through the checks in some arbitrary order, ending with some final scheduling solution $V^*$. This whole process takes a bit longer than the initial optimization of $F$, but only about a minute for the gross code cases.

\subsection{Cost estimates for T state cultivation}
\label{app:cultivation-cost}

Our goal is to estimate the cost of preparing a single T magic state by cultivation based on \cite{GSJ2024cultivation}, and our estimates are summarized in \cref{tab:Tcultivate}. 
Starting from Gidney et al.'s estimates (Figure 2 of \cite{GSJ2024cultivation}), we record the number of qubits $f$. 
The simulations in \cite{GSJ2024cultivation} use a uniform noise model that is similar to ours. 
Let $p$ be the uniform physical noise strength (which we take to be $p = 10^{-3}$ throughout this cultivation analysis), $d_1$ be the fault distance of the construction, $P_\text{cult}$ and $P'_\text{cult}$ be the logical error and discard probability up to the end of cultivation but before starting the escape process, and $P_\text{e2e}$ and $P'_\text{e2e}$ be the end-to-end logical error and discard probability. 
We further estimate the discard probability of the escape and idling in the matchable code, $P'_\text{escape}$, and the approximate mean time $\bar{\tau}_\text{e2e}$ to complete the end-to-end protocol. 
The discard probabilities in this table are consistent with the expected number of attempts per kept shot (Figures 5 and 14 of \cite{GSJ2024cultivation}).

\begin{table}[htb]
\centering
\begin{tabular}{llllll rrr}\toprule
 &\multicolumn{5}{c}{Reproduced \cite[Fig.~2]{GSJ2024cultivation}} & \multicolumn{3}{c}{Estimated} \\ \cmidrule(r){1-6} \cmidrule(l){7-9}
$p$ & $d_1$ & $P_\text{cult}$ & $P'_\text{cult}$ & $P_\text{e2e}$ & $P'_\text{e2e}$ & $P'_\text{escape}$ & $f$ & $\bar{\tau}_\text{e2e}$ \\ \midrule
$10^{-3}$ & 3 & $6\times 10^{-7}$ & $35\%$ & $3\times 10^{-6}$ & $80\%$ & $69\%$ & 454 & 351 \\
$10^{-3}$ & 5 & $6\times 10^{-10}$ & $85\%$ & $2\times 10^{-9}$ & $99\%$ & $93\%$ & 463 & 2167 \\ \bottomrule 
\end{tabular}
\caption{Logical error and time estimates for T state cultivation. The first six columns are reproduced from Fig. 2 of \cite{GSJ2024cultivation}. 
We estimate the parameters $P'_\text{escape}$, $f$, and $\bar{\tau}_\text{e2e}$. 
We read off the values of the overall cultivation protocols in \tab{Tfactories} as $\bar{\tau}_\text{factory} = \bar{\tau}_\text{e2e}$ and $P_\text{factory} = P_\text{e2e}$ respectively.
\label{tab:Tcultivate}
}
\end{table}

Ref.~\cite{GSJ2024cultivation} allows for either parallel or sequential implementations of cultivation stages. 
We assume that T states are prepared ``in-place'' by sequentially stepping through each stage of the cultivation protocol. 
This has relatively simple control requirements and low qubit overhead, but the lack of parallelism may mean somewhat longer preparation times. 

Let’s analyze the mean time to prepare a T state by assuming we discard states that fail their checks at one of two different points during the protocol, either post-cultivation before the color code grows or post-gap when the complementary gap has been computed. 
To get the post-gap discard probability $P'_\text{escape}$, we need to compute it from the end-to-end discard probability $P'_\text{e2e}$ in \tab{Tcultivate} using
\begin{equation}
P'_\text{e2e} = P'_\text{cult} + (1-P'_\text{cult})P'_\text{escape}.
\end{equation}
Rewriting, we see
\begin{equation}
P'_\text{escape} = (P'_\text{e2e} - P'_\text{cult})/(1 - P'_\text{cult}).
\end{equation}

Next, we need to compute the circuit depths of the cultivate and escape parts of the protocol for $d_1\in\{3,5\}$. 
Figure 15 \cite{GSJ2024cultivation} is relevant to understand the time needed to complete the in-place cultivation and end with the T state in a matchable code. 
The supplementary data \cite{GSJ2024cultivationdata} includes stim circuits for end-to-end cultivation protocols using the desaturation decoder. 
Each circuit is labeled by a total number of cycles $r$, a number of cycles spent idling in the grafted code during the escape stage $r_1$, and a number of cycles spend idling in the matchable code $r_2$. 
A cycle is defined to have some number of unitary gate layers and to start/end with reset/measure layers. The two relevant end-to-end parameter sets are:
\begin{itemize}
    \item $d_1=3$: $r=13$, $r_1=3$, $r_2=5$, $d_2=15$,
    \item $d_1=5$: $r=20$, $r_1=5$, $r_2=5$, $d_2=15$.
\end{itemize}
These circuits correspond to those studied in Fig. 14 of \cite{GSJ2024cultivation}.
For $d_1=3$, the circuit depth is 39 timesteps to cultivate a state and another 90 timesteps to escape and idle in the matchable code. 
For $d_1=5$, we have 89 timesteps to cultivate and another 110 timesteps to complete the protocol. 
In both cases, the total number of timesteps is close to $10r$, and the number of timesteps during the escape is close to $10(r_1+r_2)$.

We can estimate the mean time $\bar{\tau}_\text{e2e}$ to obtain a T state as follows. 
Let $t_{d_1, cult}$ and $t_{d_1, escape}$ be the durations of the two parts of the protocol at fault-distance $d_1$. 
Then we set
\begin{align}
t_{3, cult} & = 39, \\
t_{3, escape} & = 90, \\
t_{5, cult} & = 89, \\
t_{5, escape} & = 110.
\end{align}
For each case, we can estimate the mean time $\bar{t}_{d_1,p} = (1-P'_\text{cult})^{-1} t_{d_1, cult} + (1-P'_\text{escape})^{-1}t_{d_1, escape}$, omitting the second term when the state is not grown. 
Rounding to the nearest integer, we find
\begin{align}
\bar{\tau}_\text{e2e}(d_1=3) & = 1.54t_{3, cult} + 3.23t_{3, escape} =  351, \\
\bar{\tau}_\text{e2e}(d_1=5) & = 6.67t_{5, cult} + 14.3t_{5, escape} =  2167.
\end{align}

We note here the number of qubits for each of these protocols. 
The supplementary data \cite{GSJ2024cultivationdata} includes stim circuits with this information. 
The $r_1=3$ and $r_1=5$ end-to-end protocols use 454 and 463 qubits, respectively. 
The $r_1=5$ inject and cultivate protocol uses 42 qubits. 
These are consistent with the sizes of a distance-15 surface code and a distance-5 hexagonal color code.

\subsection{Simulation of bicycle instructions}
\label{app:logical-op-simulations}

Here we present our methods for estimating the logical error rates of individual bicycle instructions under circuit-level noise in \cref{sec:instruction-validation}.
For all our simulations, we make the standard modeling assumption that the system begins in an error-free codestate prior to the logical operation, and that a final cycle of noise-free stabilizer measurements are performed at the end.

For computational convenience, for simulations we further simplify the noise model in \sec{instruction-validation} by representing all faults as independent events with equal probability $q = p/15$. 
Events with probability $m \cdot p/15$ are modeled as $m$ independent faults, each occurring with probability $q$.

\paragraph{Decoding matrices} 
We do not separate the syndrome and faults into their $X$ and $Z$ components, instead we treat the full syndrome and set of faults together.
Decoding is defined by two binary matrices: a check matrix $H \in \{0,1\}^{M \times N}$ and a logical action matrix $A \in \{0,1\}^{K \times N}$. 
Each row of $H$ corresponds to a detector, formed from the parity of the same stabilizer measurement across two cycles. 
Columns index distinct faults with $H_{ij} = 1$ if and only if the $j$th fault flips the $i$th detector outcome.

The interpretation of $A$ depends on the logical operation. 
For unitary operations on a code encoding $k$ logical qubits, $K = 2k$, and the rows of $A$ specify a complete generating set of the group of logical operators for the circuit. 
For the measurement of a logical operator $\bar{L}$, $K = 2k-1$, and the rows of $A$ specify a complete generating set of that subgroup of the logical operators for the circuit which commute with $\bar{L}$.

Syndrome circuits are constructed and analyzed using stim \cite{gidney2021stim}. Since stim requires initialization of a particular stabilizer state, some additional work is required to construct an experiment that is sensitive to both $X$ and $Z$ logical errors. The method is as follows: we prepare $k$ Bell pairs and encode one qubit of each of the bell pairs into the error correction code. Then, after running the noisy syndrome circuit, we perform a logical Bell measurement. For $\bar X$ measurements we do the same thing, except that the qubit being measured is initialized in the $\ket{+}$ state, thereby ensuring a consistent measurement result. This is the same method as used in \cite{CHRY24}.

Note that the idle and shift automorphism instructions were simulated using a circuit composed of $C=d^*_\text{circ}$ repeated applications of a single idle cycle or a shift operation respectively, with the logical error rate $P(p)$ reported as the failure probability of the circuit divided by $C$. 

\paragraph{Decoding definitions}
A fault configuration is represented by a bitstring $e \in \{0,1\}^N$, producing a syndrome $\sigma = He$ and logical action $l = Ae$, both computed modulo 2. 
Error correction proceeds as follows:
\begin{enumerate}
  \item A fault $e$ produces syndrome $\sigma = He$.
  \item The decoder $\mathcal{D}$ returns a correction $c = \mathcal{D}(\sigma)$ satisfying $Hc = \sigma$.
  \item Correction succeeds if $Ac = Ae$, and fails otherwise.
\end{enumerate}
Successful correction in the third step is typically facilitated by choosing a highly probable (in the physical noise model) error $e'$ consistent with syndrome $\sigma=He=He'$ as the correction $c=e'$ in the second step. Finding the most probable correction is generically hard \cite{iyer2015hardness}, but heuristic decoders, like we use and describe below, can still give accurate results.

The \emph{circuit distance} $D$ is the minimum weight of a fault $e$ such that $He = 0$ but $Ae \neq 0$. 
It differs from the \emph{code distance} $d$ due to circuit-level and noise-model specifics. 
We define the \emph{min-failing weight} $w_0$ as the smallest weight of a fault that causes decoder failure.
For an optimal decoder, $w_0 = \lceil D/2 \rceil$.
In practice, we do not use an optimal decoder for our simulations. 

\paragraph{Relay-BP decoder}
Our numerical analyses are performed using Relay-BP, which is a novel quantum error correction decoder designed to meet the demands of real-time decoding in large-scale quantum computing~\cite{RelayPaper}. 
Relay-BP builds upon the traditional min-sum belief propagation (BP) algorithm by introducing spatially-disordered memory terms that help mitigate the algorithm’s tendency to get trapped in symmetric or oscillatory states.  
It also uses a serialized ensembling strategy, where the outputs of one run are fed as inputs of the next, using different memory terms for each run, helping to converge stubborn trapping sets, and also providing multiple candidate corrections from which the best can be selected. 
This results in a decoder that is well suited for real-time decoding with FPGAs or ASICs (it has a similar footprint to a standard BP implementation) that is capable of very accurate and fast decoding. 
Decoder performance depends on several parameters which are chosen differently depending on the circuit being decoded. We give our parameter choices in \cref{tab:decoder_parameters}.

\paragraph{Monte Carlo sampling} The simplest method is to run $T$ trials under the full noise model for a given value $q$, and count $F$ failures:
\begin{equation}
\hat{P}(q) = \frac{F}{T}. \label{eq:failure-fraction-monte-carlo}
\end{equation}

The logical error rate can equivalently be determined by the distribution of failing faults across weights. 
Let $f(w)$ denote the \emph{failure spectrum}, which is the fraction of weight-$w$ faults that lead to logical failure. 
Then:
\begin{equation}
P(q) = \sum_{w = w_0}^{N} f(w) \binom{N}{w} q^w (1 - q)^{N - w}, 
\label{eq:logical_error_by_weight}
\end{equation}
such that $f(w)$ fully captures the error correction performance. The sum is over $N$ possible faults with probability $q$ starting at $w_0 = \lceil D/2 \rceil$, the minimum weight fault. In practice this sum is evaluated not all the way up to $N$ and can be truncated at a lower value since the smaller $w$ terms dominate.

\begin{figure}[tb]
    \centering
    \includegraphics[width=\textwidth]{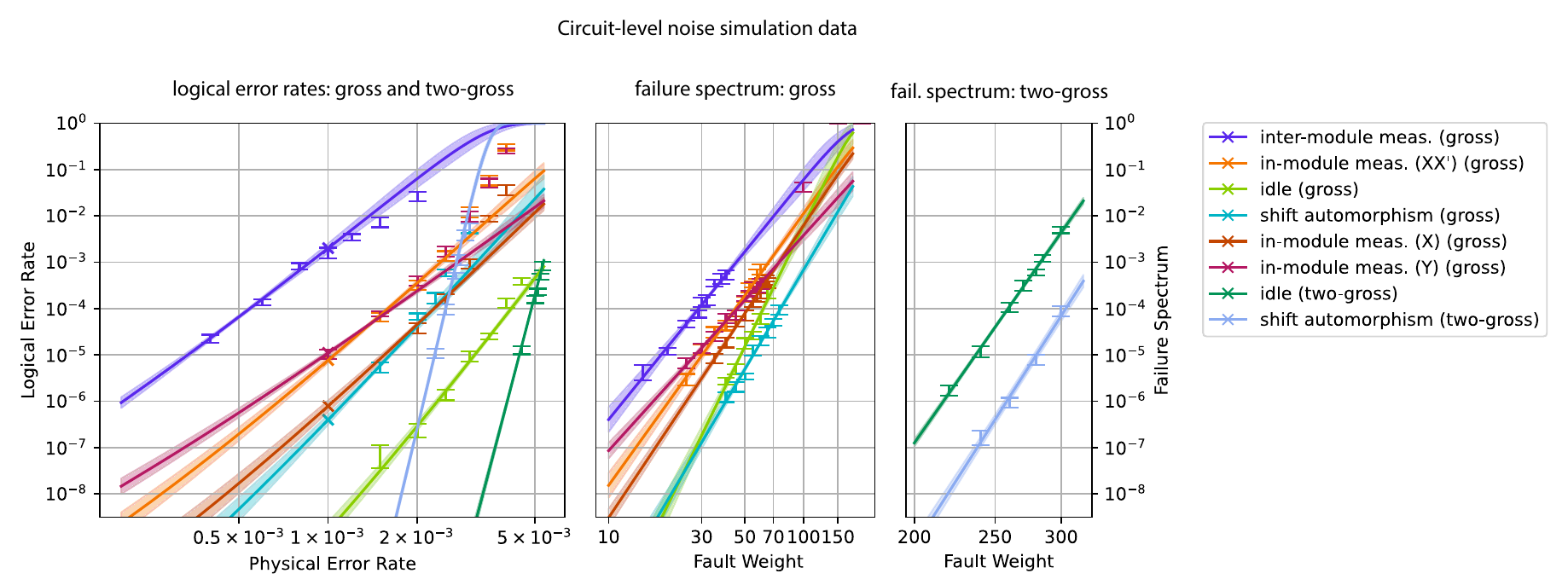}
    \caption{
        Monte Carlo estimates of $P(p)$ and $f(w)$ from circuit level noise simulations, fitting the latter with $f_\text{ansatz}(w)$ in \cref{eq:gamma-to-f}. 
        The resulting fit parameters are shown in \cref{tab:ansatz_fit_params}.
        Here, in-module meas. (X), (Y) and (XX') correspond to measuring $\bar{X}_1$, $\bar{Y}_1$ and $\bar{X}_1\bar{X}_7$ respectively, while inter-module meas. corresponds to measuring $\bar{X}_1 \otimes \bar{X}_1$. The simulations were completed using approximately $22\,700$ core hours.}
    \label{fig:all_data}
\end{figure}

\paragraph{Failure-spectrum fitting.} 
In Ref.~\cite{LowLogicalErrorPaper}, it is observed that for a wide range of QEC systems, $f(w)$ is well described by an ansatz for the full range of $w$:
\begin{equation}
f_\text{ansatz}(w) = a\left[ 1 - \exp\left(- \frac{f_0}{a} \left( \frac{w}{w_0} \right)^{\gamma} \right) \right], \qquad a = 1 - \frac{1}{2^K}, 
\label{eq:gamma-to-f}
\end{equation}
where $f_\text{ansatz}(w) = 0$ for $w < w_0$ and where parameters $\gamma$, $f_0$ and $w_0$ can be either be extracted by fitting to data, or set explicitly if known.
Note that for low $p$, this matches the known asymptotic form $f_0 \binom{N}{w_0} p^{w_0}$, and also ensures that for high $w$ (and corresponding for $p$ much greater than zero) the expected failure rate $1-1/2^K$ consistent with a random correction.
The fit parameters in \cref{tab:ansatz_fit_params} were obtained by Chi-squared minimization for the deviation between the ansatz and the data for $f(w)$, assuming $w_0 = 5$ for all gross code instructions, and $w_0 = 9$ for all two-gross instructions.  
Note that we observe some deviation from the ansatz $f_\text{ansatz}(w)$ at higher $w>80$ for $\bar{Y}_1$, and we make the conservative choice to exclude those points from the fit (including them would result in lower logical error estimates).
See \cite{LowLogicalErrorPaper} for further details and justification for this method.

Error propagation from estimates $\hat f(w)$ and $\hat{P}(q)$ are obtained through a bootstrapping approach. Data points are re-sampled according to a binomial distribution with an identical mean as the Monte Carlo average, and ansatz fits are computed according to such re-sampled data. The ensemble of fit parameters gives a distribution of extrapolation predictions for each value of $q$ or $w$ - we report the standard deviation in the shaded region around the fits in \cref{fig:data-logical-error-rates} and \cref{fig:all_data}.

\begin{table}[tb]
    \centering
    \begin{tabular}{llrrrr} \toprule
    series & $f_0$ & $\gamma$ & $w_0$ & $K$ & $N$  \\  \midrule
    gross idle & $e^{-31.02}$ & 8.66 & 5 & 24 & 210960 \\
    gross automorphism & $e^{-28.63}$ & 7.13 & 5 & 24 & 483840\\ 
    gross $X$ & $e^{-23.97}$ & 6.31 & 5 & 23 & 324526 \\
    gross $XX'$ & $e^{-22.06}$ & 5.87 & 5 & 23 & 398717\\
    gross $Y$ & $e^{-19.49}$ & 4.65 & 5 & 23 & 400117\\
    gross $X\otimes X$ & $e^{-18.32}$ & 5.19 & 5 & 23 & 743456\\ 
    two-gross idle & $e^{-95.83}$ & 25.78 & 9 & 24 & 762912  \\
    two-gross automorphism & $e^{-108.53}$ & 28.22 & 9 & 24 & 1748736\\
    \bottomrule
    \end{tabular}
    \caption{%
         Failure-spectrum ansatz fit parameters for \cref{eq:gamma-to-f}.
    }\label{tab:ansatz_fit_params}
\end{table}

\begin{table}[tb]
    \centering
\begin{tabular}{llr} \toprule
series & \texttt{gamma} & \texttt{rng\_width} \\  \midrule
gross $X\otimes X$ & 0.75 & 0.41 \\
other gross & 0.79 & 0.45  \\
all two-gross & 0.673 & 0.488 \\
\bottomrule
\end{tabular}

    \caption{%
        \label{tab:decoder_parameters} Parameters used for decoding with the Relay-BP decoder. The following parameters are the same for all gates: \texttt{ewainit\_discount\_factor} = 0.875, \texttt{num\_sets} = 600, \texttt{pre\_iter} = 80, \texttt{set\_num\_iters} = 60, \texttt{stop\_nconv} = 5, \texttt{max\_iter} = 60, \texttt{ms\_scaling\_factor} = 1.
    }
    
\end{table}

\subsection{Distance validation of bicycle instructions}\label{app:cplex}

\paragraph{Phenomenological fault-distance validation} All of our qLDPC surgery constructions boil down to measuring a sequence of check operators over time. This sequence changes with time, most prominently when we use the LPU to perform logical Pauli measurements. A simplified error model called the phenomenological noise model ignores the circuit-level implementation of these check operators and assume just two sources of noise — single-qubit depolarizing noise on each data qubit and bit flips on each check measurement result. 

In isolation, these phenomenological errors are detected---there will be a set of two (or more) measurements whose product is supposed to be $+1$ in the absence of errors, but is observed to be $-1$ instead.
Two back-to-back measurements of the same check is a common example of such a set since the measurements should agree in the error-free case. But en masse several phenomenological errors can avoid detection altogether and cause a non-trivial change in the logical information. Assuming the phenomenological noise model, the phenomenological fault-distance is the minimum number of phenomenological errors that effect a logical operation, or in other words, that cause a logical error without being detected.

We follow the method of Ref.~\cite{WY24} for verifying the phenomenological fault-distance of our qLDPC surgery gadgets. This separately considers two ways the logical Pauli measurement can fail. First, there could be a ``space-like” error, which means a logical error occurs that commutes with (but is not equivalent to) the measured logical Pauli. Space-like errors alter the logical information still in memory. Second, there could be a ``time-like” error, which means errors have occurred so as to corrupt the logical measurement result. An error can be both space- and time-like, but Ref.~\cite{WY24} shows that such errors are always the combination of a space-like error and a time-like error.

Also in Ref.~\cite{WY24}, the minimum weight $d_s$ of a space-like error is found to be the minimum of the distance of the original code $d$ and the distance of the deformed code $d^*_s$. It is often the case in practice that the deformed code is no better than the original code, $d^*_s\le d$, in which case $d_s=d^*_s$. Of course, the best one can hope for is to have $d_s=d^*_s=d$.

Similarly, time-like errors are covered by considering the distance of a subsystem code describing the code deformation \cite{vuillot2019code,CHRY24,WY24}. If $\mathcal{S}$ is the stabilizer group of the original stabilizer code, and $\mathcal{S}'$ is the stabilizer group of the deformed code, the subsystem code has gauge group $\mathcal{G}_{\text{def}}=\langle\mathcal{S},\mathcal{S}'\rangle$. In terms of qLDPC surgery specifically, this gauge group is just the stabilizer group of the deformed code plus single qubit $Z_e$ operators on each of the edge qubits $e$ (because those qubits begin in $\ket{0}$ when the code is in $\mathcal{S}$). The stabilizer group $\mathcal{Z}(\mathcal{G}_{\text{def}})$ of the subsystem code, or center of the gauge group, is generated by the checks of the deformed code minus the vertex checks. The vertex checks are not stabilizers, only gauge operators, because they anticommute with the single-qubit $Z_e$ operators from $\mathcal{S}$.

In this subsystem code $\mathcal{G}_{\text{def}}$, we are in particular interested in the minimum weight $d^*_t$ of a logical or gauge operator that anticommutes with the operator being measured (itself a product of the vertex checks). The time-like distance is $d_t=\min(C,d^*_t)$ \cite{CHRY24} where $C$ is how many syndrome cycles we perform in the deformed code, which we get to choose to be greater than $d^*_t$. Note $d^*_t$ is itself upper bounded by the code distance of the original code (any logical in the original code that anticommutes with the measured operator is also logical in the subsystem code). Thus choosing $C=d$ is sufficient.

In summary, to calculate the phenomenological fault-distance of qLDPC surgery, we need to calculate
\begin{enumerate}
\item $d^*_s$, the code distance of the deformed code, and
\item $d^*_t$, the minimum weight of a Pauli operator that both commutes with all checks of the deformed code (minus the vertex checks), but anticommutes with the operator we are measuring.
\end{enumerate}
We reduce both tasks to the same optimization problem.

Let $\mathcal{P}_n$ be the entire Pauli group on some fixed number of qubits $n$, $M$ be a set of Pauli operators, and $q$ be a particular Pauli operator. We want to find the minimum weight of a Pauli $p$ that commutes with all of $M$ but anticommutes with $q$. We imagine Paulis are presented as symplectic bit strings $p,q\in\mathbb{F}_2^{2n}$ and the set of Paulis is a matrix $M$ with each row representing one element. The symplectic inner product is implemented by a matrix $\Lambda=\left(\begin{smallmatrix}0&I\\I&0\end{smallmatrix}\right)\in\mathbb{F}_2^{2n\times2n}$. Consider the following optimization problem which enforces the commutation constraints
\begin{equation}
\label{eq:optimize}
\begin{aligned}
\min_{p\in\mathcal{P}_n} \quad & |p|\\
\textrm{s.t.} \quad & M\Lambda p^\top=0\\
  &q\Lambda p^\top=1 \\
\end{aligned}
\end{equation}
where $|p|$ is the Pauli weight of $p$.

By definition, $d^*_t$ can be calculated by solving \cref{eq:optimize} once with rows of $M$ representing generators of the stabilizer group $\mathcal{Z}(\mathcal{G}_{\text{def}})$ and $q$ the Pauli operator being measured. Now for $d_s^*$, we note that calculating the distance of any stabilizer code reduces to solving \cref{eq:optimize} $2k$ times if the code encodes $k$ qubits. For this case, $M$ always represents the stabilizer group of the code and we set $q$ to each of $2k$ logical operators that form a logical basis. The distance of the code is the minimum of the solutions of these $2k$ optimization problems. We apply this method to the deformed code to find $d_s^*$.

CPLEX \cite{cplex2022v22} is well-equipped to solve \cref{eq:optimize} when it is properly formulated as an integer linear program. The only part that is not obviously linear is the calculation of $|p|$, because if $p=(x_1,\dots,x_n|z_1,\dots,z_n)$ for bits $x_i,z_i$ then $|p|=\sum_i(x_i+z_i-x_iz_i)$. We can resolve this nonlinearity by introducing $n$ auxiliary variables $b_i\in\{0,1\}$, rewriting $|p|$ as $\sum_i(x_i+z_i-b_i)$, and adding for all $i=1,\dots,n$ additional linear inequality constraints, $x_i+z_i<b_i+2$ and $x_i+z_i>2b_i-1$. Intuitively, these constraints enforce $b_i=1$ if and only if $x_i+z_i=2$ (which itself holds if and only if $x_iz_i=1$).

Thus, we can in principle determine $d^*_s$ and $d^*_t$ for all the different in-module logical measurements that we can perform using the LPU on the gross and two-gross codes. In practice, this confirmation also works quite well, and we can quickly and efficiently verify all cases have $d^*_s=d^*_t=d$, the original code distance. See \tab{phenom_fault_distances}.

\begin{table}[t]
    \centering
    \renewcommand{\arraystretch}{1.15}
    \begin{tabular}{|cc||m{1cm}|m{1cm}|m{1cm}|m{1cm}|m{1cm}|m{1cm}|m{1cm}|m{1cm}|m{1cm}|}
        \hline
        \multicolumn{2}{|c||}{} & $\bar X_1$ & $\bar Z_1$ & $\bar Y_1$ & $\bar X_1\bar X_7$ & $\bar X_1\bar Z_7$ & $\bar Z_1\bar X_7$ & $\bar X_1\bar Y_7$ & $\bar Y_1\bar X_7$ & $\bar Y_1\bar Y_7$ \\\hline\hline
        \multirow{2}{*}{gross} & $d^*_s$ & 12 & 12 & 12 & 12 & 12 & 12 & 12 & 12 & 12 \\\cline{3-11}
        & $d^*_t$ & 12 & 12 & 12 & 12 & 12 & 12 & 12 & 12 & 12 \\\hline\hline
        \multirow{2}{*}{two-gross} & $d^*_s$ & 18 & 18 & 18 & 18 & 18 & 18 & 18 & 18 & 18 \\\cline{3-11}
        & $d^*_t$ & 18 & 18 & 18 & 18 & 18 & 18 & 18 & 18 & 18 \\\hline
    \end{tabular}
    \caption{The space- and time-like distances $d^*_s$ and $d^*_t$ evaluated using CPLEX for our two BB codes and different in-module logical measurements enabled by their LPUs. Of the 15 non-trivial logical measurements in $\langle\bar X_1,\bar X_7,\bar Z_1,\bar Z_7\rangle$ we evaluate only 9 because these also imply the other 6 cases by using the ZX-duality. For instance, the deformed code used for measuring $\bar X_7$ is dual to that for $\bar Z_1$.}
    \label{tab:phenom_fault_distances}
    \renewcommand{\arraystretch}{1}
\end{table}

The phenomenological fault-distances of inter-module measurements can in principle be verified using CPLEX as well, though in practice the bigger problem size makes it more difficult. We have verified phenomenological fault-distance directly for the gross code measuring $\bar X_1\otimes\bar X_1$ for instance. 

For some other cases, even if we cannot directly verify the fault-distance with CPLEX, we can rely on Theorem 24 in \cite{CHRY24} to imply that joining two fault-distance achieving systems will preserve the fault-distance. Making use of this theorem relies on the additional assumption that the deformed code is CSS — which applies to half of the inter-module measurements — and that the logical support in each of the individual codeblocks is minimal weight within its logical equivalence class. This latter property is automatically true for the weight-12 operators $\bar X_1,\bar X_7,\bar Z_1,\bar Z_7$ for the distance-12 gross code. For the two-gross code, these operators have weight 20, while the code has distance 18. Still, we can verify that $\bar X_1,\bar X_7,\bar Z_1,\bar Z_7$ for the two-gross code are not equivalent modulo stabilizers to any weight 18 logical operator (in the set enumerated in \cref{fig:logical_counts}, which we strongly believe to be all weight-18 logicals). Thus, they are minimum weight in their logical equivalence classes, and the theorem applies.

\paragraph{Circuit fault-distance validation} Recall the definition of circuit distance in \app{logical-op-simulations}: Given check matrix $H\in\{0,1\}^{M\times N}$ and logical action matrix $A\in\{0,1\}^{K\times N}$ describing the decoding problem, the circuit distance is the minimum weight vector $e$ satisfying $He=0$ and $Ae\neq0$. It is possible to solve this problem exactly by again solving the optimization problem in \cref{eq:optimize} $K$ times. Specifically, let $M\Lambda = H$ and $q\Lambda = A_{i}$, where $A_i$ is row $i\in\{1,2,\dots,K\}$ of $A$.

In practice however, solving the problem exactly as an integer program is rarely feasible since the circuit noise problem size $N$ is usually orders of magnitude larger than the problem size for phenomenological noise. Instead, we can employ a decoder to upper bound the circuit distance. This is the same as what is done in Ref.~\cite{bravyi2024high}. As a brief description, this works by selecting a random element $a\in\text{rowspace}(H)+(\text{rowspace}(A)\setminus0)$ and then asking the decoder to find a vector $e$ satisfying $He=0$ and $a\cdot e=1$. The weight $|e|$ is an upper bound on the circuit distance, by definition. Also, if the decoder is reasonable, the weight of $e$ will be low, perhaps even minimal. We find the BP+OSD decoder \cite{panteleev2021degenerate} to do quite well.

We upper bound the circuit distances of our bicycle instruction circuits by decoding several thousand times for the in-module instructions and more than 700 times for the $\bar{X}_1\otimes\bar{X}_1$ measurement. The reported circuit distance upper bounds are reported in \tab{distances}.

\clearpage
\subsection{Pauli-based computation}%
\label{app:PBC}

\emph{Pauli-based computation} (PBC) broadly refers to several classes of circuits and compilation techniques based on gates defined by multi-qubit Paulis.
In this work, we employ PBC as an intermediate representation to compile general circuits into a form that is more amenable for further compilation onto the bicycle architecture.
Specifically, we transform input circuits into ones composed solely of Pauli-generated rotations and multi-qubit Pauli measurements.

Pauli-based computation was introduced by \textcite{bravyi2016trading}, building on earlier frameworks for magic state distillation and injection developed by \textcite{bravyi2005universal,knill2005quantum}.
These works considered gate sets consisting solely of Pauli measurements and magic state preparations.
Subsequent work developed efficient compilation techniques for surface codes~\cite{litinski2019game,chamberland2022universal,beverland2022assessing}, extending PBC to include Pauli-generated rotations such as Clifford $e^{i \frac{\pi}{4} P}$ and non-Clifford $e^{i \frac{\pi}{8} P}$ rotations.

In our compilation strategy (\cref{sec:compiling}), our first compilation step is a simple extension of the transformation illustrated by \cite{litinski2019game} to input circuits with arbitrary angles $\frac{\phi_j}{2}$ (rather than $\pi/4$) and multi-qubit Paulis $P_j$ (rather than single-qubit $Z$). 
Any such circuit acting on $n$ logical qubits can be written as a sequence, for some $k \in \mathbb N$, as
\begin{align}
C_{k+1} \prod_{j=1}^k e^{i \phi_j P_j} C_j \ket{0^n}
\end{align}
of Clifford gates $C_j$, multi-qubit Pauli matrices $P_j \in \mathcal P_n$ and angles $\phi_j \in \mathbb R$.
To extract computation results, the final state may be measured using one or more commuting Pauli operators $Q_j$.

Our goal is to eliminate all Clifford unitaries from the circuit, leaving only Pauli-generated rotations and Pauli measurements.
This is achieved by commuting Clifford gates toward the end of the circuit and absorbing their action into the subsequent Pauli operators, using the identity
\begin{equation} 
C^\dagger e^{i \phi P} C = e^{i \phi (C^\dagger PC)},
\end{equation}
for unitary $C$.
Explicitly, the transformation proceeds as: 
\begin{equation}
C_{k+1} \prod_{j=1}^k e^{i \phi_j P_j} C_j \ket{0^n} = \paren*{\prod_{\ell = 1}^{k+1}C_\ell} \prod_{j=1}^k e^{i \phi_j P_j'} \ket{0^n},
\end{equation}  
where 
\begin{equation}
P_j' \coloneqq \paren*{\prod_{\ell = 1}^j C_\ell}^\dagger P_j \paren*{\prod_{\ell = 1}^j C_\ell}
\end{equation} 
Finally, the global Clifford $C' \coloneqq \prod_j C_j$ is commuted past the final measurements, modifying them according to $Q_j' = (C')^\dagger Q_j C'$.
Since only the measurement outcomes are of interest and $C'$ merely relabels the observables,
the Clifford gate $C'$ need not be physically implemented.

\clearpage
\subsection{Logical capability comparison details}
\label{app:logical-capability}
\sisetup{%
    exponent-mode=scientific,
    round-mode=places,
    round-precision=2,
}%

Here we describe how we estimated the regions which indicate the class of algorithm which can be implemented with a given capability (used in \cref{fig:logical-circuit-capabilities}),
then how we obtained our capability estimates for the surface code architecture,
and finally how the resources for simulating the transverse-field Ising model (TFIM) were computed.

\begin{figure}[htb]
\centering
\includegraphics[width=0.9\columnwidth]{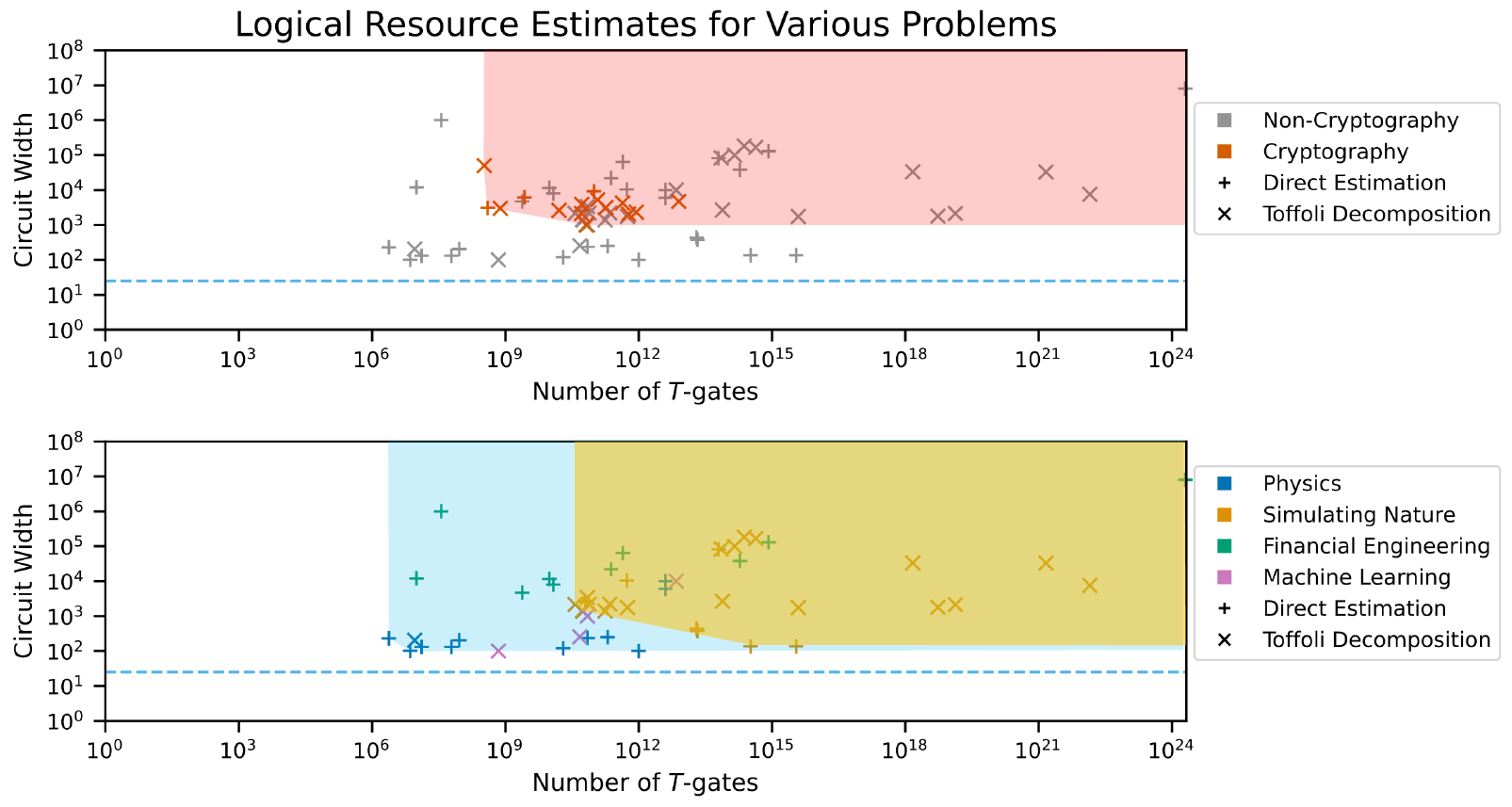}
\caption{%
We reproduce \cite[Figure~2]{scholten2024assessing}
and crudely draw the convex hulls (colored overlay) relevant for different classes of algorithm that are used in \cref{fig:logical-circuit-capabilities}.
The dashed line can be ignored.
}\label{fig:assessing-benefits-data}
\end{figure}

\paragraph{Application class regions}
Here we describe our approach to identify the application regions depicted in \cref{fig:logical-circuit-capabilities}.
\textcite{scholten2024assessing} produce a scatter plot of various algorithm instances of different classes of problem, and label these points by algorithm class.
We take the convex hull of the data points for each of three classes: cryptography, simulating nature, and physics (which we call cryptography, quantum chemistry, and scientific toy models, respectively) in \cref{fig:assessing-benefits-data}.
We further extend each convex hull to infinity for more qubits or T gates since having access to more logical qubits or T gates cannot decrease the computational capability.
The names of the regions are justified by the class of problems the extreme data points of the convex hull represent:
for Physics the extreme points are spin-system simulations and a 2D transverse-field Ising model simulation;
for Simulating Nature these are quantum chemistry applications.

We also mark the region with fewer than 75 qubits or fewer than 75 T gates as being accessible to classical simulation.
As for all the regions we highlight, the boundary of this is somewhat approximate as classical algorithms continue to improve.
For example, in 2016 Ref.~\cite{bravyi2016improved} reported a simulation of 40 qubits and 50 T gates, while in 2019 Ref.~\cite{bravyi2019simulation} reported a simulation of 50 qubits and 60 T gates. 
Drawing the limit of 75 qubits and 75 T gates seems to be in the ballpark of what can expected to be accessible by classically simulations for the foreseeable future without major algorithmic advances or further structure in the quantum circuit.

\paragraph{Capability estimates for the surface code architecture}
To complement the capability estimates for the gross and two-gross codes, we estimate the capabilities of a surface code architecture.
Similar to \cref{sec:results}, we compute the length, $k$, of a circuit of random Pauli-generated rotations, $\mathcal C$, (with angle $\frac{\pi}{4}$)
such that the surface code architecture has circuit failure probability less than $\frac{1}{3}$.

We make similar assumptions to those described in \cite{beverland2022assessing}, namely that each patch requires $2d^2$ physical qubits, that $P_1=0.03 \left(\frac{p}{0.01}\right)^{d/2}$ predicts the per-cycle error rate of each surface code patch~\cite{fowler2012surface,wang2011surface}, and that an arbitrary multi-qubit logical Pauli measurement (as required for one Pauli-generated rotation in $\mathcal C$) can be implemented in $d$ QEC cycles, using about half the patches for routing.
For simplicity, we assume a single T factory equal to the comparable bicycle architecture that uses $f$ physical qubits (see \cref{tab:Tfactories}),
and we will overestimate the logical capability of the surface code architecture by neglecting any noise introduced by (waiting for) the T factory.
Given these assumptions, all $q' = q-f$ physical qubits are used to build $\floor{q'/(2d^2)}$ surface code patches of distance $d$, encoding $n = \floor{q'/(4d^2)}$ logical qubits (with half the patches reserved for routing). 
As all $2n$ patches are involved in implementing a Pauli-generated rotation, we expect a logical failure rate of $2 n P_1 d$ for each T state consumed.
Therefore, the logical failure probability when producing $N_T$ T states is approximately $2 N_T n P_1 d$
and the surface code architecture succeeds with logical failure probability at most $\frac{1}{3}$ if
\begin{equation}\label{eq:surfaceCodeCapability}
    N_T \le \frac{\frac{1}{3}}{2 n P_1 d}.
\end{equation}
We compute surface code entries in \cref{tab:failure} using \cref{eq:surfaceCodeCapability},
choosing $d$ to minimize the difference with the T gate capabilities with the gross or two-gross code in a logarithmic scale,
as in
\begin{equation}
    \min_d \lvert \log(N_T^{\text{surface}}(d)) - \log(N_T^{\text{bicycle}}) \rvert,
\end{equation}
with $N_T^{\text{surface}}(d)$ following from \cref{eq:surfaceCodeCapability} and $N_T^{\text{bicycle}}$ from our results for the gross and two-gross code in \cref{tab:failure}.
Selected data points from \cref{tab:failure} are plotted in \cref{fig:logical-circuit-capabilities}.

We associate the $10^{-3}$ and $10^{-4}$ gross factories (\cref{tab:Tfactories}) with surface codes of distance $d<11$ and
the two-gross factories for the surface code architectures with larger distance.
The time to per rotation is computed as the duration of the T factory plus the time it takes to perform lattice surgery, $\bar \tau_{\text{factory}} + 6d$.

\begin{figure}[tb]
    \centering
    \includegraphics{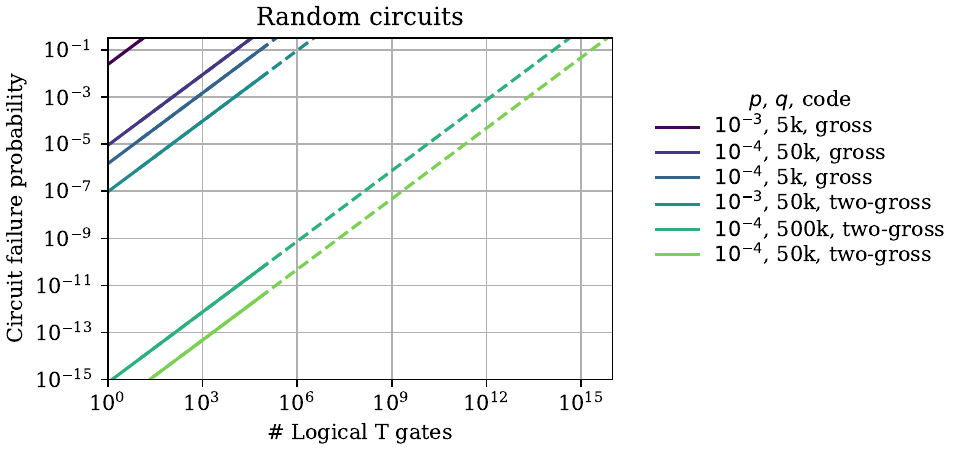}
    \caption{
        Estimating the bicycle architecture capabilities using random circuits.
        We plot the cumulative mean failure probability (8 trials) after generating at most $10^5$ random Pauli-generated rotations with angle $\frac{\pi}{4}$.
        The standard deviation is visually indistinguishable, so is not plotted.
        For larger numbers of T gates, we extrapolate using a linear fit (dashed) whose intersection with $\frac{1}{3}$ is given in \cref{tab:failure}.
    }
    \label{fig:errorRate}
\end{figure}

\paragraph{Resource estimation: transverse-field Ising model}
The transverse-field Ising model (TFIM) for a $\sqrt{n_L}\times\sqrt{n_L}$ sized system with nearest neighbor interactions, has a Hamiltonian
\begin{equation}
    H \coloneqq \underbrace{-J\sum_{j \sim \ell} Z_j Z_\ell}_{A} + \underbrace{g\sum_{j \in [n_L]} X_j}_{B},
\end{equation}
where $j \sim \ell$ ranges over indices $j, \ell \in \mathbb [n_L]$ such that $j$ and $\ell$ are adjacent in the $\sqrt{n_L} \times \sqrt{n_L}$ 2D-grid.
The problem of simulating the TFIM equals implementing an evolution by $e^{iHt}$, for time $t$.

Following \cite{beverland2020lower},
we apply a product formula decomposition (or Trotter-Suzuki decomposition) to approximate $e^{iHt}$.
We first split the evolution into $T \in \mathbb N$ steps
such that, for $\Delta \coloneqq \frac{t}{T}$, we have $e^{iHt} = \paren{e^{iH\Delta}}^T$.
Then, we choose fourth order product formulas, $U_4(\Delta)$, to simulate the Hamiltonian
that are defined as
\begin{align}
    U_2(\Delta) &\coloneqq e^{-iA \Delta/2}e^{-iB \Delta}e^{-iA \Delta/2}, \\
    U_4(\Delta) &\coloneqq U_2(\gamma \Delta) U_2(\gamma \Delta) U_2((1-4\gamma) \Delta) U_2(\gamma \Delta) U_2(\gamma \Delta) = e^{-iH \Delta} + \bigo*{\Delta^5}, \\
    \text{for } \gamma &\coloneq (4 - 4^{1/3})^{-1},
\end{align}
and approximate the evolution by $H$ for time $\Delta$.
Since the terms in $A$ and $B$ commute, we can exactly decompose $U_4(\Delta)$ into Pauli-generated rotations using
\begin{align}
    e^{-iA \frac{\Delta}{2}} &= \prod_{j \sim \ell} e^{i J \frac{\Delta}{2} Z_jZ_\ell} & &\text{and} & e^{-iB \Delta} &= \prod_{j \in [n_L]} e^{-i \Delta g X_j}.
\end{align}

We choose the same concrete parameters as~\cite{microsoft2022resource}.
We take $n_L = 100$ sites and a total evolution time of $t=20$ with $\Delta = 0.25$ (so $T = 80$ steps),
as well as TFIM parameters $J=g=1$.\footnote{%
    Note that there is a discrepency between the paper \cite{beverland2020lower} (which states $T=20$ steps are used) and the associated code~\cite{microsoft2022resource} (which states $T=80$ steps are used).
    We follow the code~\cite{microsoft2022resource}, which in addition to $T=80$ also specifies $J=g=1$.
}
The resulting circuit $\mathcal C$ consists of $N_R \coloneqq 184\,000$ Pauli-generated rotations.
We apply the compilation algorithm described in \cref{sec:results} to $\mathcal C$
with the small-angle synthesis error set to $5 \times 10^{-9} \approx 10^{-3} / N_R$.
We augment the compiled circuit assuming the worst-case timing information (\cref{tab:logical_ops}),
which is $10^{-3}$ two-gross,
since it is not a significant contributor to the logical error rate.
We list the number of gates and idle cycles in \cref{tab:tfimGateCounts}.
{%
\begin{table}[htb]
    \centering
    \sisetup{
        table-format=1.4e1,
        table-auto-round,
    }
    \begin{tabular}{lS[table-format=1.4e2]SSSS} \toprule
        code & $N_I$ & $N_U$ & $N_M$ & $N_C$ & $N_T$ \\ \midrule
        gross &  41634378765 & 2270400 & 4215240 & 946800 & 15542400 \\
        two-gross & 41636714298 & 2294400 & 4238828 & 946800 & 15542400 \\ \bottomrule
    \end{tabular}
    \caption{Gate counts of TFIM for each code where $N_i$ ranges over idle and bicycle instructions, $\mathcal I \cup \set{I}$.
        The gate count differs between the two codes because the decomposition of Pauli measurements is different, see \cref{fig:overhead}.}
    \label{tab:tfimGateCounts}
\end{table}%
}

We then compute the resulting failure probability for the bicycle architecture using gross and two-gross codes at various physical error rates, $p$, and compare to the surface code architecture.
We use the fits in \cref{app:logical-op-simulations} to obtain the logical error rates
and, for the two-gross, supplement missing values by using the assumed logical error rates listed in \cref{tab:logical_ops}.
The resulting logical error rates and failure probabilities rates of the augmented circuit (using \cref{eq:error-budget}) are listed in \cref{tab:logicalErrorRate}.
We then compute the intersection of the interpolating curve (of the form $a x^b$, for $a,b \in\mathbb R$) with the target failure probability $10^{-3}$.
For the bicycle architecture using the gross code, we use the data points at $p=3\times 10^{-5}$ and $p=10^{-5}$
and obtain a necessary physical error rate of $p \le \num{2.0855E-05}$.
By interpolating, we see that the bicycle architecture with two-gross codes achieves a failure probability of $10^{-3}$
when the physical error rate is bounded by $p \le \num{7.3220E-04}$.

\begin{table}[htb]
    \centering
    \sisetup{
        exponent-mode=scientific,
        table-format = 1.1e-2,
        table-auto-round,
    }
    \begin{tabular}{lS[table-format=1e-1]SSSSS[table-format=1.2e-2]} \toprule%
        code & $p$ & $P_I$ & {$P_U$} & $P_M$ & $P_C$ & {failure probability} \\ \midrule
        gross & 1e-3 & 1.61e-09 & 4.01e-07 & 1.11e-05 & 0.00201 & 33258.93337461165 \\
         & 1e-4 & 1.44e-15 & 6.07e-13 & 1.01e-09 & 4.81e-08 & 0.7974506221710217 \\
         & 5e-5 & 4.46e-17 & 2.16e-14 & 5.74e-11 & 2.79e-09 & 0.04624877775057292 \\
         & 3e-5 & 3.54e-18 & 1.92e-15 & 5.9e-12 & 3.18e-10 & 0.005268591620036828 \\
         & 1e-5 & 1.51e-20 & 9.53e-18 & 3.29e-14 & 2.1e-12 & 3.4766673348943354e-05 \\
        two-gross & 1e-3 & 8.2e-21 & 3.25e-15 & 1e-11 & 1e-09 & 0.016531603535021058 \\
         & 1e-4 & 5.39e-39 & 1.34e-37 & 1e-20 & 1e-18 & 1.6531588280000003e-11 \\ \bottomrule
    \end{tabular}
    \caption{%
        Estimated logical error rate of the bicycle architecture using gross and two-gross codes for each of the bicycle instructions at various $p$
        and the resulting circuit failure probability for TFIM simulation.
        We extrapolate the worst-case error rate, $P_i$, for the idle, $I$, and most bicycle instructions $i \in \mathcal I$
        using \cref{app:logical-op-simulations},
        otherwise we use values from \cref{tab:logical_ops}.
        We set $P_T = P_C$ since the T factory can be chosen to distill states of sufficient fidelity.
        Finally, we multiply with the instruction counts, $N_i$, from \cref{tab:tfimGateCounts} to obtain the logical failure probability.
    }%
    \label{tab:logicalErrorRate}
\end{table}

We obtain the resource estimates for the surface code architecture as follows.
First, we note that the same method for synthesizing a single-qubit Pauli-generated rotation $Q(\varphi)$
as in \cref{sec:small-angle-rotations}
can be applied once the surface code architecture has prepared a global entangled state.
Therefore, we set the requirement for the physical error rate, $p$, for the surface code architecture to be slightly different from \cref{eq:surfaceCodeCapability}
and is the solution to
\begin{equation}
    10^{-3} \ge P_1d(2nN_R + N_T - N_R)
\end{equation}
where we assume two surface code patches need to perform lattice surgery to inject a T state
and neglect the contributions from the T factory (just as we do for the bicycle architecture).

We list the surface code architectures that can implement TFIM simulation at physical error rates comparable to the bicycle architecture.
The surface codes architectures with a sufficient physical error rate comparable to $\num{2.0855E-05}$ (from gross codes) are
distance $d=7$ with a physical error rate bounded by
$p \le \num{1.3290E-05}$,
and
distance $d=9$ with a physical error rate bounded by
$p \le \num{5.4762E-05}$.
Whereas surface code architectures with a sufficient physical error rate comparable to $\num{7.3220E-04}$ (from two-gross codes)
are distance $d=17$
with a physical error rate bounded by $p \le \num{5.8915E-04}$ 
and distance $d=19$ with physical error rate bounded by $p \le \num{7.8450E-04}$.

\end{document}